\documentclass[prd,showpacs,floatfix,amsmath,amssymb,floatfix]{revtex4}
\usepackage{amsfonts}
\usepackage{mathrsfs}
\usepackage{amssymb}
\usepackage{amsmath}
\usepackage{pifont}
\usepackage{extarrows}
\usepackage{slashed}
\usepackage{curves}
\usepackage{color}
\usepackage{lscape}
\usepackage{multirow}
\usepackage{makecell}
\usepackage{indentfirst}

\begin{document}

\title{Mass and axial charge of heavy baryons}

\author{Nan Jiang}\email{Jiangn@pku.edu.cn}
\author{Xiao-Lin Chen}\email{chenxl@pku.edu.cn}
\author{Shi-Lin Zhu}\email{zhusl@pku.edu.cn}

\affiliation
{$^1$Department of Physics and State Key Laboratory of Nuclear Physics
and Technology, Peking University, Beijing 100871, China\\
$^2$Collaborative Innovation Center of Quantum Matter, Beijing 100871, China}



\begin{abstract}

We investigate the antitriplet and sextet heavy baryon systems with
$J^P= \frac{1}{2}^+, \frac{3}{2}^+$ in the framework of the heavy
baryon chiral perturbation theory. We first calculate the
chiral corrections to the heavy baryon mass from the SU(3) flavor
breaking effect up to $O(p^3)$. Then we extend the same formalism to
calculate the chiral corrections to the axial charges of the heavy
baryons in the isospin symmetry limit.

\end{abstract}

\maketitle

\section{Introduction}\label{sec1}

With the discovery of the Higgs Bosons, all particles within the
standard model were established. However, the low energy behavior of
the strong interaction remains extremely challenging due to the
complicated infrared structure of quantum chromodynamics (QCD).

The discovery of the $J/\psi$ in 1974
\cite{Aubert:1974js,Augustin:1974xw} opened the door to a new world
of the heavy hadrons, which was accompanied by observation of the
$\Upsilon$ family of mesons in 1977 \cite{Herb:1977ek,Innes:1977ae}.
Decades of research leads a wealth of experimental data on heavy
baryons. The discovery of charm and bottom baryon states greatly
enriches our knowledge of heavy quarks and heavy baryons. Meanwhile,
the heavy baryons are not the simple copies of the known light
baryons. The properties of the heavy flavor baryons and light flavor
baryons are quite distinctive because the heavy baryon mass is much
larger than the QCD scale $\Lambda_{\textrm{QCD}}\approx400$ MeV.

Compared with the light mesons and baryons, the heavy flavored
hadron systems containing a single heavy quark are particularly
interesting. The structure of a heavy quark meson is very similar to a hydrogen atom in QED, which contains a heavy particle and a light one. A heavy meson can be regarded as a hydrogen atom of QCD. Likewise, a
heavy baryon is similar to a helium atom. In this sense, the research of heavy baryons will provide
us a more accurate test for QCD. In fact, there exists the additional heavy quark spin
and flavor symmetry when the heavy quark mass goes to infinity. The
observables can be expanded in terms of $1/m_Q$ where $m_Q$ is the
heavy quark mass. In addition, another motivation of the present work is that the nonanalytic corrections derived from the loop diagrams might reduce the error of extraction in the lattice calculation. We hope our results will be useful in the chiral extrapolation of the lattice simulation data.

In this work we focus on the heavy baryons containing a single heavy
quark with either C=1 or B=1. The ground states satisfy the SU(4)
symmetry in flavor space and form multiplets 20 (for
spin-$\frac{1}{2}$ baryons) or 20'(for spin-$\frac{3}{2}$ baryons).
The spin-$\frac{1}{2}$ baryons include the octet (C=0),
antitriplet(C=1), sextet (C=1), and triplet (C=2). The
spin-$\frac{3}{2}$ baryons include the decuplet (C=0), sextet (C=1),
triplet (C=2) and a singlet (C=3). The investigations of the mass,
lifetime, and axial charge of the heavy baryons will help us
understand the underlying structure of heavy baryons.

When the two light quarks within the ground state heavy baryon are
in the flavor antitriplet, the quantum number of the heavy baryon
is $J^P=\frac{1}{2}^+$. When the two light quarks are in the
symmetric flavor sextet, the quantum number of the heavy baryon can
be either $J^P=\frac{1}{2}^+$ or $\frac{3}{2}^+$. In recent
years, many charmed and bottomed baryons were observed
experimentally \cite{Nakamura:2010zzi}. The spectroscopic properties of some charm baryons were explored by CDF Collaboration \cite{Wick:2011jn} \cite{Aaltonen:2011sf}. On the other hand, the
scattering lengths of heavy baryons with Goldstone bosons were
calculated in Refs. \cite{Liu:2012uw,Liu:2012sw}. The possible
deuteronlike hadronic molecular states composed of two heavy
baryons were investigated in Ref. \cite{Li:2012bt}. The pionic
coupling constants of the heavy baryons played an important role in
the above work.

In this work, we will investigate the heavy antitriplet and sextet
systems. We will calculate the chiral corrections to the heavy
baryon mass and axial charge from the SU(3) flavor breaking effects
by employing HBChPT. We adopt the heavy quark limit and discard all
the recoil corrections. We include the corrections up to $O(p^3)$
from both the strong and electromagnetic interaction. There were
many references on the chiral corrections to the axial currents of
the nucleon octet
\cite{Jenkins:1990jv,Jenkins:1991es,Zhu:2000zf,Zhu:2002tn}. We adopt
the same approach to study the axial charges of the heavy baryons.

This paper is organized as follows: In Sec. II, we introduce the
effective chiral Lagrangians at the leading order. In Sec. III, we
calculate the chiral corrections to the masses of the antitriplet
and sextet baryon systems. In Sec. IV, we calculate the chiral
corrections to the axial charges in the isospin symmetry limit. The
last section is a short summary.

\section{The HBChPT Formalism}\label{sec2}

The approximate chiral symmetry and its spontaneous breaking play an
important role in the low energy hadron interaction. Chiral
perturbation theory (ChPT) \cite{Weinberg:1978kz} provides a
systematic expansion of the physical observables in terms of small
momentum $p$ and the mass of Goldstone bosons $m$. In fact, ChPT has
been widely used to study the lowenergy hadron interaction.

In the early stage, ChPT was employed to study the purely mesonic
system \cite{Gasser:1983yg,Gasser:1984gg}. Later it was extended to
discuss the baryon-meson system
\cite{Gasser:1987rb,Jenkins:1990jv,Bernard:1992qa,Bernard:1992nc}.
At the lowest order, the couplings between the baryon and
pseudoscalar mesons ($\pi$,K,$\eta$) are solely governed by chiral
dynamics. With the consistent power counting scheme, one can
construct the effective Lagrangians of the meson baryon system and
calculate physical quantities order by order
\cite{Scherer:2002tk,Bernard:1995dp}.

In order to deal with the heavy baryon system, the heavy baryon
chiral perturbation theory (HBChPT) was developed
\cite{Bernard:1996gq,Mojzis:1997tu,Fettes:1998ud,Fettes:2001cr},
which provides a convenient framework to make a dual expansion in
terms of both the small momentum and $\frac{1}{M}$, where $M$ is the
heavy baryon mass. On the other hand, the infrared regularization
scheme was introduced to preserve both the power counting and
analyticity in the framework of the relativistic baryon ChPT
\cite{Becher:1999he}. In this work, we use HBChPT to investigate the
heavy baryon systems.

For the heavy baryons multiplet (Qqq), only the two light quarks
participate in the flavor transformation with the heavy quark Q
acting as a spectator. The pseudoscalar meson fields and
spin-$\frac{1}{2}$ baryon multiplets are defined as follows
\[\phi=\left(
  \begin{array}{ccc}
  \pi^0+\frac{1}{\sqrt{3}}\eta&\sqrt{2}\pi^+&\sqrt{2}K^+\\
  \sqrt{2}\pi^-&-\pi^0+\frac{1}{\sqrt{3}}\eta&\sqrt{2}K^0\\
  \sqrt{2}K^-&\sqrt{2}\bar{K}^0&-\frac{2}{\sqrt{3}}\eta\\
  \end{array}
\right)\]
\begin{eqnarray}\label{eq1}
B_{\bar{3}}=\left(
                \begin{array}{ccc}
                0&\Lambda_c^+&\Xi_c^+\\
                -\Lambda_c^+&0&\Xi_c^0\\
                -\Xi_c^+&-\Xi_c^0&0\\
                \end{array}
              \right),\quad
B_{6}=\left(
                \begin{array}{ccc}
                \Sigma_c^{++}&\frac{1}{\sqrt{2}}\Sigma_c^+&\frac{1}{\sqrt{2}}\Xi_c'^+\\
                \frac{1}{\sqrt{2}}\Sigma_c^+&\Sigma_c^0&\frac{1}{\sqrt{2}}\Xi_c'^0\\
                \frac{1}{\sqrt{2}}\Xi_c'^+&\frac{1}{\sqrt{2}}\Xi_c'^0&\Omega_c^0\\
                \end{array}
              \right)
\end{eqnarray}
The spin-$\frac{3}{2}$ baryons $B_6^{*\mu}$ are the so-called
Rarita-Schwinger vector-spinor fields \cite{Johnson:1960vt}, which
are similar to $B_6$. We adopt the nonlinear realization of the
chiral symmetry and its spontaneous breaking and introduce the
following building blocks.
\begin{eqnarray}\label{eq2}
U(x)&=&e^{\frac{i}{F_0}\phi(x)},\quad u^2=U\nonumber\\
\Gamma_\mu&=&\frac{1}{2}(u^\dagger\partial_\mu u+u\partial_\mu u^\dagger)\nonumber\\
u_\mu&=&\frac{i}{2}(u^\dagger\partial_\mu u-u\partial_\mu u^\dagger)\nonumber\\
D_\mu B&=&\partial_\mu B+\Gamma_\mu B+B\Gamma_\mu^T
\end{eqnarray}
The superscript $T$ denotes the transpose in the flavor space. The
pion decay constant $F_0\approx92.4$ MeV. The leading order
pseudoscalar meson and heavy baryon Lagrangian is $O(p^1)$, which
reads \cite{Yan:1992gz}
\begin{eqnarray}
\mathcal{L}_0^{(1)}&=&\frac{1}{2}tr[\bar{B}_{\bar{3}}(i\slashed D-M_{\bar{3}})B_{\bar{3}}]+tr[\bar{B}_6(i\slashed D-M_6)B_6]\nonumber\\
&&+tr\{\bar{B}_6^{*\mu}[-g_{\mu\nu}(i\slashed D-M_6^*)+i(\gamma_\mu D_\nu+\gamma_\nu D_\mu)-\gamma_\mu(i\slashed D+M_6^*)\gamma_\nu]B_6^{*\nu}\}\label{eq3}\\
\mathcal{L}_{\textrm{int}}^{(1)}&=&g_1tr(\bar{B}_6\slashed u\gamma_5B_6)+g_2[tr(\bar{B}_6\slashed u\gamma_5B_{\bar{3}})+h.c.]+g_3[tr(\bar{B}_6^{*\mu}u_\mu B_6)+h.c.]\nonumber\\
&&+g_4[tr(\bar{B}_6^{*\mu}u_\mu
B_{\bar{3}})+h.c.]+g_5tr(\bar{B}_6^{*\nu}\slashed
u\gamma_5B_{6\nu}^*)+g_6tr(\bar{B}_{\bar{3}}\slashed
u\gamma_5B_{\bar{3}})\label{eq4}
\end{eqnarray}
where $\mathcal{L}_0^{(1)}$ is the free part and
$\mathcal{L}_{\textrm{int}}^{(1)}$ contains the interaction at
$O(p^1)$. From the quark model and flavor SU(3) symmetry, the axial
coupling constants $g_1=0.98$ \cite{Liu:2012uw,Liu:2012sw}, $g_1=-\sqrt{\frac{8}{3}}g_2$
\cite{Yan:1992gz}. The heavy quark spin flavor symmetry leads to the
following relations among these coupling constants, i.e.,
$g_3=\frac{\sqrt{3}}{2}g_1,g_5=-\frac{3}{2}g_1,g_4=-\sqrt{3}g_2$
\cite{Yan:1992gz}. Within the antitriplet, the total angular
momentum of the two light quarks is zero. The conservation of the
angular moment and parity forbids the coupling of pseudoscalar
mesons with the antitriplet heavy baryons, hence $g_6=0$. However,
we keep the $g_6$-related terms in formulas till the numerical
analysis.

The heavy baryon formulation of ChPT consists in an expansion in terms of
$\frac{k}{4\pi F_0}$ and $\frac{k}{\overset{\circ}{M_B}}$, where $k$ is the
small residual component of external baryon.
In the framework of HBChPT, the baryon field $B$ is decomposed into
the large (or light) component $\mathcal{N}$ and the small (or
heavy) component $\mathcal{H}$. By using the path integral theory,
the small component can be integrated out. Thus, the reduced
effective Lagrangian only relies on the large component
\cite{Scherer:2002tk}. Their relationships are
\begin{eqnarray*}
&&B=e^{-imv\cdot x}(\mathcal{N}+\mathcal{H})\\
&&\mathcal{N}=e^{imv\cdot x}\frac{1+\slashed v}{2}B,\quad\mathcal{H}=e^{imv\cdot x}\frac{1-\slashed v}{2}B,
\end{eqnarray*}
where $v^\mu$ is the on shell velocity. For the spin-$\frac{3}{2}$ baryon, the large
component is denoted as $\mathcal{T}^\mu$. In the heavy quark limit,
the nonrelativistic Lagrangian reads
\begin{eqnarray}\label{eq5}
\hat{\mathcal{L}}^{(1)}&=&\frac{1}{2}tr[\bar{\mathcal{N}}_{\bar{3}}(iv\cdot D)\mathcal{N}_{\bar{3}}]+tr[\bar{\mathcal{N}}_6(iv\cdot D-\delta_1)\mathcal{N}_6]+tr\{\bar{\mathcal{T}}^\rho[-g_{\rho\sigma}(iv\cdot D-\delta_2)]\mathcal{T}^\sigma\}\nonumber\\
&&+2g_1tr(\bar{\mathcal{N}}_6S\cdot u\mathcal{N}_6)+2g_2[tr(\bar{\mathcal{N}}_6S\cdot u\mathcal{N}_{\bar{3}})+h.c.]+g_3[tr(\bar{\mathcal{T}}^\mu u_\mu \mathcal{N}_6)+h.c.]\nonumber\\
&&+g_4[tr(\bar{\mathcal{T}}^\mu u_\mu
\mathcal{N}_{\bar{3}})+h.c.]+2g_5tr(\bar{\mathcal{T}}^\nu S\cdot
u\mathcal{T}_\nu)+2g_6tr(\bar{\mathcal{N}}_{\bar{3}}S\cdot
u\mathcal{N}_{\bar{3}})
\end{eqnarray}
The mass difference parameters are defined as
$\delta_1=M_6-M_{\bar3}, \delta_2=M_{6^*}-M_6$. In the isospin
symmetry limit, $\delta_1=126.52$ MeV, $\delta_2=67.03$ MeV. Here,
we choose the average mass of the spin-$\frac{1}{2}$ antitriplet,
sextet and spin-$\frac{3}{2}$ sextet baryons as $M_{\bar3}=2286.46$
MeV, $M_6=2454.02$ MeV, $M_{6^*}=2518.4$ MeV respectively
\cite{Nakamura:2010zzi}.

The SU(3) flavor symmetry-breaking (SB) Lagrangian at $O(p^2)$ reads
\begin{eqnarray}\label{eq6}
\mathcal{L}^{(2)}_{bc}&=&b_1\textrm{tr}(\bar{B}_6\chi_+B_6)+b_5g_{\mu\nu}\textrm{tr}(\bar{B}^{*\mu}_6\chi_+B^{*\nu}_6)+b_6\textrm{tr}(\bar{B}_{\bar3}\chi_+B_{\bar3})\nonumber\\
&&+c_1\textrm{tr}(\bar{B}_6B_6)\textrm{tr}\chi_++c_5g_{\mu\nu}\textrm{tr}(\bar{B}^{*\mu}_6B^{*\nu}_6)\textrm{tr}\chi_++c_6\textrm{tr}(\bar{B}_{\bar3}B_{\bar3})\textrm{tr}\chi_+
\end{eqnarray}
where
\[\chi=2B_0\mathcal{M},\quad\mathcal{M}=\textrm{diag}(m_u,m_d,m_s)\]
\[\chi_+=u^\dagger\chi u^\dagger+u\chi^\dagger u=2\chi+O(\phi^2)\]
$m_u,m_d,m_s$ are the $u$, $d$, $s$ quark mass. The constant $B_0$
is related to the quark condensate. The method of constructing the
chiral effective Lagrangian can be found in Ref.
\cite{Fettes:2000gb}.

The mass splitting not only arises from the up and down quark mass
difference but also from the different heavy baryon electric charges
within an isospin multiplet. The QED Lagrangian at $O(p^2)$ reads
\begin{eqnarray}
\mathcal{L}^{(2)66}_{\textrm{QED}}&=&e^{66}_1\textrm{tr}(\bar B_6Q_+^2B_6)+e^{66}_2\textrm{tr}(\bar B_6Q_+B_6)\textrm{tr}Q_++e^{66}_3\textrm{tr}(\bar B_6B_6)\textrm{tr}Q_+^2\nonumber\\
&&+e^{66}_4\textrm{tr}(\bar B_6Q_+B_6Q_+^T)\label{eq9}\\
\mathcal{L}^{(2)\bar3\bar3}_{\textrm{QED}}&=&e^{\bar3\bar3}_1\textrm{tr}(\bar B_{\bar3}Q_+^2B_{\bar3})+e^{\bar3\bar3}_2\textrm{tr}(\bar B_{\bar3}Q_+B_{\bar3})\textrm{tr}Q_++e^{\bar3\bar3}_3\textrm{tr}(\bar B_{\bar3}B_{\bar3})\textrm{tr}Q_+^2\label{eq10}\\
\mathcal{L}^{(2)6^*6^*}_{\textrm{QED}}&=&e^{6^*6^*}_1g_{\rho\sigma}\textrm{tr}(\bar B_6^{*\rho}Q_+^2B_6^{*\sigma})+e^{6^*6^*}_2g_{\rho\sigma}\textrm{tr}(\bar B_6^{*\rho}Q_+B_6^{*\sigma})\textrm{tr}Q_++e^{6^*6^*}_3g_{\rho\sigma}\textrm{tr}(\bar B_6^{*\rho}B_6^{*\sigma})\textrm{tr}Q_+^2\nonumber\\
&&+e^{6^*6^*}_4g_{\rho\sigma}\textrm{tr}(\bar
B_6^{*\rho}Q_+B_6^{*\sigma}Q_+^T)\label{eq11}
\end{eqnarray}
where
\[Q_+=\frac{1}{2}(u^\dagger Qu+uQu^\dagger)\]
\[Q=2q_l+q_cI=e\ \textrm{diag}(2,0,0)\]
\[q_l=e\ \textrm{diag}(\frac{2}{3},-\frac{1}{3},-\frac{1}{3}),\quad q_c=\frac{2}{3}e\]

To some extent, the above effective Lagrangians mimic the
electromagnetic spin-flavor interaction in the quark model, which
arises from the hard photon exchange between two constituent quarks
and is the important source of the isospin symmetry breaking.

\section{The Heavy Baryon Mass}\label{sec3}

In the framework of HBChPT, the correction to the self energy of the
heavy baryons from the explicit flavor SU(3) breaking terms is
$O(p^2)$. The one-loop chiral correction appears at $O(p^3)$.

\subsection{The Counterterms}\label{sec3.1}

In the ChPT framework, the
divergence from the loop diagram is absorbed by the counterterms at
the same or lower orders. These counterterms come from Eq.(\ref{eq6}). All the self-energy functions $\Sigma$ of counterterms are listed as follows:
\begin{eqnarray*}
 \Sigma_{bc}(\Sigma_c^0)&=&4 b_1 B_0 m_d+4 B_0 c_1 m_d+4 B_0 c_1 m_s+4 B_0 c_1 m_u \\
 \Sigma_{bc}(\Xi_c'^0)&=&2 b_1 B_0 m_d+4 B_0 c_1 m_d+2 b_1 B_0 m_s+4 B_0 c_1 m_s+4 B_0 c_1 m_u \\
 \Sigma_{bc}(\Omega_c^0)&=&4 B_0 c_1 m_d+4 b_1 B_0 m_s+4 B_0 c_1 m_s+4 B_0 c_1 m_u \\
 \Sigma_{bc}(\Sigma_c^+)&=&2 b_1 B_0 m_d+4 B_0 c_1 m_d+4 B_0 c_1 m_s+2 b_1 B_0 m_u+4 B_0 c_1 m_u \\
 \Sigma_{bc}(\Xi_c'^+)&=&4 B_0 c_1 m_d+2 b_1 B_0 m_s+4 B_0 c_1 m_s+2 b_1 B_0 m_u+4 B_0 c_1 m_u \\
 \Sigma_{bc}(\Sigma_c^{++})&=&4 B_0 c_1 m_d+4 B_0 c_1 m_s+4 b_1 B_0 m_u+4 B_0 c_1 m_u
\end{eqnarray*}
\begin{eqnarray*}
 \Sigma_{bc}(\Xi_c^0)&=&4 b_6 B_0 m_d+8 B_0 c_6 m_d+4 b_6 B_0 m_s+8 B_0 c_6 m_s+8 B_0 c_6 m_u \\
 \Sigma_{bc}(\Lambda_c^+)&=&4 b_6 B_0 m_d+8 B_0 c_6 m_d+8 B_0 c_6 m_s+4 b_6 B_0 m_u+8 B_0 c_6 m_u \\
 \Sigma_{bc}(\Xi_c^+)&=&8 B_0 c_6 m_d+4 b_6 B_0 m_s+8 B_0 c_6 m_s+4 b_6 B_0 m_u+8 B_0 c_6 m_u
\end{eqnarray*}
\begin{eqnarray*}
 \Sigma_{bc}(\Sigma_c^{*0})&=&-4 b_5 B_0 m_d-4 B_0 c_5 m_d-4 B_0 c_5 m_s-4 B_0 c_5 m_u \\
 \Sigma_{bc}({\Xi_c^*}'^0)&=&-2 b_5 B_0 m_d-4 B_0 c_5 m_d-2 b_5 B_0 m_s-4 B_0 c_5 m_s-4 B_0 c_5 m_u \\
 \Sigma_{bc}(\Omega_c^{*0})&=&-4 B_0 c_5 m_d-4 b_5 B_0 m_s-4 B_0 c_5 m_s-4 B_0 c_5 m_u \\
 \Sigma_{bc}(\Sigma_c^{*+})&=&-2 b_5 B_0 m_d-4 B_0 c_5 m_d-4 B_0 c_5 m_s-2 b_5 B_0 m_u-4 B_0 c_5 m_u \\
 \Sigma_{bc}({\Xi_c^*}'^+)&=&-4 B_0 c_5 m_d-2 b_5 B_0 m_s-4 B_0 c_5 m_s-2 b_5 B_0 m_u-4 B_0 c_5 m_u \\
 \Sigma_{bc}(\Sigma_c^{*++})&=&-4 B_0 c_5 m_d-4 B_0 c_5 m_s-4 b_5 B_0 m_u-4 B_0 c_5 m_u
\end{eqnarray*}

The mass splitting of the isospin multiplets mainly arises from QED
effects and the mass difference between up and down quarks. Light
quarks have different charges. In fact, the tree-level QED
correction starts at $O(p^2)$. They also act as the counterterms.
All the self-energy functions $\Sigma$ of QED counterterms are listed as follows:
\begin{eqnarray*}
 \Sigma_{\textrm{QED}}(\Xi_c'^0)&=&4 e^2 e_3^{66} \\
 \Sigma_{\textrm{QED}}(\Sigma_c^0)&=&4 e^2 e_3^{66} \\
 \Sigma_{\textrm{QED}}(\Omega_c^0)&=&4 e^2 e_3^{66} \\
 \Sigma_{\textrm{QED}}(\Xi_c'^+)&=&2 e^2 e_1^{66}+2 e^2 e_2^{66}+4 e^2 e_3^{66} \\
 \Sigma_{\textrm{QED}}(\Sigma_c^+)&=&2 e^2 e_1^{66}+2 e^2 e_2^{66}+4 e^2 e_3^{66} \\
 \Sigma_{\textrm{QED}}(\Sigma_c^{++})&=&4 e^2 e_1^{66}+4 e^2 e_2^{66}+4 e^2 e_3^6+4 e^2 e_4^{66}
\end{eqnarray*}
\begin{eqnarray*}
 \Sigma_{\textrm{QED}}(\Xi_c^0)&=&8 e^2 e_3^{\bar{3}\bar{3}} \\
 \Sigma_{\textrm{QED}}(\Lambda_c^+)&=&4 e^2 e_1^{\bar{3}\bar{3}}+4 e^2 e_2^{\bar{3}\bar{3}}+8 e^2 e_3^{\bar{3}\bar{3}}\\
 \Sigma_{\textrm{QED}}(\Xi_c^+)&=&4 e^2 e_1^{\bar{3}\bar{3}}+4 e^2 e_2^{\bar{3}\bar{3}}+8 e^2 e_3^{\bar{3}\bar{3}}
\end{eqnarray*}
\begin{eqnarray*}
 \Sigma_{\textrm{QED}}({\Xi_c^*}'^0)&=&-4 e^2 e_3^{6^*6^*} \\
 \Sigma_{\textrm{QED}}(\Sigma_c^{*0})&=&-4 e^2 e_3^{6^*6^*} \\
 \Sigma_{\textrm{QED}}(\Omega_c^{*0})&=&-4 e^2 e_3^{6^*6^*} \\
 \Sigma_{\textrm{QED}}({\Xi_c^*}'^+)&=&-2 e^2 e_1^{6^*6^*}-2 e^2 e_2^{6^*6^*}-4 e^2 e_3^{6^*6^*} \\
 \Sigma_{\textrm{QED}}(\Sigma_c^{*+})&=&-2 e^2 e_1^{6^*6^*}-2 e^2 e_2^{6^*6^*}-4 e^2 e_3^{6^*6^*} \\
 \Sigma_{\textrm{QED}}(\Sigma_c^{*++})&=&-4 e^2 e_1^{6^*6^*}-4 e^2 e_2^{6^*6^*}-4 e^2 e_3^{6^*6^*}-4 e^2 e_4^{6^*6^*}
\end{eqnarray*}

\subsection{Loop contribution}\label{sec3.2}

The lowest order loop correction is $O(p^3)$ where the interaction
vertex in Fig. 1 arises from $\hat{\mathcal{L}}^{(1)}$. The
single line represents a spin-$\frac{1}{2}$ baryon and double line a
spin-$\frac{3}{2}$ baryon. In the computation of the Feynman
diagrams, we need the spin projection operators
$P^{\frac{3}{2}}_{(33)\mu\nu}$ of the spin-$\frac{3}{2}$ heavy
baryons \cite{Benmerrouche:1989uc,Hemmert:1997ye,Pilling:2004cu}.
Some properties of the spin projection operator and the
Pauli-Lubanski spin operator $S^\mu$ are collected in Appendix \ref{sec6.4}.

The self-energy function can be written as
\begin{eqnarray}
\Sigma_{6,\bar3,\textrm{I}}&=&C_{6,\bar3}(A_{6,\bar3,\textrm{I}}+B_{6,\bar3,\textrm{I}}\epsilon)f(m,\omega)\nonumber\\
\Sigma_{6,\bar3,\textrm{II}}&=&C_{6,\bar3}(A_{6,\bar3,\textrm{II}}+B_{6,\bar3,\textrm{II}}\epsilon)f(m,\omega)\\
\Sigma_{6^*,\textrm{I}}&=&C_{6^*}(A_{6^*,\textrm{I}}+B_{6^*,\textrm{I}}\epsilon)f(m,\omega)\nonumber\\
\Sigma_{6^*,\textrm{II}}&=&C_{6^*}(A_{6^*,\textrm{II}}+B_{6^*,\textrm{II}}\epsilon)f(m,\omega)
\end{eqnarray}
The function $f(m,\omega)$ is defined in Appendix \ref{sec6.3}.
The parameters $A$ and $B$ are related to the dimension
$d=4-\epsilon$ in the dimensional regularization. For the
spin-$\frac{1}{2}$ particles
$A_\textrm{I}+B_\textrm{I}\epsilon=-\frac{1}{4},A_\textrm{II}+B_\textrm{II}\epsilon=\frac{d-2}{d-1}$.
For the spin-$\frac{3}{2}$ particles,
$A_\textrm{I}+B_\textrm{I}\epsilon=\frac{-(d+1)(d-3)}{4(d-1)^2},A_\textrm{II}+B_\textrm{II}\epsilon=\frac{1}{d-1}$.
The coefficients C are listed in Table \ref{Coeff.sel} in the
appendix.

\begin{figure}[h]
\caption{The one-loop Feynman diagrams that contribute to the
self energy.}\label{Figure-self}
\begin{center}
\includegraphics{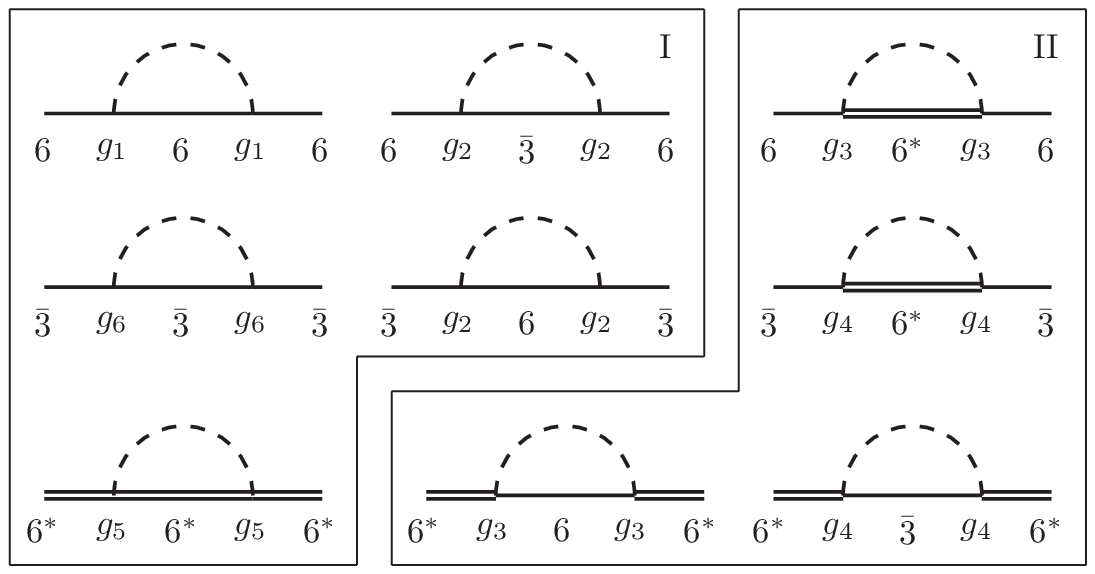}
\end{center}
\end{figure}

We calculate the loop contribution for each type of diagram listed
in Fig. \ref{Figure-self} separately. The intermediate and
external baryons have the same spin for the type-I loops while their
spin is different for the type-II loops. Throughout our calculation,
we use the $\overline{\textrm{MS}}$ (modified minimal subtraction)
scheme. The simplest case corresponds to $\delta_1=\delta_2=0$,
where the self-energy correction has a very simple form,
\[\Sigma^{\textrm{loop}}\propto m_\phi^3\]
where $m_\phi$ is the pseudoscalar meson mass.

The QED correction may also appear at $O(p^3)$ from the photon loop.
With
\[r_\mu=l_\mu=-Q\mathcal{A}_\mu\]
where $Q$ is the charge operator and $\mathcal{A}_\mu$ represents
the photon field, the chiral connection $\Gamma_\mu$ is modified as
follows
\begin{eqnarray}\label{eq12}
\Gamma_\mu&=&\frac{1}{2}(u^\dagger\partial_\mu u+u\partial_\mu u^\dagger)-\frac{i}{2}(u^\dagger r_\mu u+ul_\mu u^\dagger)\\
&=&\Gamma_\mu^0+\Gamma_\mu^{r,l}\nonumber
\end{eqnarray}
At the lowest order
\[\Gamma_\mu^{\textrm{QED}}=i\mathcal{A}_\mu Q+O(\phi^2),\]
\begin{eqnarray}\label{eq13}
\hat{\mathcal{L}}_{\textrm{con,QED}}=-v^\mu \mathcal{A}_\mu
tr\bar{\mathcal{N}}(Q\mathcal{N}+\mathcal{N}Q^T)
\end{eqnarray}
The correction to the self energy from the following photon loop
vanishes with the infrared regularization as pointed out in Ref.
\cite{Guo:2008ns}.
\begin{center}
\includegraphics{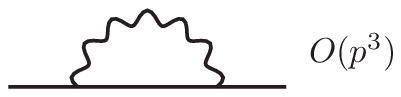}
\end{center}
The charge operator may also enter $u_\mu$.
\begin{eqnarray}\label{eq14}
u_\mu&=&\frac{i}{2}(u^\dagger\partial_\mu u-u\partial_\mu u^\dagger)+\frac{i}{2}[u^\dagger(-ir_\mu)u-u(-il_\mu)u^\dagger]\\
&=&u_\mu^0+u_\mu^{r,l}\nonumber
\end{eqnarray}
However, its contribution is of higher order. Numerically, the QED effects are very small since the QED Lagrangian can be expanded both in terms of the chiral order and the fine-structure constant $\alpha$, which is also a small number. If we calculate the correction up to $O(p^4)$, the higher order of $\alpha$ should also be considered, whose contributions will be much smaller.

To sum up, the heavy baryon mass reads
\begin{equation}\label{eq15}
M=\overset{\circ}{M}+\Sigma_{bc}+\Sigma_{\textrm{QED}}+\Sigma^{\textrm{loop}}
\end{equation}
where $\overset{\circ}{M}$ is the bare mass without the chiral
corrections.

\subsection{Numerical results}\label{sec3.3}

\begin{table}[!h]
  \centering
  \caption{Chiral loop corrections to the heavy baryon masses in unit of MeV with $\delta=0$.}\label{sel1}
  \textrm{The marked entries are the predictions.}
\begin{tabular}{c|ccc|c|cc}
\Xhline{1.2pt}
&\multicolumn{3}{c|}{Loop contribution}&\multicolumn{3}{c}{Heavy Baryon Masses}\\
\cline{2-7}
&case I&case II&case I+II&Experimental data&Fit 1 for case I&Fit 2 for case I+II\\
\Xhline{1.2pt}
$M_{\Sigma_c^{++}}$&$-305.00$&$-100.87$&$-405.87$&$2454.02\pm0.18$&$2449.04\pm2.1^\ddag$&$2448.38\pm2.1^\ddag$\\
$M_{\Sigma_c^+}$&   $-307.91$&$-101.83$&$-409.74$&$2452.9 \pm0.4 $&$2452.90\pm0.4$      &$2452.90\pm0.4$\\
$M_{\Sigma_c^0}$&   $-310.68$&$-102.50$&$-413.18$&$2453.76\pm0.18$&$2456.90\pm2.1^\ddag$&$2457.85\pm2.1^\ddag$\\
$M_{\Xi_c'^+}$&     $-512.58$&$-177.86$&$-690.44$&$2575.6 \pm3.1 $&$2573.14\pm2.1^\ddag$&$2574.83\pm2.1^\ddag$\\
$M_{\Xi_c'^0}$&     $-516.23$&$-180.30$&$-696.53$&$2577.9 \pm2.9 $&$2576.25\pm2.1^\ddag$&$2577.13\pm2.1^\ddag$\\
$M_{\Omega_c^0}$&   $-722.19$&$-258.91$&$-981.09$&$2695.2 \pm1.7 $&$2695.20\pm1.7$      &$2695.20\pm1.7$\\
\hline\hline
$M_{\Lambda_c^+}$&$-108.81$&$-217.63$&$ -326.44$&$2286.46\pm0.14$         &$2286.46\pm0.14$       &$2286.46\pm0.14$\\
$M_{\Xi_c^+}$&    $-358.79$&$-717.57$&$-1076.36$&$2467.8 ^{+0.4}_{-0.6}$  &$2467.80^{+0.4}_{-0.6}$&$2467.80^{+0.4}_{-0.6}$\\
$M_{\Xi_c^0}$&    $-362.44$&$-724.89$&$-1087.33$&$2470.88^{+0.34}_{-0.80}$&$2473.14\pm0.65^\ddag$ &$2476.23\pm0.65^\ddag$\\
\hline\hline
$M_{\Sigma_c^{*++}}$&$-252.18$&$-153.69$&$-405.87$&$2518.4\pm0.6$        &$2513.29\pm4.3^\ddag$&$2512.92\pm4.3^\ddag$\\
$M_{\Sigma_c^{*+}}$& $-254.58$&$-155.17$&$-409.74$&$2517.5\pm2.3$        &$2517.50\pm2.3$      &$2517.50\pm2.3$\\
$M_{\Sigma_c^{*0}}$& $-256.25$&$-156.93$&$-413.18$&$2518.0\pm0.5$        &$2522.44\pm4.3^\ddag$&$2522.51\pm4.3^\ddag$\\
$M_{\Xi_c'^{*+}}$&   $-444.65$&$-245.78$&$-690.44$&$2645.9^{+0.5}_{-0.6}$&$2644.76\pm4.3^\ddag$&$2642.45\pm4.3^\ddag$\\
$M_{\Xi_c'^{*0}}$&   $-450.75$&$-245.78$&$-696.53$&$2645.9\pm0.5$        &$2645.18\pm4.3^\ddag$&$2644.82\pm4.3^\ddag$\\
$M_{\Omega_c^{*0}}$& $-647.27$&$-333.83$&$-981.10$&$2765.9\pm2.0$        &$2765.90\pm2.0$      &$2765.90\pm2.0$\\
\Xhline{1.2pt}
$b_1'$&$ $&$ $&$ $&$ $&$-1.69$                   &$-2.10$\\
$c_1'$&$ $&$ $&$ $&$ $&$-6.27\pm4.8\times10^{-3}$&$-6.49\pm4.8\times10^{-3}$\\
\hline\hline
$b_6'$&$ $&$ $&$ $&$ $&$-1.12$                   &$-2.43$\\
$c_6'$&$ $&$ $&$ $&$ $&$-2.71\pm0.7\times10^{-3}$&$-2.91\pm0.7\times10^{-3}$\\
\hline\hline
$b_5'$&$ $&$ $&$ $&$ $&$-1.65$                   &$-2.11$\\
$c_5'$&$ $&$ $&$ $&$ $&$-6.30\pm9.9\times10^{-3}$&$-6.64\pm9.9\times10^{-3}$\\
\Xhline{1.2pt}
\end{tabular}
\end{table}

In our numerical analysis, the LECs $b$ and $c$ are replaced by the
dimensionless parameters $b'$ and $c'$, which are defined in Eqs.
(\ref{eq16})--(\ref{eq17}),
\begin{eqnarray}
&&b_{1,5}'=B_0b_{1,5},\quad c_{1,5}'=B_0c_{1,5}+\frac{\overset{\circ}{M}_{6,6^*}}{4(m_u+m_d+m_s)}\label{eq16}\\
&&b_{6}'=B_0b_{6},\quad
c_{6}'=B_0c_{6}+\frac{\overset{\circ}{M}_{\bar3}}{8(m_u+m_d+m_s)}\label{eq17}
\end{eqnarray}

With the experimental values of the heavy baryon masses as input
\cite{Nakamura:2010zzi}, we extract the values of the LECs $b'$s
and $c'$s. Fit 1 corresponds to the case of including the type-I
loop corrections only.
$b_1'=-1.69,c_1'=-6.27,b_6'=-1.12,c_6'=-2.71,b_5'=-1.65,c_5'=-6.30$.
Fit 2 contains both the type-I and type-II loop corrections.
$b_1'=-2.10,c_1'=-6.49,b_6'=-2.43,c_6'=-2.91,b_5'=-2.11,c_5'=-6.64$.
We also list the contribution of the type-I loop and sum of type-I
and type-II loops in the second to fourth columns of Table \ref{sel1}
explicitly. The heavy baryon masses with the notation $\ddag$ in the
fourth to fifth columns are the predicted values. The errors are also listed in the table. Some of them are numerically
very small, which are omitted in the table. The mass splitting between
the spin-$\frac{1}{2}$ charmed baryons was also discussed in Ref.
\cite{Guo:2008ns}.

The LECs at $O(p^2)$ were estimated in Refs.
\cite{Liu:2012uw,Liu:2012sw}, where the SU(3) flavor symmetry-breaking Lagrangian reads
\[\mathcal{L}^{(2)}=c_1\textrm{tr}\bar B_6\tilde{\chi}_+B_6+\bar{c}_1\textrm{tr}\bar B_{\bar{3}}\tilde{\chi}_+B_{\bar{3}}+\cdots\]
with
\[\tilde{\chi}_+=\chi_+-\frac{1}{3}\textrm{tr}\chi_+\]
The authors first constructed the flavor SU(4) Lagrangian. Then they
reduced the SU(4) Lagrangian into the SU(3) form. In this way, they
estimated the LECs. Using the Gell-Mann-Okubo relation, the LECs
extracted in Refs. \cite{Liu:2012uw,Liu:2012sw} correspond to the
following values of $b$'s
\[b'_{1,5}=B_0b_{1,5}\approx-2.30\]
\[b'_6=B_0b_6\approx-1.22\]
These values are consistent with the above values extracted from
fitting to experimental data in this work.

The spin and flavor representation of the external and intermediate
baryons may be different. Their mass splitting will contribute to
the self energy through the chiral loop. Such corrections are quite
important in the nucleon octet and $\Delta$ decuplet case. We
consider three cases. Fit 3 corresponds to the case when the type-I
loop correction is included with $\delta\neq0$. Fit 4 includes both
types of loop corrections with $\delta\neq0$. Fit 5 corresponds to
the inclusion of both types of loop corrections with $\delta\neq0$
and QED effects.

We collect the fit results Table \ref{sel2}. Comparing the fourth
column in Table \ref{sel1}--\ref{sel2}, we notice that the
loop contributions are suppressed after considering the mass
difference $\delta$. On the other hand, the absolute value of parameters $b$ and
$c$ becomes slightly smaller. $b_1'=-2.05$, $c_1'=-6.44$, $b_6'=-2.37$, $c_6'=-2.87$, $b_5'=-1.95$, $c_5'=-6.50$.

Even if the QED effects are not considered, the up and down quark mass difference will cause the isospin breaking. The experimental baryon masses always contain the isospin breaking. When we consider QED corrections, more LECs contribute to the
self energy. The LECs $e^{66}_3,e^{6^*6^*}_3,e^{\bar3\bar3}_3$ can
be absorbed by $c_1'$, $c_5'$, $c_6'$.
\begin{eqnarray}
&&c_{1,5}'=B_0c_{1,5}+\frac{\overset{\circ}{M}_{6,6^*}+e^2e^{66,6^*6^*}_3}{4(m_u+m_d+m_s)}\label{eq18}\\
&&c_{6}'=B_0c_{6}+\frac{\overset{\circ}{M}_{\bar3}+e^2e^{\bar3\bar3}_3}{8(m_u+m_d+m_s).}\label{eq19}
\end{eqnarray}
In this case,
$b_1'=-2.06,c_1'=-6.43,b_6'=-2.37,c_6'=-2.86,b_5'=-1.97,c_5'=-6.49$.
The values of $e^2(e_1+e_2)$ and $e^2e_4$ are given in Table
\ref{sel2}. The heavy baryon masses with $\ddag$ in the fifth to seventh
columns are the predicted values.

In the isospin symmetry limit, the divergences from the loop
diagrams can be absorbed by the LECs (or counterterm) at $O(p^2)$.
However, the low energy constants at $O(p^2)$ are not enough to
cancel the divergences from the loop diagrams at $O(p^3)$ when we
consider the SU(2)-breaking corrections, which was also pointed out
in Ref. \cite{Guo:2008ns}. Chiral symmetry ensures that the
divergences will be absorbed by the counterterms at the higher
order if we treat the SU(2) symmetry-breaking terms as the higher
order correction. In Ref. \cite{Guo:2008ns}, the authors studied the spin-$\frac{1}{2}$ baryons. In our work, we studied both the spin-$\frac{1}{2}$ and -$\frac{3}{2}$ heavy baryons. In Ref. \cite{Guo:2008ns}, the authors focused on the mass splitting only and used the experimental mass splitting as input with the infrared regularization scheme. In our work, we calculated all the possible chiral corrections to the heavy baryon masses up to $O(p^3)$ and used the values of experimental mass as input within the framework of heavy baryon ChPT.

\begin{table}[!h]
  \centering
  \caption{Chiral loop corrections to the heavy baryon masses in units of MeV
with $\delta\neq0$ and QED effects.}\label{sel2}
\textrm{The marked entries are the predictions.}
\begin{tabular}{c|ccc|cc|c}
\Xhline{1.2pt}
&\multicolumn{3}{c|}{Loop contribution $\Sigma^{\textrm{loop}}$}&\multicolumn{3}{c}{Mass of baryons}\\
\cline{2-7}
&case I&case II&case I+II&Fit 3 for case I&Fit 4 for case I+II&Fit 5 for case I+II with QED\\
\Xhline{1.2pt}
$M_{\Sigma_c^{++}}$&$-278.36$&$-103.34$&$-381.70$&$2449.56\pm2.1^\ddag$&$2448.91\pm2.1^\ddag$&$2454.02\pm0.18$\\
$M_{\Sigma_c^+}$&   $-281.59$&$-104.32$&$-385.91$&$2452.90\pm0.4$      &$2452.90\pm0.4$      &$2452.90\pm0.4$\\
$M_{\Sigma_c^0}$&   $-283.95$&$-104.93$&$-388.88$&$2457.11\pm2.1^\ddag$&$2458.13\pm2.1^\ddag$&$2453.76\pm0.18$\\
$M_{\Xi_c'^+}$&     $-480.54$&$-181.43$&$-661.97$&$2569.42\pm2.1^\ddag$&$2570.48\pm2.1^\ddag$&$2572.66\pm2.46^\ddag$\\
$M_{\Xi_c'^0}$&     $-484.24$&$-183.82$&$-668.06$&$2572.30\pm2.1^\ddag$&$2572.59\pm2.1^\ddag$&$2570.40\pm2.46^\ddag$\\
$M_{\Omega_c^0}$&   $-676.81$&$-262.29$&$-939.09$&$2695.20\pm1.7$      &$2695.20\pm1.7$      &$2695.20\pm1.7$\\
\hline\hline
$M_{\Lambda_c^+}$&$-111.09$&$-177.57$&$ -288.66$&$2286.46\pm0.14$        &$2286.46\pm0.14$       &$2286.46\pm0.14$\\
$M_{\Xi_c^+}$&    $-367.68$&$-650.46$&$-1018.14$&$2467.80^{+0.4}_{-0.6}$ &$2467.80^{+0.4}_{-0.6}$&$2467.80^{+0.4}_{-0.6}$\\
$M_{\Xi_c^0}$&    $-371.24$&$-657.02$&$-1028.26$&$2473.36\pm0.65^\ddag  $&$2476.66\pm0.65^\ddag$ &$2470.88^{+0.38}_{-0.80}$\\
\hline\hline
$M_{\Sigma_c^{*++}}$&$-252.18$&$ -92.06$&$-344.25$&$2513.29\pm4.3^\ddag$&$2513.88\pm4.3^\ddag$&$2518.40\pm0.6$\\
$M_{\Sigma_c^{*+}}$& $-254.58$&$ -93.87$&$-348.45$&$2517.50\pm2.3$      &$2517.50\pm2.3$      &$2517.50\pm2.3$\\
$M_{\Sigma_c^{*0}}$& $-256.25$&$ -94.84$&$-351.09$&$2522.44\pm4.3^\ddag$&$2522.68\pm4.3^\ddag$&$2518.0 \pm0.5$\\
$M_{\Xi_c'^{*+}}$&   $-444.65$&$-161.97$&$-606.63$&$2644.67\pm4.3^\ddag$&$2634.50\pm4.3^\ddag$&$2636.83\pm5.4^\ddag$\\
$M_{\Xi_c'^{*0}}$&   $-450.75$&$-162.15$&$-612.89$&$2645.18\pm4.3^\ddag$&$2636.05\pm4.3^\ddag$&$2633.71\pm5.4^\ddag$\\
$M_{\Omega_c^{*0}}$& $-647.27$&$-210.96$&$-858.22$&$2765.90\pm2.0$      &$2765.90\pm2.0$      &$2765.90\pm2.0$\\
\Xhline{1.2pt}
$b_1'$                  &$ $&$ $&$ $&$-1.64$                   &$-2.05$                   &$-2.06$\\
$c_1'$                  &$ $&$ $&$ $&$-6.21\pm4.8\times10^{-3}$&$-6.44\pm4.8\times10^{-3}$&$-6.43\pm5.6\times10^{-3}$\\
$e^2(e^{66}_1+e^{66}_2)$&$ $&$ $&$ $&$ $&$ $&$-2.21$\\
$e^2e^{66}_4$           &$ $&$ $&$ $&$ $&$ $&$-0.19$\\
\hline\hline
$b_6'$                                  &$ $&$ $&$ $&$-1.14$&$-2.37$&$-2.37$\\
$c_6'$                                  &$ $&$ $&$ $&$-2.71\pm0.7\times10^{-3}$&$-2.87\pm0.7\times10^{-3}$&$-2.86\pm1.4\times10^{-3}$\\
$e^2(e^{\bar3\bar3}_1+e^{\bar3\bar3}_2)$&$ $&$ $&$ $&$ $&$ $&$-1.44$\\
\hline\hline
$b_5'$                          &$ $&$ $&$ $&$-1.65$&$-1.95$&$-1.97$\\
$c_5'$                          &$ $&$ $&$ $&$-6.30\pm9.9\times10^{-3}$&$-6.50\pm9.9\times10^{-3}$&$-6.49\pm0.0124$\\
$e^2(e^{6^*6^*}_1+e^{6^*6^*}_2)$&$ $&$ $&$ $&$ $&$ $&$-2.36$\\
$e^2e^{6^*6^*}_4$               &$ $&$ $&$ $&$ $&$ $&$0.04$\\
\Xhline{1.2pt}
\end{tabular}
\end{table}
\begin{table}[!h]
  \centering
  \caption{The decay width of the heavy baryons in units of MeV.}\label{width}
\begin{tabular}{cccc}
\Xhline{1.2pt}
&Experimental data&   case I loop & case II loop \\
\Xhline{1.2pt}
$\Gamma_{\Sigma_c^{++}}$&$2.23$&$2.60$&$$\\
$\Gamma_{\Sigma_c^+}$&$<4.6$&$3.20$&$$\\
$\Gamma_{\Sigma_c^0}$&$2.2$&$2.60$&$$\\
\Xhline{1.2pt}
$\Gamma_{\Sigma_c^{*++}}$&$14.9$&$$&$8.10$\\
$\Gamma_{\Sigma_c^{*+}}$&$<17$&$$&$8.96$\\
$\Gamma_{\Sigma_c^{*0}}$&$16.1$&$$&$8.10$\\
$\Gamma_{\Xi_c'^{*+}}$&$<3.1$&$$&$6.28$\\
$\Gamma_{\Xi_c'^{*0}}$&$<5.5$&$$&$6.28$\\
\Xhline{1.2pt}
\end{tabular}
\end{table}

In the above analysis, $\delta$ is the difference of the average
mass between baryons in the different representations. In the
derivation of the imaginary part of the loop diagrams, one should be
cautious about the choice of $\delta$. For example, the process
$\Sigma_c\rightarrow\Lambda_c^++\pi$ is forbidden if we choose the
average value:
$\delta_{\Sigma_c\Lambda_c^+}=M_{\Sigma_c}-M_{\Lambda_c^+}=126.52$MeV
$<M_{\pi}$. To avoid such a paradox, we used the experimental mass
as input to calculate $\delta_{\Sigma_c\Lambda_c^+}$ and the imaginary
part of the self energy. For all the other processes, $\delta$ takes
the average value. We collect the experimental and theoretical width
$\Gamma$ in Table \ref{width}. These values are consistent with
experimental data.

\section{The Axial Charge of the Heavy Baryon}\label{sec4}

The baryon axial charge is a very important physical observable,
which can be measured through semileptonic decays. In this section,
we will explore the chiral corrections to the axial charges $g_1$ to
$g_6$ in Eq. (\ref{eq4}). At the leading order, the axial currents
are determined by chiral symmetry entirely. At $O(p^2)$, the loop
contributions arise from the vertex correction and wave function
renormalization while the correction from the chiral connection
vanishes in the heavy quark limit $M_c\rightarrow\infty$.

\subsection{The axial currents on tree level}\label{sec4.1}

The axial currents at the tree level can be obtained from Eq.
(\ref{eq5}). With the external source $r^\mu, l^\mu$ in Eqs.
(\ref{eq12}) and (\ref{eq14})
\[r_\mu=\frac{\lambda^a}{2}r^a_\mu,\quad l_\mu=\frac{\lambda^a}{2}l^a_\mu\]
where $\lambda^a$ is the Gell-Mann generator in the flavor space,
the difference of the chiral currents $R^{a,\mu}$ and $L^{a,\mu}$
leads to the axial current
\[A^{a,\mu}=R^{a,\mu}-L^{a,\mu}\]
The axial currents arising from the chiral connection and $O(p)$
interaction terms are
\begin{eqnarray}
A^{a,\mu}_{\textrm{con}}(\bar3)&=&\frac{1}{8}v^\mu\textrm{tr}[\bar{\mathcal{N}}_{\bar3}(u^\dagger\lambda^au-u\lambda^au^\dagger)\mathcal{N}_{\bar3}+\bar{\mathcal{N}}_{\bar3}\mathcal{N}_{\bar3}(u^\dagger\lambda^au-u\lambda^au^\dagger)^T]\nonumber\\
A^{a,\mu}_{\textrm{con}}(6)&=&\frac{1}{4}v^\mu\textrm{tr}[\bar{\mathcal{N}}_6(u^\dagger\lambda^au-u\lambda^au^\dagger)\mathcal{N}_6+\bar{\mathcal{N}}_6\mathcal{N}_6(u^\dagger\lambda^au-u\lambda^au^\dagger)^T]\nonumber\\
A^{a,\mu}_{\textrm{con}}(6^*)&=&\frac{1}{4}v^\mu\textrm{tr}\{-g_{\rho\sigma}[\bar{\mathcal{T}}^\rho(u^\dagger\lambda^au-u\lambda^au^\dagger)\mathcal{T}^\sigma+\bar{\mathcal{T}}^\rho\mathcal{T}^\sigma(u^\dagger\lambda^au-u\lambda^au^\dagger)^T]\}\label{eq20}
\end{eqnarray}
\begin{eqnarray}
A^{a,\mu}(g_1)&=&\frac{1}{2}g_1\textrm{tr}[\bar{\mathcal{N}}_6S^\mu(u^\dagger\lambda^au+u\lambda^au^\dagger)\mathcal{N}_6]\nonumber\\
A^{a,\mu}(g_2)&=&\frac{1}{2}g_2\textrm{tr}[\bar{\mathcal{N}}_6S^\mu(u^\dagger\lambda^au+u\lambda^au^\dagger)\mathcal{N}_{\bar3}+h.c.]\nonumber\\
A^{a,\mu}(g_3)&=&\frac{1}{4}g_3\textrm{tr}[\bar{\mathcal{T}}^\mu(u^\dagger\lambda^au+u\lambda^au^\dagger)\mathcal{N}_6+h.c.]\nonumber\\
A^{a,\mu}(g_4)&=&\frac{1}{4}g_4\textrm{tr}[\bar{\mathcal{T}}^\mu(u^\dagger\lambda^au+u\lambda^au^\dagger)\mathcal{N}_{\bar3}+h.c.]\nonumber\\
A^{a,\mu}(g_5)&=&\frac{1}{2}g_5\textrm{tr}[\bar{\mathcal{T}}^\nu
S^\mu(u^\dagger\lambda^au+u\lambda^au^\dagger)\mathcal{T}_\nu]\label{eq21}
\end{eqnarray}
The lowest order axial charges arising from the $g_1-g_5$ terms of
the sextet and antitriplet are collected in Table \ref{g0}, where
we only list the channels allowing the semileptonic decays.

\begin{table}[!h]
  \centering
  \caption{The axial charge $g_{(ij)}^{(0)}$ at the tree level.}\label{g0}
\begin{tabular}{cc|cc}
\Xhline{1.2pt}
\multicolumn{2}{c|}{Flavor $a=1+i2$}&\multicolumn{2}{c}{Flavor $a=4+i5$}\\
\hline\hline
$g_{\Xi_c'^+\Xi_c'^0}$&$g_1$&$g_{\Xi_c'^+\Omega_c^0}$&$\sqrt{2}g_1$\\
$g_{\Sigma_c^+\Sigma_c^0}$&$\sqrt{2}g_1$&$g_{\Sigma_c^+\Xi_c'^0}$&$g_1$\\
&$ $&$g_{\Sigma_c^{++}\Xi_c'^+}$&$\sqrt{2}g_1$\\
\hline\hline
$g_{\Lambda_c^+\Sigma_c^0}$&$2g_2$&$g_{\Lambda_c^+\Xi_c'^0}$&$\sqrt{2}g_2$\\
$g_{\Xi_c^+\Xi_c'^0}$&$\sqrt{2}g_2$&$g_{\Xi_c^+\Omega_c^0}$&$2g_2$\\
&$ $&$g_{\Sigma_c^+\Xi_c^0}$&$-\sqrt{2}g_2$\\
&$ $&$g_{\Sigma_c^{++}\Xi_c^+}$&$-2g_2$\\
\hline\hline
$g_{{\Xi_c^*}'^+{\Xi_c^*}'^0}$&$g_5$&$g_{{\Xi_c^*}'^+\Omega_c^{*0}}$&$\sqrt{2}g_5$\\
$g_{\Sigma_c^{*+}\Sigma_c^{*0}}$&$\sqrt{2}g_5$&$g_{\Sigma_c^{*+}{\Xi_c^*}'^0}$&$g_5$\\
&$ $&$g_{\Sigma_c^{*++}{\Xi_c^*}'^+}$&$\sqrt{2}g_5$\\
\hline\hline
$g_{\Xi_c'^+{\Xi_c^*}'^0}$&$\frac{g_3}{2}$&$g_{\Xi_c'^+\Omega_c^{*0}}$&$\frac{g_3}{\sqrt{2}}$\\
$g_{\Sigma_c^+\Sigma_c^{*0}}$&$\frac{g_3}{\sqrt{2}}$&$g_{\Sigma_c^+{\Xi_c^*}'^0}$&$\frac{g_3}{2}$\\
$g_{\Sigma_c^{++}\Sigma_c^{*+}}$&$\frac{g_3}{\sqrt{2}}$&$g_{\Sigma_c^{++}{\Xi_c^*}'^+}$&$\frac{g_3}{\sqrt{2}}$\\
&$ $&$g_{{\Xi_c^*}'^+\Omega_c^0}$&$\frac{g_3}{\sqrt{2}}$\\
&$ $&$g_{\Sigma_c^{*+}\Xi_c'^0}$&$\frac{g_3}{2}$\\
&$ $&$g_{\Sigma_c^{*++}\Xi_c'^+}$&$\frac{g_3}{\sqrt{2}}$\\
\hline\hline
$g_{\Lambda_c^+\Sigma_c^{*0}}$&$g_4$&$g_{\Lambda_c^+{\Xi_c^*}'^0}$&$\frac{g_4}{\sqrt{2}}$\\
$g_{\Xi_c^+{\Xi_c^*}'^0}$&$\frac{g_4}{\sqrt{2}}$&$g_{\Xi_c^+\Omega_c^{*0}}$&$g_4$\\
\Xhline{1.2pt}
\end{tabular}
\end{table}

The $O(p^0)$ axial current arises from the $O(p)$ Lagrangian. The
$O(p^3)$ SU(3) symmetry-breaking Lagrangian
$\mathcal{L}_{\textrm{counter}}^{(3)}$ contributes to the $O(p^2)$
corrections to the axial current. Moreover, these new vertices will
cancel the infinity from the loop corrections.
\begin{eqnarray}\label{eq22}
&&\mathcal{L}_{\textrm{counter}}^{(3)}\nonumber\\
&=&d_1\textrm{tr}(\bar{B}_6\gamma^\mu\gamma_5\{u_\mu,\chi_+\}B_6)+f_1\textrm{tr}(\bar{B}_6\gamma^\mu\gamma_5u_\mu B_6\chi_+^T)+h_1\textrm{tr}(\bar{B}_6\gamma^\mu\gamma_5u_\mu B_6)\textrm{tr}\chi_+\nonumber\\
&&+d_2\textrm{tr}(\bar{B}_6\gamma^\mu\gamma_5\{u_\mu,\chi_+\}B_{\bar3})+f_2\textrm{tr}(\bar{B}_6\gamma^\mu\gamma_5u_\mu B_{\bar3}\chi_+^T)+h_2\textrm{tr}(\bar{B}_6\gamma^\mu\gamma_5u_\mu B_{\bar3})\textrm{tr}\chi_++h.c.\nonumber\\
&&+d_6\textrm{tr}(\bar{B}_{\bar3}\gamma^\mu\gamma_5\{u_\mu,\chi_+\}B_{\bar3})+f_6\textrm{tr}(\bar{B}_{\bar3}\gamma^\mu\gamma_5u_\mu B_{\bar3}\chi_+^T)+h_6\textrm{tr}(\bar{B}_{\bar3}\gamma^\mu\gamma_5u_\mu B_{\bar3})\textrm{tr}\chi_+\nonumber\\
&&+d_5g_{\rho\sigma}\textrm{tr}(\bar{B}_6^{*\rho}\gamma^\mu\gamma_5\{u_\mu,\chi_+\}B_6^{*\sigma})+f_5g_{\rho\sigma}\textrm{tr}(\bar{B}_6^{*\rho}\gamma^\mu\gamma_5u_\mu B_6^{*\sigma}\chi_+^T)+h_5g_{\rho\sigma}\textrm{tr}(\bar{B}_6^{*\rho}\gamma^\mu\gamma_5u_\mu B_6^{*\sigma})\textrm{tr}\chi_+\nonumber\\
&&+d_3\textrm{tr}(\bar{B}_6^{*\mu}\{u_\mu,\chi_+\}B_6)+f_3\textrm{tr}(\bar{B}_6^{*\mu}u_\mu B_6\chi_+^T)+h_3\textrm{tr}(\bar{B}_6^{*\mu}u_\mu B_6)\textrm{tr}\chi_++h.c.\nonumber\\
&&+d_4\textrm{tr}(\bar{B}_6^{*\mu}\{u_\mu,\chi_+\}B_{\bar3})+f_4\textrm{tr}(\bar{B}_6^{*\mu}u_\mu
B_{\bar3}\chi_+^T)+h_4\textrm{tr}(\bar{B}_6^{*\mu}u_\mu
B_{\bar3})\textrm{tr}\chi_++h.c.
\end{eqnarray}
In HBChPT, the $O(p^2)$ axial current arising from
$\mathcal{L}_{\textrm{counter}}^{(3)}$ is
\begin{eqnarray}\label{eq23}
A_{dfh}^{a,\mu}&=&\frac{d_1}{2}\textrm{tr}(\bar{\mathcal{N}}_6S^\mu\{w^a_+,\chi_+\}\mathcal{N}_6)+\frac{f_1}{2}\textrm{tr}(\bar{\mathcal{N}}_6S^\mu w^a_+ \mathcal{N}_6\chi_+^T)+\frac{h_1}{2}\textrm{tr}(\bar{\mathcal{N}}_6S^\mu w^a_+ \mathcal{N}_6)\textrm{tr}\chi_+\nonumber\\
&&+\frac{d_2}{2}\textrm{tr}(\bar{\mathcal{N}}_6S^\mu\{w^a_+,\chi_+\}\mathcal{N}_{\bar3})+\frac{f_2}{2}\textrm{tr}(\bar{\mathcal{N}}_6S^\mu w^a_+ \mathcal{N}_{\bar3}\chi_+^T)+\frac{h_2}{2}\textrm{tr}(\bar{\mathcal{N}}_6S^\mu w^a_+ \mathcal{N}_{\bar3})\textrm{tr}\chi_++h.c.\nonumber\\
&&+\frac{d_5}{2}g_{\rho\sigma}\textrm{tr}(\bar{\mathcal{T}}^\rho S^\mu\{w^a_+,\chi_+\}\mathcal{T}^\sigma)+\frac{f_5}{2}g_{\rho\sigma}\textrm{tr}(\bar{\mathcal{T}}^\rho S^\mu w^a_+ \mathcal{T}^\sigma\chi_+^T)+\frac{h_5}{2}g_{\rho\sigma}\textrm{tr}(\bar{\mathcal{T}}^\rho S^\mu w^a_+ \mathcal{T}^\sigma)\textrm{tr}\chi_+\nonumber\\
&&+\frac{d_3}{4}\textrm{tr}(\bar{\mathcal{T}}^\mu\{w^a_+,\chi_+\}\mathcal{N}_6)+\frac{f_3}{4}\textrm{tr}(\bar{\mathcal{T}}^\mu w^a_+ \mathcal{N}_6\chi_+^T)+\frac{h_3}{4}\textrm{tr}(\bar{\mathcal{T}}^\mu w^a_+ \mathcal{N}_6)\textrm{tr}\chi_++h.c.\nonumber\\
&&+\frac{d_4}{4}\textrm{tr}(\bar{\mathcal{T}}^\mu\{w^a_+,\chi_+\}\mathcal{N}_{\bar3})+\frac{f_4}{4}\textrm{tr}(\bar{\mathcal{T}}^\mu
w^a_+
\mathcal{N}_{\bar3}\chi_+^T)+\frac{h_4}{4}\textrm{tr}(\bar{\mathcal{T}}^\mu
w^a_+ \mathcal{N}_{\bar3})\textrm{tr}\chi_++h.c.
\end{eqnarray}
where
\[w^a_\pm=u^\dagger\lambda^au\pm u\lambda^au^\dagger\]
The axial charges $g_{(ij)}^{(2)}$ in terms of the coefficients
$d,f,h$ are listed in Table \ref{g2}.

\begin{table}[!h]
  \centering
  \caption{The axial charges $g_{(ij)}^{(2)}$ from the counterterms.}\label{g2}
  \renewcommand{\arraystretch}{1.0}
\begin{tabular}{cc|cc}
\Xhline{1.2pt}
\multicolumn{2}{c|}{Flavor $a=1+i2$}&\multicolumn{2}{c}{Flavor $a=4+i5$}\\
\hline\hline
$g_{\Xi_c'^+\Xi_c'^0}$&$2md_1+f_1m_s+h_1(2m+m_s)$&$g_{\Xi_c'^+\Omega_c^0}$&$\sqrt{2}f_1m_s+d_1(\sqrt{2}m+\sqrt{2}m_s)+h_1(2 \sqrt{2}m+\sqrt{2}m_s)$\\
$g_{\Sigma_c^+\Sigma_c^0}$&$2\sqrt{2}md_1+\sqrt{2}mf_1+h_1(2\sqrt{2}m+\sqrt{2}m_s)$&$g_{\Sigma_c^+\Xi_c'^0}$&$mf_1+d_1 (m+m_s)+h_1(2m+m_s)$\\
&$ $&$g_{\Sigma_c^{++}\Xi_c'^+}$&$\sqrt{2}mf_1+d_1(\sqrt{2}m+\sqrt{2}m_s)+h_1(2\sqrt{2}m+\sqrt{2}m_s)$\\
\hline\hline
$g_{\Lambda_c^+\Sigma_c^0}$&$4md_2+2mf_2+h_2(4m+2m_s)$&$g_{\Lambda_c^+\Xi_c'^0}$&$\sqrt{2}mf_2+d_2(\sqrt{2}m+\sqrt{2}m_s)+h_2 (2\sqrt{2}m+\sqrt{2}m_s)$\\
$g_{\Xi_c^+\Xi_c'^0}$&$2\sqrt{2}md_2+\sqrt{2}f_2m_s+h_2(2\sqrt{2}m+\sqrt{2}m_s)$&$g_{\Xi_c^+\Omega_c^0}$&$\sqrt{2}f_5m_s+d_5 (\sqrt{2}m+\sqrt{2}m_s)+h_5(2\sqrt{2}m+\sqrt{2}m_s)$\\
&$ $&$g_{\Sigma_c^+\Xi_c^0}$&$-\sqrt{2}mf_2-d_2(\sqrt{2}m+\sqrt{2}m_s)-h_2(2\sqrt{2}m+\sqrt{2}m_s)$\\
&$ $&$g_{\Sigma_c^{++}\Xi_c^+}$&$-2mf_2-d_2(2m+2m_s)-h_2(4m+2m_s)$\\
\hline\hline
$g_{{\Xi_c^*}'^+{\Xi_c^*}'^0}$&$2md_5+f_5m_s+h_5(2m+m_s)$&$g_{{\Xi_c^*}'^+\Omega_c^{*0}}$&$\sqrt{2}f_5m_s+d_5(\sqrt{2} m+\sqrt{2}m_s)+h_5(2\sqrt{2}m+\sqrt{2}m_s)$\\
$g_{\Sigma_c^{*+}\Sigma_c^{*0}}$&$2\sqrt{2}md_5+\sqrt{2}mf_5+h_5(2\sqrt{2}m+\sqrt{2}m_s)$&$g_{\Sigma_c^{*+}{\Xi_c^*}'^0}$&$m f_5+d_5(m+m_s)+h_5(2m+m_s)$\\
&$ $&$g_{\Sigma_c^{*++}{\Xi_c^*}'^+}$&$\sqrt{2}mf_5+d_5(\sqrt{2}m+\sqrt{2}m_s)+h_5(2\sqrt{2}m+\sqrt{2}m_s)$\\
\hline\hline
$g_{\Xi_c'^+{\Xi_c^*}'^0}$&$md_3+h_3(m+\frac{m_s}{2})+\frac{f_3m_s}{2}$&$g_{\Xi_c'^+\Omega_c^{*0}}$&$\frac{f_3
m_s}{\sqrt{2}}+d_3(\frac{m}{\sqrt{2}}+\frac{m_s}{\sqrt{2}})+h_3(\sqrt{2}
m+\frac{m_s}{\sqrt{2}})$\\
$g_{\Sigma_c^+\Sigma_c^{*0}}$&$\sqrt{2}md_3+\frac{mf_3}{\sqrt{2}}+h_3(\sqrt{2} m+\frac{m_s}{\sqrt{2}})$&$g_{\Sigma_c^+{\Xi_c^*}'^0}$&$\frac{1}{2}mf_3+d_3(\frac{m}{2}+\frac{m_s}{2})+h_3(m+\frac{ m_s}{2})$\\
$g_{\Sigma_c^{++}\Sigma_c^{*+}}$&$\sqrt{2}md_3+\frac{mf_3}{\sqrt{2}}+h_3(\sqrt{2}
m+\frac{m_s}{\sqrt{2}})$&$g_{\Sigma_c^{++}{\Xi_c^*}'^+}$&$\frac{mf_3}{\sqrt{2}}+d_3
(\frac{m}{\sqrt{2}}+\frac{m_s}{\sqrt{2}})+h_3(\sqrt{2}
m+\frac{m_s}{\sqrt{2}})$\\
&$ $&$g_{{\Xi_c^*}'^+\Omega_c^0}$&$\frac{f_3m_s}{\sqrt{2}}+d_3(\frac{m}{\sqrt{2}}+\frac{m_s}{\sqrt{2}})+h_3(\sqrt{2}
m+\frac{m_s}{\sqrt{2}})$\\
&$ $&$g_{\Sigma_c^{*+}\Xi_c'^0}$&$\frac{mf_3}{2}+d_3(\frac{m}{2}+\frac{m_s}{2})+h_3(m+\frac{m_s}{2})$\\
&$ $&$g_{\Sigma_c^{*++}\Xi_c'^+}$&$\frac{mf_3}{\sqrt{2}}+d_3(\frac{m}{\sqrt{2}}+\frac{m_s}{\sqrt{2}})+h_3(\sqrt{2}
m+\frac{m_s}{\sqrt{2}})$\\
\hline\hline
$g_{\Lambda_c^+\Sigma_c^{*0}}$&$2md_4+mf_4+h_4(2m+m_s)$&$g_{\Lambda_c^+{\Xi_c^*}'^0}$&$\frac{mf_4}{\sqrt{2}}+d_4
(\frac{m}{\sqrt{2}}+\frac{m_s}{\sqrt{2}})+h_4(\sqrt{2}
m+\frac{m_s}{\sqrt{2}})$\\
$g_{\Xi_c^+{\Xi_c^*}'^0}$&$\sqrt{2}md_4+\frac{f_4m_s}{\sqrt{2}}+h_4(\sqrt{2}
m+\frac{m_s}{\sqrt{2}})$&$g_{\Xi_c^+\Omega_c^{*0}}$&$-\frac{mf_4}{\sqrt{2}}-d_4
(\frac{m}{\sqrt{2}}+\frac{m_s}{\sqrt{2}})-h_4(\sqrt{2}
m+\frac{m_s}{\sqrt{2}})$\\
\Xhline{1.2pt}
\end{tabular}
\end{table}

The renormalized matrix elements of the axial currents can be
written as
\begin{eqnarray}
\langle\mathcal{N}_i|A^{a,\mu}(g_{1,2})|\mathcal{N}_j\rangle&=&\bar{u}_iS^\mu u_j(g_{1,2(ij)}^{(0)}+g_{1,2(ij)}^{(2)}+g_{1,2(ij)}^{a}+g_{1,2(ij)}^{b}+g_{1,2(ij)}^{\textrm{Re}})\label{eq24}\\
\langle\mathcal{T}_i^\rho|A^{a,\mu}(g_5)|\mathcal{T}_j^\sigma\rangle&=&g_{\rho\sigma}\bar{u}_i^\rho S^\mu u_j^\sigma(g_{5(ij)}^{(0)}+g_{5(ij)}^{(2)}+g_{5(ij)}^{a}+g_{5(ij)}^{b}+g_{5(ij)}^{\textrm{Re}})\label{eq25}\\
\langle\mathcal{N}_i|A^{a,\mu}(g_{3,4})|\mathcal{T}_j^\mu\rangle&=&\bar{u}_iu_j^\mu(g_{3,4(ij)}^{(0)}+g_{3,4(ij)}^{(2)}+g_{3,4(ij)}^{a}+g_{3,4(ij)}^{b}+g_{3,4(ij)}^{\textrm{Re}})\label{eq26}
\end{eqnarray}
$g_{(ij)}^{a,b}$ etc are the corrections at the one-loop level in
Fig. \ref{Figure-axialabcd}. $g_{(ij)}^{\textrm{Re}}$ etc. arise
from the wave function renormalization.

\subsection{The axial currents correction on loop level}\label{sec4.2}

\begin{figure}[!h]
\caption{Vertex correction}\label{Figure-axialabcd}
\begin{center}
\includegraphics{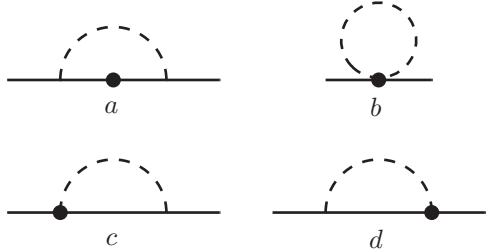}
\end{center}
\end{figure}

At the one-loop level, there are four Feynman diagrams as shown in
Fig. \ref{Figure-axialabcd}, where the filled circle represents the
axial current vertex. Diagrams c and d arise from the chiral
connection in Eq. (\ref{eq20}).

The vertex correction diagram (a) can be classified into three or
four different types according to the Lorentz structure in the loop
integrals, which are displayed in Figs. \ref{fig-g1}--\ref{fig-g4}
in Appendix \ref{sec6.1}. For the
vertex corrections to the axial charges $g_1$ and $g_2$, type I
denotes the case that only the spin-$\frac{1}{2}$ baryons
participate in the intermediate process. Type III contains only the
spin-$\frac{3}{2}$ baryons as intermediate states. Type II contains
both the spin-$\frac{1}{2}$ and spin-$\frac{3}{2}$ baryons. For the
other axial charges, the classification of the vertex correction
diagrams is similar.

With the contraction formulas between the spin projection operator
$P^{\frac{3}{2}}_{(33)\mu\nu}$, Pauli-Lubanski vector $S^\mu$, and
the metric $g_{\mu\nu}$ listed in Appendix \ref{sec6.4}, we obtain the
expressions of the axial currents from the vertex correction diagram
(a).
\begin{eqnarray}\label{eq27}
A^{a,\mu}_{ij}(g_1,g_2)_{\textrm{I}}&=&g^a_{1,2(ij)}\bar u_iS^\mu u_j(a_{1,2,\textrm{I}}+b_{1,2,\textrm{I}}\epsilon)\frac{\Delta f}{\Delta\omega},\quad a_{1,2,\textrm{I}}+b_{1,2,\textrm{I}}\epsilon=\frac{d-3}{4}\frac{-1}{d-1}\nonumber\\
A^{a,\mu}_{ij}(g_1,g_2)_{\textrm{II}}&=&g^a_{1,2(ij)}\bar u_iS^\mu u_j(a_{1,2,\textrm{II}}+b_{1,2,\textrm{II}}\epsilon)\frac{\Delta f}{\Delta\omega},\quad a_{1,2,\textrm{II}}+b_{1,2,\textrm{II}}\epsilon=\frac{2(d-2)}{d-1}\frac{-1}{d-1}\nonumber\\
A^{a,\mu}_{ij}(g_1,g_2)_{\textrm{III}}&=&g^a_{1,2(ij)}\bar u_iS^\mu
u_j(a_{1,2,\textrm{III}}+b_{1,2,\textrm{III}}\epsilon)\frac{\Delta f}{\Delta\omega},\nonumber\\
&&\quad\quad\quad\quad a_{1,2,\textrm{III}}+b_{1,2,\textrm{III}}\epsilon=\frac{(d-3)(d-2)(d+1)}{(d-1)^2}\frac{-1}{d-1}
\end{eqnarray}
\begin{eqnarray}\label{eq28}
A^{a,\mu}(g_5)_{\textrm{I}}&=&g^a_{5(ij)}g_{\rho\sigma}\bar u^\rho_jS^\mu u^\sigma_j(a_{5,\textrm{I}}+b_{5,\textrm{I}}\epsilon)\frac{\Delta f}{\Delta\omega},\quad a_{5,\textrm{I}}+b_{5,\textrm{I}}\epsilon=\frac{-1}{d-1}\\
A^{a,\mu}(g_5)_{\textrm{II}}&=&g^a_{5(ij)}g_{\rho\sigma}\bar u^\rho_jS^\mu u^\sigma_j(a_{5,\textrm{II}}+b_{5,\textrm{II}}\epsilon)\frac{\Delta f}{\Delta\omega},\quad a_{5,\textrm{II}}+b_{5,\textrm{II}}\epsilon=\frac{-2}{d-1}\frac{-1}{d-1}\nonumber\\
A^{a,\mu}(g_5)_{\textrm{III}}&=&g^a_{5(ij)}g_{\rho\sigma}\bar
u^\rho_jS^\mu
u^\sigma_j(a_{5,\textrm{III}}+b_{5,\textrm{III}}\epsilon)\frac{\Delta
f}{\Delta\omega},\quad
a_{5,\textrm{III}}+b_{5,\textrm{III}}\epsilon=\frac{d^3-5d^2+3d-7}{4(d-1)^2}\frac{-1}{d-1}\nonumber
\end{eqnarray}
\begin{eqnarray}\label{eq29}
A^{a,\mu}(g_3,g_4)_{\textrm{I}}&=&g^a_{3,4(ij)}\bar u_iu^\mu_j(a_{3,4,\textrm{I}}+b_{3,4,\textrm{I}}\epsilon)\frac{\Delta f}{\Delta\omega},\quad a_{3,4,\textrm{I}}+b_{3,4,\textrm{I}}\epsilon=(-\frac{1}{2})\frac{-1}{d-1}\\
A^{a,\mu}(g_3,g_4)_{\textrm{II}}&=&g^a_{3,4(ij)}\bar u_iu^\mu_j(a_{3,4,\textrm{II}}+b_{3,4,\textrm{II}}\epsilon)\frac{\Delta f}{\Delta\omega},\quad a_{3,4,\textrm{II}}+b_{3,4,\textrm{II}}\epsilon=\frac{d-3}{d-1}\frac{-1}{d-1}\nonumber\\
A^{a,\mu}(g_3,g_4)_{\textrm{III}}&=&g^a_{3,4(ij)}\bar u_iu^\mu_j(a_{3,4,\textrm{III}}+b_{3,4,\textrm{III}}\epsilon)\frac{\Delta f}{\Delta\omega},\quad a_{3,4,\textrm{III}}+b_{3,4,\textrm{III}}\epsilon=(\frac{1-d}{4}+\frac{1}{d-1})\frac{-1}{d-1}\nonumber\\
A^{a,\mu}(g_3,g_4)_{\textrm{IV}}&=&g^a_{3,4(ij)}\bar
u_iu^\mu_j(a_{3,4,\textrm{IV}}+b_{3,4,\textrm{IV}}\epsilon)\frac{\Delta
f}{\Delta\omega},\quad
a_{3,4,\textrm{IV}}+b_{3,4,\textrm{IV}}\epsilon=\frac{(d-3)(d+1)}{2(d-1)^2}\frac{-1}{d-1}\nonumber
\end{eqnarray}
The parameters $a, b$ arise from the loop integration with the
dimensional regularization scheme. The coefficients $g^a_{(ij)}$ are
listed in Tables \ref{loopa-g1}--\ref{loopa-g4} while the function $\frac{\Delta f}{\Delta
\omega}$ is defined in Appendix \ref{sec6.3}.

$A^{b,\mu}$ is the correction from the vertex diagram (b):
\begin{eqnarray}
A^{b,\mu}_{ij}(g_1,g_2)&=&g^b_{1,2(ij)}\bar u_iS^\mu u_jI(m)\nonumber\\
A^{b,\mu}(g_5)&=&g^b_{5(ij)}g_{\rho\sigma}\bar u^\rho_jS^\mu u^\sigma_jI(m)\nonumber\\
A^{b,\mu}(g_3,g_4)&=&g^b_{3,4(ij)}\bar u_iu^\mu_jI(m)
\end{eqnarray}
where the function $I(m)$ is defined in Appendix \ref{sec6.3}. The
corresponding coefficients $g^b_{(ij)}$ are collected in Table
\ref{loopb}. The loop corrections from diagrams (c) and (d) vanish
in the heavy baryon limit $M_B\rightarrow\infty$. Their
contributions are of higher order in the $\frac{1}{M_B}$ expansion.
The analogous situation occurs in the nucleon octet case. Interested
readers may refer to Refs. \cite{Zhu:2000zf,Zhu:2002tn}.

The composite axial current operator also receives the correction
from the wave function renormalization \cite{Manohar:2000dt}. The
renormalization factor can be derived from the self-energy function
\[Z=\frac{1}{1-\Sigma'}\approx1+\Sigma',\quad\Sigma'=\left.\frac{\partial\Sigma}{\partial(v\cdot k)}\right|_{v\cdot k,k^2=0}\]
The matrix elements of the renormalized axial current between the
initial and final states read
\begin{eqnarray}\label{eq31}
^r\langle B_i|A^{a,\mu}|B_j\rangle^r&=&\langle B_i|A^{a,\mu}\sqrt{Z_iZ_j}|B_j\rangle\nonumber\\
&=&\langle B_i|A^{a,\mu}|B_j\rangle+\frac{1}{2}(\Sigma_i'+\Sigma_j')]\langle B_i|A^{a,\mu}|B_j\rangle\nonumber\\
&=&\bar u_iS^\mu u_jg^{(0)}_{ij}(1+\lambda_{ij})
\end{eqnarray}
where $g^{(0)}_{ij}$ is the axial charge at the tree level. The
coefficients $\lambda_{ij}$ are collected in Tables
\ref{wav-1}--\ref{wav-4}. Comparing with Eqs.
(\ref{eq24})--(\ref{eq26}), we have
$g_{(ij)}^{\textrm{Re}}=g^{(0)}_{ij}\lambda_{ij}$. The real part of
$\Sigma'$ reads
\begin{eqnarray}\label{eq32}
\Sigma'_{\textrm{Re}}&=&(A+B\epsilon)\left.\left[(m^2-\omega^2)\frac{\partial J}{\partial\omega}-2\omega J(m,\omega)\right]\right|_{\textrm{Re},\omega=\delta}\nonumber\\
&=&A\left[(m^2-\omega^2)\left.\frac{\partial
J}{\partial\omega}\right|_{\textrm{Re}}-2\omega
J(m,\omega)|_{\textrm{Re}}\right]-B\frac{1}{4\pi^2}(m^2-3\omega^2).
\end{eqnarray}
The function $J$ is defined in Appendix \ref{sec6.3}.

\subsection{Numerical results of the chiral correction to the axial charge}\label{sec4.3}

In principle, the axial charges of the heavy baryons can be
extracted from the measurement of their semileptonic decays.
However, there do not exist any experimental data now. The lack of
the data renders the determination of the low energy constants
$d,f,h$ etc very difficult.

\begin{table}[!h]
  \centering
  \caption{The chiral corrections to the axial charges.}\label{gvalue}
  \textrm{The marked entries are the predictions.}
\begin{tabular}{cp{40mm}<{\centering}ccp{40mm}<{\centering}c}
\Xhline{1.2pt}
&Loop-a [with the same spin states in the loop only]&loop-a (full)&loop-b&Wave function renormalization effect&Fit values\\
\hline\hline
$g_{\Lambda_c^+\Sigma_c^0}$   &$-0.06$&$-0.38$&$0.22$&$0.31$&$1.46\pm0.44$\\
$g_{\Sigma_c^+\Sigma_c^0}$    &$-0.04$&$-0.27$&$0.32$&$0.32$&$1.46\pm0.44$\\
$g_{\Xi_c'^+\Omega_c^0}$      &$-0.09$&$-0.60$&$0.45$&$0.53$&$1.71\pm0.62^\ddag$\\
$g_{\Sigma_c^+\Xi_c'^0}$      &$-0.03$&$-0.24$&$0.32$&$0.27$&$1.32\pm0.31^\ddag$\\
$g_{\Sigma_c^{++}\Xi_c'^+}$   &$-0.04$&$-0.34$&$0.45$&$0.38$&$1.46\pm0.44$\\
\hline\hline
$g_{\Lambda_c^+\Sigma_c^0}$&$-0.04$&$-0.29$&$-0.27$&$ 0.51$&$-0.93\pm0.28$\\
$g_{\Xi_c^+\Xi_c'^0}$      &$-0.08$&$-0.34$&$-0.19$&$ 0.63$&$-0.93\pm0.28$\\
$g_{\Lambda_c^+\Xi_c'^0}$  &$-0.05$&$-0.34$&$-0.28$&$ 0.33$&$-0.93\pm0.28$\\
$g_{\Xi_c^+\Omega_c^0}$    &$-0.15$&$-0.66$&$-0.39$&$ 0.82$&$-1.21\pm0.51^\ddag$\\
$g_{\Sigma_c^+\Xi_c^0}$    &$ 0.07$&$ 0.34$&$ 0.28$&$-0.67$&$-0.61\pm0.28^\ddag$\\
$g_{\Sigma_c^{++}\Xi_c^+}$ &$ 0.10$&$ 0.48$&$ 0.39$&$-0.95$&$-0.62\pm0.39^\ddag$\\
\hline\hline
$g_{{\Xi_c^*}'^+{\Xi_c^*}'^0}$  &$1.84$&$3.32$&$-0.33$&$-0.55$&$-2.19\pm0.66$\\
$g_{\Sigma_c^{*+}\Sigma_c^{*0}}$&$1.27$&$2.13$&$-0.47$&$-0.53$&$-2.19\pm0.66$\\
$g_{{\Xi_c^*}'^+\Omega_c^{*0}}$ &$2.37$&$5.09$&$-0.67$&$-1.00$&$-2.19\pm0.66$\\
$g_{\Sigma_c^{*+}{\Xi_c^*}'^0}$ &$0.82$&$1.84$&$-0.48$&$-0.46$&$-1.64\pm0.27^\ddag$\\
$g_{\Sigma_c^{*++}{\Xi_c^*}'^+}$&$1.16$&$2.60$&$-0.67$&$-0.65$&$-1.71\pm0.38^\ddag$\\
\hline\hline
$g_{\Xi_c'^+{\Xi_c^*}'^0}$      &$ $&$-0.29$&$0.10$&$0.15$&$1.26\pm0.38$\\
$g_{\Sigma_c^+\Sigma_c^{*0}}$   &$ $&$-0.21$&$0.14$&$0.15$&$1.26\pm0.38$\\
$g_{\Sigma_c^{++}\Sigma_c^{*+}}$&$ $&$-0.21$&$0.14$&$0.15$&$1.26\pm0.38^\ddag$\\
$g_{\Xi_c'^+\Omega_c^{*0}}$     &$ $&$-0.40$&$0.19$&$0.27$&$1.26\pm0.38$\\
$g_{\Sigma_c^+{\Xi_c^*}'^0}$    &$ $&$-0.14$&$0.14$&$0.13$&$1.01\pm0.16^\ddag$\\
$g_{\Sigma_c^{++}{\Xi_c^*}'^+}$ &$ $&$-0.20$&$0.19$&$0.18$&$1.06\pm0.22^\ddag$\\
$g_{{\Xi_c^*}'^+\Omega_c^0}$    &$ $&$-0.45$&$0.19$&$0.25$&$1.19\pm0.38^\ddag$\\
$g_{\Sigma_c^{*+}\Xi_c'^0}$     &$ $&$-0.21$&$0.14$&$0.12$&$0.93\pm0.16^\ddag$\\
$g_{\Sigma_c^{*++}\Xi_c'^+}$    &$ $&$-0.30$&$0.19$&$0.17$&$0.95\pm0.22^\ddag$\\
\hline\hline
$g_{\Lambda_c^+\Sigma_c^{*0}}$&$ $&$ 0.01$&$0.24$&$-0.43$&$1.61\pm1.34$\\
$g_{\Xi_c^+{\Xi_c^*}'^0}$     &$ $&$-0.11$&$0.17$&$-0.53$&$1.61\pm1.05$\\
$g_{\Lambda_c^+{\Xi_c^*}'^0}$ &$ $&$ 0.07$&$0.24$&$-0.26$&$1.14\pm1.19^\ddag$\\
$g_{\Xi_c^+\Omega_c^{*0}}$    &$ $&$-0.07$&$0.34$&$-0.63$&$1.61\pm1.16$\\
\Xhline{1.2pt}
\end{tabular}
\end{table}

In Ref. \cite{Li:2012bt}, the authors calculated the pseudoscalar
couplings of the heavy baryons. Within the framework of the chiral
quark model, both the pseudoscalar couplings of the nucleons and
heavy baryons can be expressed in terms of the pseudoscalar
couplings of the constituent quarks. Since there exist plenty of
nucleon nucleon scattering data, the pseudoscalar couplings of the
nucleons can be determined very well experimentally. With the pion
nucleon coupling as input, the authors first extracted the
pseudoscalar couplings of the constituent quarks, and then
determined the pseudoscalar couplings of the heavy baryons
\cite{Li:2012bt}. The axial charges are related to the coupling
constants $g_{pBB}$
\[g=\frac{2F_0}{M_a+M_b}g_{pBB}\]
From the values listed in Ref. \cite{Li:2012bt}, we have $g_1=1.46$
and $g_2=-0.93$. With the relationship among various $g$'s in
Sec. \ref{sec2}, we get $g_5=-2.19,g_3=1.26,g_4=1.61$. In the
following analysis, we regard the above values of the axial charge
as the pseudoexperimental data and use them as input to extract
various low energy constants. The values of LECs
$d'=B_0d,f'=B_0f,h'=B_0h$ (MeV$^{-1}$) are
\[d_1'=-0.9\times10^{-3},f_1'=2.6\times10^{-3},h_1'=0.7\times10^{-3}\]
\[d_2'=1.2\times10^{-3},f_2'=-1.7\times10^{-3},h_2'=-1.3\times10^{-3}\]
\[d_5'=2.4\times10^{-3},f_5'=-19.1\times10^{-3},h_5'=-11.5\times10^{-3}\]
\[d_3'=0.3\times10^{-3},f_3'=4.5\times10^{-3},h_3'=4.1\times10^{-3}\]
\[d_4'=-5.6\times10^{-3},f_4'=7.4\times10^{-3},h_4'=7.0\times10^{-3}\]

In our numerical analysis we also need the values of the axial
charges at $O(p)$
$g^{(0)}_1=0.98,g_2^{(0)}=-0.60,g_5^{(0)}=-1.47,g_3^{(0)}=0.85,g_4^{(0)}=1.04$.
We collect the numerical results of the chiral corrections to the
axial charges in Table \ref{gvalue}. We also list the separate contributions
from the vertex correction and wave function renormalization.
In the calculation of the self energy, we considered the isospin breaking effects because there exist plenty of data on the heavy baryon masses. However, in the case of the chiral correction to the axial charge, we have to work in the isospin symmetry limit because of the scarce data.
Actually, to calculate the contributions of the wave function renormalization effects, we could not use the results in Sec. \ref{sec4} directly because what we needed was the wave function renormalization factor, which is the derivative of the self-energy function ($\Sigma'$), not the self-energy function ($\Sigma$) itself. The expression of $\Sigma'$ can be seen in Eq. (\ref{eq32}).

The second column in Table \ref{gvalue} corresponds to the vertex
corrections from diagram (a) where the intermediate and external
heavy baryons have the same spin. The third column contains the
contribution from all types of diagram (a). For the corrections from diagram (a), comparing the second and third columns, one notices that the values increase with the interactions between baryons with different spin. The contributions from diagram (b) and wave function renormalization are listed in the
fourth and fifth columns. The last column is the fit value of the axial
charge.
From Table \ref{gvalue}, we can see that the chiral
expansion converges well. The axial charges with the notation
$\ddag$ in the last column are the predicted values. To show the sensitivity of the axial charges, we varied the input data by 10\%. The errors of all fit values are listed in the last column of Table \ref{gvalue}.

We have calculated the flavor SU(3) breaking chiral corrections to
the axial charges of the heavy baryons in the exact isospin limit.
We notice that the divergences from diagram (a) for the flavor
structure (1+i2) can be absorbed by the counterterms completely. In
contrast, the divergences from diagram (a) for the flavor structure
(4+i5) cannot be absorbed by the counterterms completely with the
explicit SU(3) breaking. Only in the exact SU(3) flavor symmetry
limit, both divergences can be absorbed by the counterterms.

For example, let us consider the axial currents with the flavor
(1+i2) $A_{\Xi_c^+\Xi_c'^0}^\mu$ and $A_{\Xi_c'^+\Xi_c^0}^\mu$. The
two processes occupy the same position in the weight diagram and
their counterterms are the same due to the SU(3) symmetry at the
tree level. The corrections to $A_{\Xi_c^+\Xi_c'^0}^\mu$ from
diagram (a) contain the $\pi^0$ loop and $K^+$ loop. The corrections
to $A_{\Xi_c'^+\Xi_c^0}^\mu$ contain the $\pi^0$ loop and $K^0$
loop. In the isospin symmetry limit even with explicit SU(3)
symmetry breaking, $m_{K^+}=m_{K^0}$. So the divergences are the
same and can be canceled exactly.

On the other hand, for instance, the axial currents with the flavor
(4+i5) $A_{\Lambda_c^+\Xi_c'^0}^\mu$ and $A_{\Sigma_c^0\Xi_c^0}^\mu$
also have the same form of counterterms. The corrections to
$A_{\Lambda_c^+\Xi_c'^0}^\mu$ contain the $\pi^0$ loop and $\pi^+$
loop. The corrections to $A_{\Sigma_c^0\Xi_c^0}^\mu$ contain the
$\pi^0$ loop and $K^+$ loop. In the isospin limit but with the
explicit SU(3) flavor breaking, the divergence from the $\pi^+$ and
$K^+$ loops cannot be canceled by the same counterterms. But in
the SU(3) limit, both divergences are the same.
That is to say, there are not enough counterterms to absorb the
divergence in the SU(3) flavor breaking situation. However, the
redundant divergences will be absorbed by the new LECs at higher
order. We can drop the redundant infinities safely since the chiral
symmetry ensures that the divergences can be absorbed into the
higher order counterterms. The underlying reason is the asymmetry
between the triplet and sextet representations in the weight
diagram.

The axial currents between two sextet representations or two octet
representations do not suffer from the above problems. The same
situation occurs to the wave function renormalization. However, once
again, the chiral symmetry ensures that the divergences can be
absorbed into the higher order counterterms if the SU(3)-breaking
terms are regarded as higher order.

\section{Summary}\label{sec5}

In short summary, we have calculated the one-loop chiral corrections
to the masses and axial charges of the charmed antitriplet and
sextet heavy baryon systems in the HBChPT framework.

After introducing the chiral Lagrangians, we have systematically
calculated the baryon masses to the $O(p^3)$ due to the explicit
SU(3) breaking. The mass splitting of the heavy baryons is related
to up and down quark mass difference, and electric charge. Both
strong interaction and QED effects are involved in our calculation.
The resulting charmed baryon masses and decay widths are in good
agreement with experimental data. The LECs are consistent with the
values in Ref. \cite{Liu:2012sw,Liu:2012uw}.

We have also calculated the chiral loop contributions from the
vertex corrections and wave function renormalization to the axial
charges of the heavy baryons in the isospin symmetry limit but with
explicit SU(3) breaking. The convergence of the chiral expansion is
quite good. In the future, the axial charges of the heavy baryons
may be measured through their semileptonic and nonleptonic decays
experimentally. The ongoing LHCb experiment and the future B factories will enrich the data of heavy baryons. Moreover, the axial
charges play an important role in the study of the loosely bound
molecular states composed of two heavy baryons.

The mass spectrum of the charmed and bottom baryons with different quark content and isospin has been computed with the Lattice NRQCD formalism \cite{Mathur:2002ce}. The axial current of the bottom hadrons was explored by using partially quenched chiral
perturbation theory in Lattice QCD \cite{Detmold:2011rb}. Hopefully the expressions of the chiral loop corrections to the
masses and axial charges of the heavy baryons will be useful in the
chiral extrapolation of the lattice simulation data of these two
quantities where the pion mass on the lattice is larger than its
experimental value.

\section*{ACKNOWLEDGMENTS}

N. Jiang is very grateful to Zhan-Wei Liu and
Zhi-Feng Sun for very helpful discussions. This project is supported
by the National Natural Science Foundation of China under Grant No.
11261130311.

\newpage

\section{Appendix}\label{sec6}

\subsection{CATEGORIES OF THE VERTEX CORRECTION DIAGRAM (a).}\label{sec6.1}

\begin{figure}[!h]
\begin{center}
\caption{The flavor and Lorentz structures of diagram (a) for
$g_1$.}\label{fig-g1}
\includegraphics{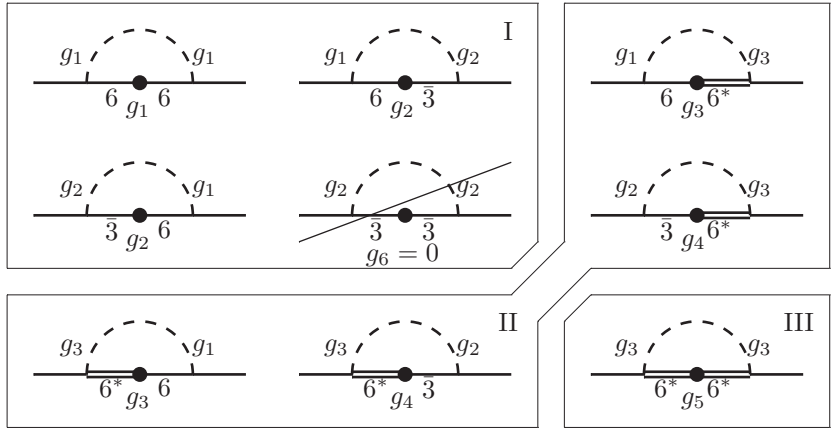}
\end{center}
\end{figure}
\begin{figure}[!h]
\begin{center}
\caption{The flavor and Lorentz structures of diagram (a) for
$g_2$.}\label{fig-g2}
\includegraphics{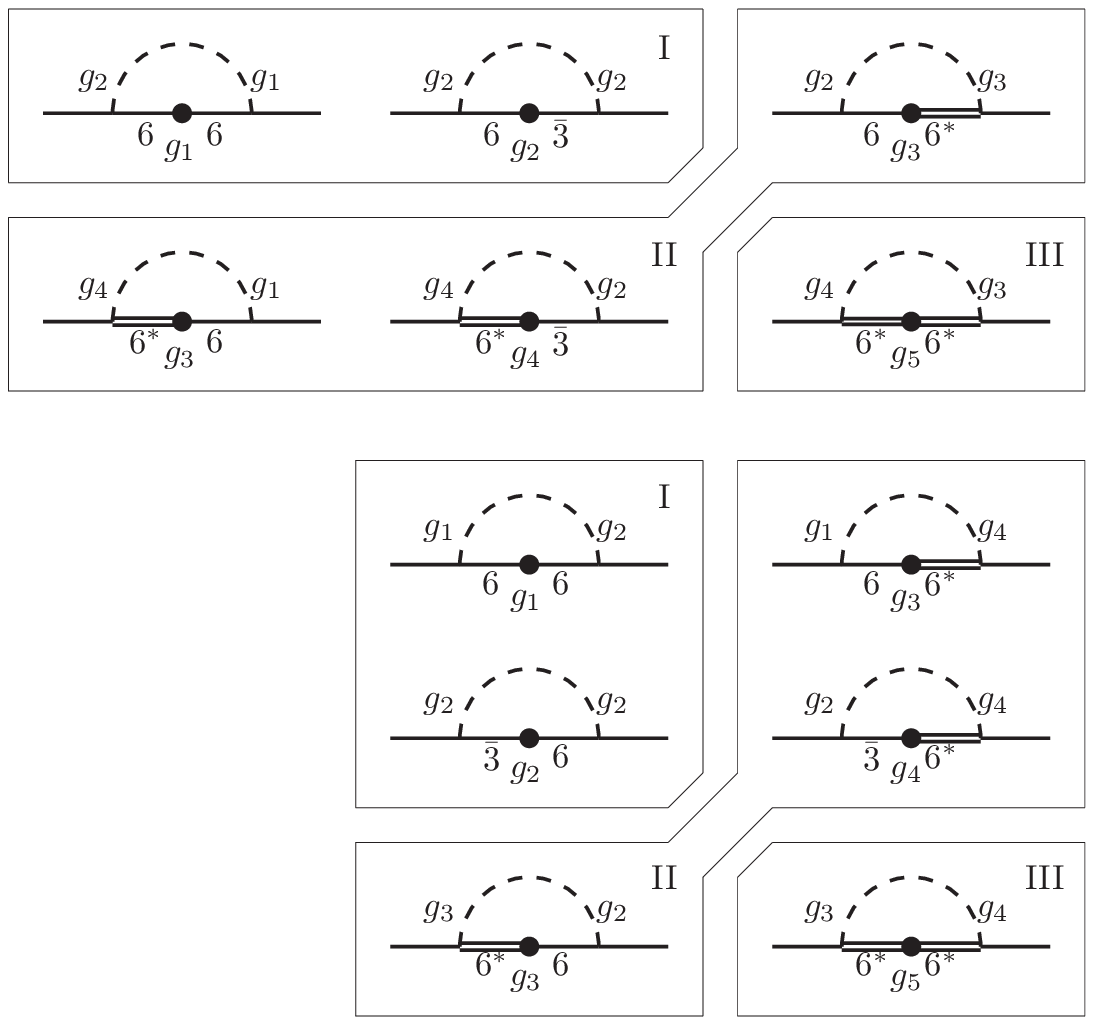}
\end{center}
\end{figure}
\begin{figure}[!h]
\begin{center}
\caption{The flavor and Lorentz structures of diagram (a) for $g_5$
.}\label{fig-g5}
\includegraphics{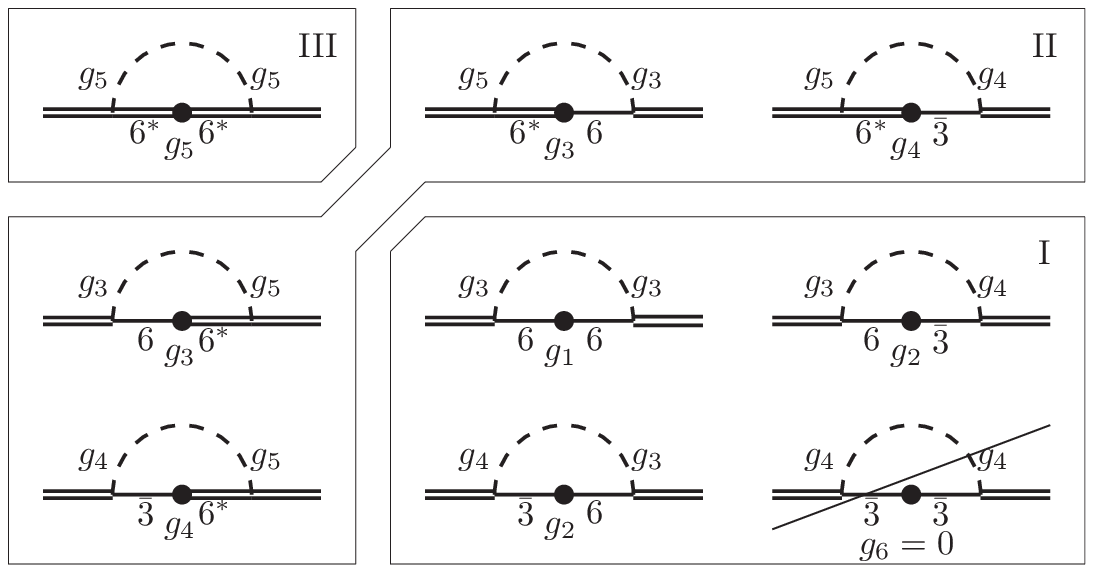}
\end{center}
\end{figure}
\begin{figure}[!h]
\begin{center}
\caption{The flavor and Lorentz structures of diagram (a) for $g_3$
.}\label{fig-g3}
\includegraphics{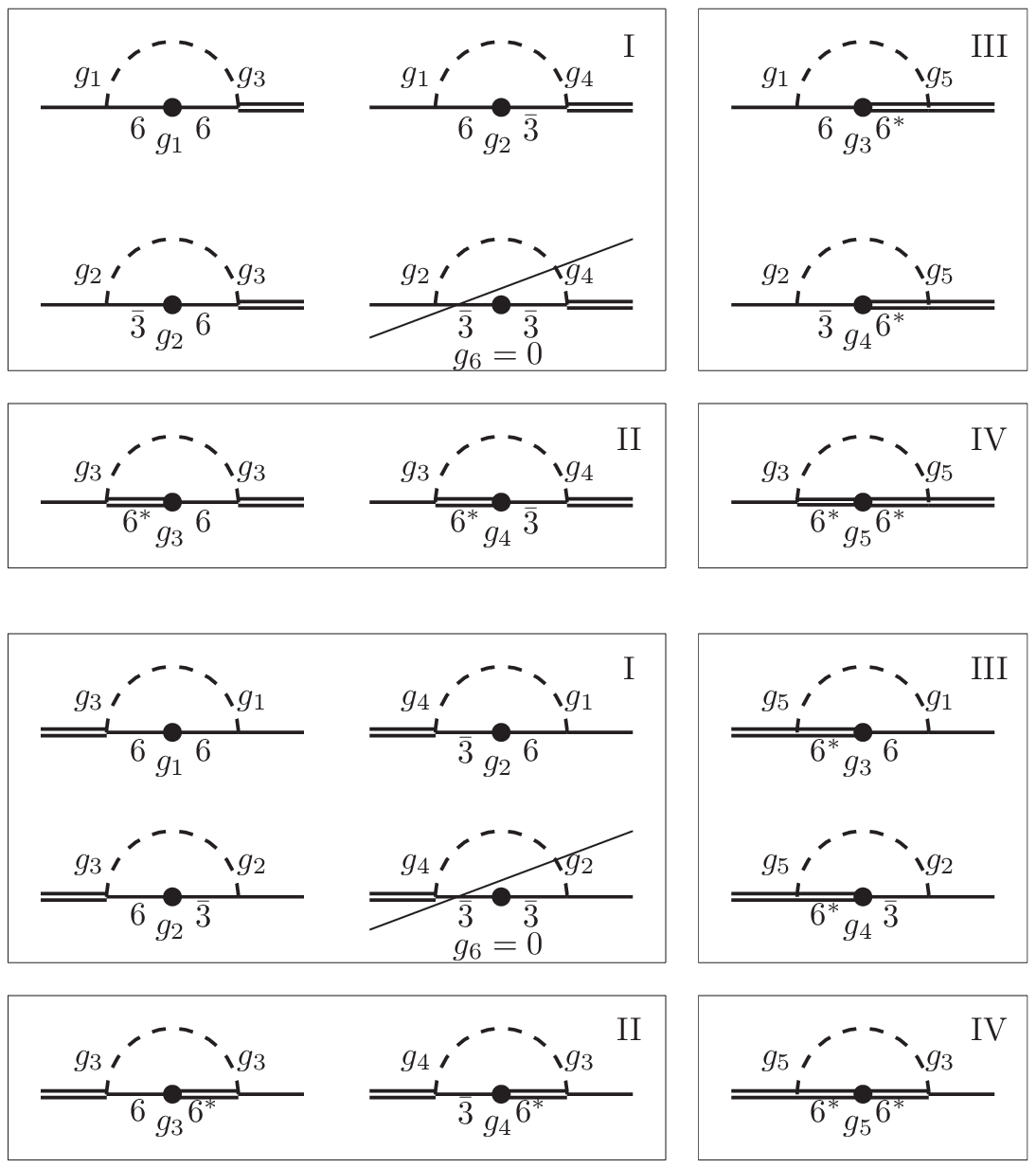}
\end{center}
\end{figure}
\begin{figure}[!h]
\begin{center}
\caption{The flavor and Lorentz structures of diagram (a) for $g_4$
.}\label{fig-g4}
\includegraphics{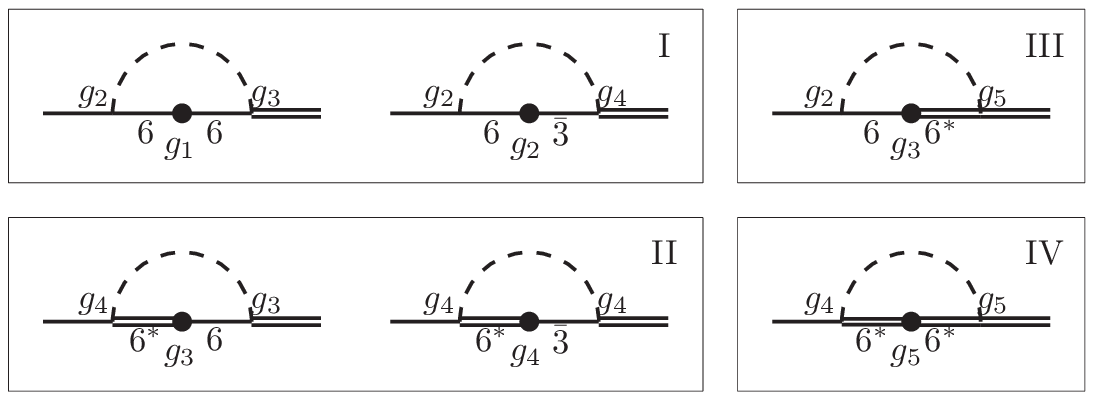}
\end{center}
\end{figure}

\begin{table}
\subsection{TABLES}\label{sec6.2}
  \centering
  \caption{The coefficients $C$ in the self-energy function.}\label{Coeff.sel}
  \renewcommand{\arraystretch}{1.5}
\begin{tabular}{c|ccccc|ccccc}
\Xhline{1.2pt}
&\multicolumn{5}{c|}{Case I meson loop}&\multicolumn{5}{c}{Case II meson loop}\\
\cline{2-11}
&$K^\pm$&$K^0/\bar K^0$&$\eta$&$\pi^\pm$&$\pi^0$&$K^\pm$&$K^0/\bar K^0$&$\eta$&$\pi^\pm$&$\pi^0$\\
\hline\hline
$C_{\Sigma_c^{++}}$&$\frac{g_1^2}{F_0^2}+\frac{2 g_2^2}{F_0^2}$&$-$&$\frac{g_1^2}{3 F_0^2}$&$\frac{g_1^2}{F_0^2}+\frac{2 g_2^2}{F_0^2}$&$\frac{g_1^2}{F_0^2}$&$-\frac{g_3^2}{4 F_0^2}$&$-$&$-\frac{g_3^2}{12 F_0^2}$&$-\frac{g_3^2}{4 F_0^2}$&$-\frac{g_3^2}{4 F_0^2}$\\
$C_{\Sigma_c^+}$&$\frac{g_1^2}{2 F_0^2}+\frac{g_2^2}{F_0^2}$&$\frac{g_1^2}{2 F_0^2}+\frac{g_2^2}{F_0^2}$&$\frac{g_1^2}{3 F_0^2}$&$\frac{2 g_1^2}{F_0^2}$&$\frac{2 g_2^2}{F_0^2}$&$-\frac{g_3^2}{8 F_0^2}$&$-\frac{g_3^2}{8 F_0^2}$&$-\frac{g_3^2}{12 F_0^2}$&$-\frac{g_3^2}{2 F_0^2}$&$-$\\
$C_{\Sigma_c^0}$&$-$&$\frac{g_1^2}{F_0^2}+\frac{2 g_2^2}{F_0^2}$&$\frac{g_1^2}{3 F_0^2}$&$\frac{g_1^2}{F_0^2}+\frac{2 g_2^2}{F_0^2}$&$\frac{g_1^2}{F_0^2}$&$-$&$-\frac{g_3^2}{4 F_0^2}$&$-\frac{g_3^2}{12 F_0^2}$&$-\frac{g_3^2}{4 F_0^2}$&$-\frac{g_3^2}{4 F_0^2}$\\
$C_{\Xi_c'^+}$&$\frac{2 g_1^2}{F_0^2}$&$\frac{g_1^2}{2 F_0^2}+\frac{g_2^2}{F_0^2}$&$\frac{g_1^2}{12 F_0^2}+\frac{3 g_2^2}{2 F_0^2}$&$\frac{g_1^2}{2 F_0^2}+\frac{g_2^2}{F_0^2}$&$\frac{g_1^2}{4 F_0^2}+\frac{g_2^2}{2 F_0^2}$&$-\frac{g_3^2}{2 F_0^2}$&$-\frac{g_3^2}{8 F_0^2}$&$-\frac{g_3^2}{48 F_0^2}$&$-\frac{g_3^2}{8 F_0^2}$&$-\frac{g_3^2}{16 F_0^2}$\\
$C_{\Xi_c'^0}$&$\frac{g_1^2}{2 F_0^2}+\frac{g_2^2}{F_0^2}$&$\frac{2 g_1^2}{F_0^2}$&$\frac{g_1^2}{12 F_0^2}+\frac{3 g_2^2}{2 F_0^2}$&$\frac{g_1^2}{2 F_0^2}+\frac{g_2^2}{F_0^2}$&$\frac{g_1^2}{4 F_0^2}+\frac{g_2^2}{2 F_0^2}$&$-\frac{g_3^2}{8 F_0^2}$&$-\frac{g_3^2}{2 F_0^2}$&$-\frac{g_3^2}{48 F_0^2}$&$-\frac{g_3^2}{8 F_0^2}$&$-\frac{g_3^2}{16 F_0^2}$\\
$C_{\Omega_c^0}$&$\frac{g_1^2}{F_0^2}+\frac{2 g_2^2}{F_0^2}$&$\frac{g_1^2}{F_0^2}+\frac{2 g_2^2}{F_0^2}$&$\frac{4 g_1^2}{3 F_0^2}$&$-$&$-$&$-\frac{g_3^2}{4 F_0^2}$&$-\frac{g_3^2}{4 F_0^2}$&$-\frac{g_3^2}{3 F_0^2}$&$-$&$-$\\
\hline\hline
$C_{\Lambda_c^+}$&$\frac{g_2^2}{F_0^2}+\frac{2 g_6^2}{F_0^2}$&$\frac{g_2^2}{F_0^2}+\frac{2 g_6^2}{F_0^2}$&$\frac{4 g_6^2}{3 F_0^2}$&$\frac{4 g_2^2}{F_0^2}$&$\frac{2 g_2^2}{F_0^2}$&$-\frac{g_4^2}{4 F_0^2}$&$-\frac{g_4^2}{4 F_0^2}$&$-\frac{g_4^2}{F_0^2}$&$-$&$-\frac{g_4^2}{2 F_0^2}$\\
$C_{\Xi_c^+}$&$\frac{4 g_2^2}{F_0^2}$&$\frac{g_2^2}{F_0^2}+\frac{2 g_6^2}{F_0^2}$&$\frac{3 g_2^2}{2 F_0^2}+\frac{g_6^2}{3 F_0^2}$&$\frac{g_2^2}{F_0^2}+\frac{2 g_6^2}{F_0^2}$&$\frac{g_2^2}{2 F_0^2}+\frac{g_6^2}{F_0^2}$&$-\frac{g_4^2}{F_0^2}$&$-\frac{g_4^2}{4 F_0^2}$&$-\frac{3 g_4^2}{8 F_0^2}$&$-\frac{g_4^2}{4 F_0^2}$&$-\frac{g_4^2}{8 F_0^2}$\\
$C_{\Xi_c^0}$&$\frac{g_2^2}{F_0^2}+\frac{2 g_6^2}{F_0^2}$&$\frac{4 g_2^2}{F_0^2}$&$\frac{3 g_2^2}{2 F_0^2}+\frac{g_6^2}{3 F_0^2}$&$\frac{g_2^2}{F_0^2}+\frac{2 g_6^2}{F_0^2}$&$\frac{g_2^2}{2 F_0^2}+\frac{g_6^2}{F_0^2}$&$-\frac{g_4^2}{4 F_0^2}$&$-\frac{g_4^2}{F_0^2}$&$-\frac{3 g_4^2}{8 F_0^2}$&$-\frac{g_4^2}{4 F_0^2}$&$-\frac{g_4^2}{8 F_0^2}$\\
\Xhline{1.2pt}
&\multicolumn{5}{c|}{Case II meson loop}&\multicolumn{5}{c}{Case I meson loop}\\
\cline{2-11}
&$K^\pm$&$K^0/\bar K^0$&$\eta$&$\pi^\pm$&$\pi^0$&$K^\pm$&$K^0/\bar K^0$&$\eta$&$\pi^\pm$&$\pi^0$\\
\hline\hline
$C_{\Sigma_c^{*++}}$&$-\frac{g_3^2}{4 F_0^2}-\frac{g_4^2}{2 F_0^2}$&$-$&$-\frac{g_3^2}{12 F_0^2}$&$-\frac{g_3^2}{4 F_0^2}-\frac{g_4^2}{2 F_0^2}$&$-\frac{g_3^2}{4 F_0^2}$&$\frac{g_5^2}{F_0^2}$&$-$&$\frac{g_5^2}{3 F_0^2}$&$\frac{g_5^2}{F_0^2}$&$\frac{g_5^2}{F_0^2}$\\
$C_{\Sigma_c^{*+}}$&$-\frac{g_3^2}{8 F_0^2}-\frac{g_4^2}{4 F_0^2}$&$-\frac{g_3^2}{8 F_0^2}-\frac{g_4^2}{4 F_0^2}$&$-\frac{g_3^2}{12 F_0^2}$&$-\frac{g_3^2}{2 F_0^2}$&$-\frac{g_4^2}{2 F_0^2}$&$\frac{g_5^2}{2 F_0^2}$&$\frac{g_5^2}{2 F_0^2}$&$\frac{g_5^2}{3 F_0^2}$&$\frac{2 g_5^2}{F_0^2}$&$-$\\
$C_{\Sigma_c^{*0}}$&$-$&$-\frac{g_3^2}{4 F_0^2}-\frac{g_4^2}{2 F_0^2}$&$-\frac{g_3^2}{12 F_0^2}$&$-\frac{g_3^2}{4 F_0^2}-\frac{g_4^2}{2 F_0^2}$&$-\frac{g_3^2}{4 F_0^2}$&$-$&$\frac{g_5^2}{F_0^2}$&$\frac{g_5^2}{3 F_0^2}$&$\frac{g_5^2}{F_0^2}$&$\frac{g_5^2}{F_0^2}$\\
$C_{\Xi_c'^{*+}}$&$-\frac{g_3^2}{2 F_0^2}$&$-\frac{g_3^2}{8 F_0^2}-\frac{g_4^2}{4 F_0^2}$&$-\frac{g_3^2}{48 F_0^2}-\frac{3 g_4^2}{8 F_0^2}$&$-\frac{g_3^2}{8 F_0^2}-\frac{g_4^2}{4 F_0^2}$&$-\frac{g_3^2}{16 F_0^2}-\frac{g_4^2}{8 F_0^2}$&$\frac{2 g_5^2}{F_0^2}$&$\frac{g_5^2}{2 F_0^2}$&$\frac{g_5^2}{12 F_0^2}$&$\frac{g_5^2}{2 F_0^2}$&$\frac{g_5^2}{4 F_0^2}$\\
$C_{\Xi_c'^{*0}}$&$-\frac{g_3^2}{8 F_0^2}-\frac{g_4^2}{4 F_0^2}$&$-\frac{g_3^2}{2 F_0^2}$&$-\frac{g_3^2}{48 F_0^2}-\frac{3 g_4^2}{8 F_0^2}$&$-\frac{g_3^2}{8 F_0^2}-\frac{g_4^2}{4 F_0^2}$&$-\frac{g_3^2}{16 F_0^2}-\frac{g_4^2}{8 F_0^2}$&$\frac{g_5^2}{2 F_0^2}$&$\frac{2 g_5^2}{F_0^2}$&$\frac{g_5^2}{12 F_0^2}$&$\frac{g_5^2}{2 F_0^2}$&$\frac{g_5^2}{4 F_0^2}$\\
$C_{\Omega_c^{*0}}$&$-\frac{g_3^2}{4 F_0^2}-\frac{g_4^2}{2 F_0^2}$&$-\frac{g_3^2}{4 F_0^2}-\frac{g_4^2}{2 F_0^2}$&$-\frac{g_3^2}{3 F_0^2}$&$-$&$-$&$\frac{g_5^2}{F_0^2}$&$\frac{4 g_5^2}{3 F_0^2}$&$\frac{4 g_5^2}{3 F_0^2}$&$-$&$-$\\
\Xhline{1.2pt}
\end{tabular}
\end{table}

\begin{table}[!h]
  \centering
  \renewcommand{\arraystretch}{1.5}
  \caption{The coefficients $g_{1(ij)}^a$ of the axial current from diagram (a).}\label{loopa-g1}
\begin{tabular}{c|ccp{35mm}<{\centering}p{30mm}<{\centering}}
\Xhline{1.2pt}
&&Type I&Type II&Type III\\
\hline\hline
\multirow{3}{*}{$g_{\Xi_c'^+\Xi_c'^0}$}
    &$K\textrm{-loop}$&$\frac{2 g_1^3}{F_0^2}-\frac{4 g_1 g_2^2}{F_0^2}$&$\frac{g_2 g_3 g_4}{F_0^2}-\frac{g_1 g_3^2}{F_0^2}$&$\frac{g_3^2 g_5}{2 F_0^2}$\\
    &$\eta\textrm{-loop}$&$\frac{g_1^3}{12 F_0^2}-\frac{g_1 g_2^2}{F_0^2}$&$\frac{g_2 g_3 g_4}{4 F_0^2}-\frac{g_1 g_3^2}{24 F_0^2}$&$\frac{g_3^2 g_5}{48 F_0^2}$\\
    &$\pi\textrm{-loop}$&$-\frac{g_1^3}{4 F_0^2}-\frac{g_2^2 g_1}{F_0^2}$&$\frac{g_1 g_3^2}{8 F_0^2}+\frac{g_2 g_4 g_3}{4 F_0^2}$&$-\frac{g_3^2 g_5}{16 F_0^2}$\\
\hline
\multirow{3}{*}{$g_{\Sigma_c^+\Sigma_c^0}$}
    &$K\textrm{-loop}$&$\frac{g_1^3}{\sqrt{2} F_0^2}-\frac{2
   \sqrt{2} g_1 g_2^2}{F_0^2}$&$\frac{g_2 g_3 g_4}{\sqrt{2} F_0^2}-\frac{g_1 g_3^2}{2 \sqrt{2}
   F_0^2}$&$\frac{g_3^2 g_5}{4 \sqrt{2} F_0^2}$\\
    &$\eta\textrm{-loop}$&$\frac{\sqrt{2} g_1^3}{3 F_0^2}$&$-\frac{g_1 g_3^2}{3 \sqrt{2} F_0^2}$&$\frac{g_3^2 g_5}{6 \sqrt{2} F_0^2}$\\
    &$\pi\textrm{-loop}$&$\frac{\sqrt{2} g_1^3}{F_0^2}-\frac{4 \sqrt{2} g_1 g_2^2}{F_0^2}$&$\frac{\sqrt{2} g_2 g_3 g_4}{F_0^2}-\frac{g_1 g_3^2}{\sqrt{2} F_0^2}$&$\frac{g_3^2 g_5}{2 \sqrt{2} F_0^2}$\\
\hline
\multirow{3}{*}{$g_{\Xi_c'^+\Omega_c^0}$}
    &$K\textrm{-loop}$&$\frac{3 g_1^3}{\sqrt{2} F_0^2}-\frac{4 \sqrt{2} g_1 g_2^2}{F_0^2}$&$\frac{\sqrt{2} g_2 g_3 g_4}{F_0^2}-\frac{3 g_1 g_3^2}{2 \sqrt{2}
   F_0^2}$&$\frac{3 g_3^2 g_5}{4 \sqrt{2} F_0^2}$\\
    &$\eta\textrm{-loop}$&$\frac{\sqrt{2} g_1^3}{3 F_0^2}-\frac{2 \sqrt{2} g_1 g_2^2}{F_0^2}$&$\frac{g_2 g_3 g_4}{\sqrt{2} F_0^2}-\frac{g_1 g_3^2}{3 \sqrt{2}
   F_0^2}$&$\frac{g_3^2 g_5}{6 \sqrt{2} F_0^2}$\\
    &$\pi\textrm{-loop}$&$-$&$-$&$-$\\
\hline
\multirow{3}{*}{$g_{\Sigma_c^+\Xi_c'^0}$}
    &$K\textrm{-loop}$&$\frac{g_1^3}{F_0^2}-\frac{2 g_1 g_2^2}{F_0^2}$&$\frac{g_2 g_3 g_4}{2 F_0^2}-\frac{g_1 g_3^2}{2 F_0^2}$&$\frac{g_3^2 g_5}{4 F_0^2}$\\
    &$\eta\textrm{-loop}$&$-\frac{g_1^3}{6 F_0^2}-\frac{g_2^2 g_1}{F_0^2}$&$\frac{g_1 g_3^2}{12 F_0^2}+\frac{g_2 g_4 g_3}{4 F_0^2}$&$-\frac{g_3^2 g_5}{24 F_0^2}$\\
    &$\pi\textrm{-loop}$&$\frac{g_1^3}{F_0^2}-\frac{3 g_1 g_2^2}{F_0^2}$&$\frac{3 g_2 g_3 g_4}{4 F_0^2}-\frac{g_1 g_3^2}{2 F_0^2}$&$\frac{g_3^2 g_5}{4 F_0^2}$\\
\hline
\multirow{3}{*}{$g_{\Sigma_c^{++}\Xi_c'^+}$}
    &$K\textrm{-loop}$&$\frac{\sqrt{2} g_1^3}{F_0^2}-\frac{2 \sqrt{2} g_1 g_2^2}{F_0^2}$&$\frac{g_2 g_3 g_4}{\sqrt{2} F_0^2}-\frac{g_1 g_3^2}{\sqrt{2} F_0^2}$&$\frac{g_3^2 g_5}{2 \sqrt{2} F_0^2}$\\
    &$\eta\textrm{-loop}$&$-\frac{g_1^3}{3 \sqrt{2} F_0^2}-\frac{\sqrt{2} g_2^2 g_1}{F_0^2}$&$\frac{g_1 g_3^2}{6 \sqrt{2} F_0^2}+\frac{g_2 g_4 g_3}{2 \sqrt{2}
   F_0^2}$&$-\frac{g_3^2 g_5}{12 \sqrt{2} F_0^2}$\\
    &$\pi\textrm{-loop}$&$\frac{\sqrt{2} g_1^3}{F_0^2}-\frac{3 \sqrt{2} g_1 g_2^2}{F_0^2}$&$\frac{3 g_2 g_3 g_4}{2 \sqrt{2} F_0^2}-\frac{g_1 g_3^2}{\sqrt{2}
   F_0^2}$&$\frac{g_3^2 g_5}{2 \sqrt{2} F_0^2}$\\
\Xhline{1.2pt}
\end{tabular}
\bigskip
  \centering
  \renewcommand{\arraystretch}{1.5}
  \caption{The coefficients $g_{2(ij)}^a$ of the axial current from diagram (a).}\label{loopa-g2}
\begin{tabular}{c|cccp{30mm}<{\centering}}
\Xhline{1.2pt}
&&Type I&Type II&Type III\\
\hline\hline
\multirow{3}{*}{$g_{\Lambda_c^+\Sigma_c^0}$}
    &$K\textrm{-loop}$&$\frac{2 g_2^3}{F_0^2}-\frac{g_1^2 g_2}{F_0^2}$&$\frac{g_2 g_3^2}{4 F_0^2}+\frac{g_1 g_4 g_3}{4 F_0^2}-\frac{g_2
   g_4^2}{2 F_0^2}$&$-\frac{g_3 g_4 g_5}{4 F_0^2}$\\
    &$\eta\textrm{-loop}$&$-$&$-$&$-$\\
    &$\pi\textrm{-loop}$&$\frac{4 g_2^3}{F_0^2}-\frac{4 g_1^2 g_2}{F_0^2}$&$\frac{g_2 g_3^2}{F_0^2}+\frac{g_1 g_4 g_3}{F_0^2}-\frac{g_2
   g_4^2}{F_0^2}$&$-\frac{g_3 g_4 g_5}{F_0^2}$\\
\hline
\multirow{3}{*}{$g_{\Xi_c^+\Xi_c'^0}$}
    &$K\textrm{-loop}$&$\frac{2 \sqrt{2} g_2^3}{F_0^2}-\frac{2 \sqrt{2} g_1^2 g_2}{F_0^2}$&$\frac{g_2 g_3^2}{\sqrt{2} F_0^2}+\frac{g_1 g_4 g_3}{\sqrt{2}
   F_0^2}-\frac{g_2 g_4^2}{\sqrt{2} F_0^2}$&$-\frac{g_3 g_4 g_5}{\sqrt{2} F_0^2}$\\
    &$\eta\textrm{-loop}$&$\frac{3 g_2^3}{\sqrt{2} F_0^2}-\frac{g_1^2 g_2}{2 \sqrt{2} F_0^2}$&$\frac{g_2 g_3^2}{8 \sqrt{2} F_0^2}+\frac{g_1 g_4 g_3}{8 \sqrt{2}
   F_0^2}-\frac{3 g_2 g_4^2}{4 \sqrt{2} F_0^2}$&$-\frac{g_3 g_4 g_5}{8 \sqrt{2} F_0^2}$\\
    &$\pi\textrm{-loop}$&$-\frac{g_2^3}{\sqrt{2} F_0^2}-\frac{g_1^2 g_2}{2 \sqrt{2} F_0^2}$&$\frac{g_2 g_3^2}{8 \sqrt{2} F_0^2}+\frac{g_1 g_4 g_3}{8 \sqrt{2}
   F_0^2}+\frac{g_2 g_4^2}{4 \sqrt{2} F_0^2}$&$-\frac{g_3 g_4 g_5}{8 \sqrt{2} F_0^2}$\\
\hline
\multirow{3}{*}{$g_{\Lambda_c^+\Xi_c'^0}$}
    &$K\textrm{-loop}$&$-\frac{\sqrt{2} g_1^2 g_2}{F_0^2}$&$\frac{g_2 g_3^2}{2 \sqrt{2} F_0^2}+\frac{g_1 g_4 g_3}{2 \sqrt{2}
   F_0^2}$&$-\frac{g_3 g_4 g_5}{2 \sqrt{2} F_0^2}$\\
    &$\eta\textrm{-loop}$&$-$&$-$&$-$\\
    &$\pi\textrm{-loop}$&$\frac{3 \sqrt{2} g_2^3}{F_0^2}-\frac{3 g_1^2 g_2}{\sqrt{2} F_0^2}$&$\frac{3 g_2 g_3^2}{4 \sqrt{2} F_0^2}+\frac{3 g_1 g_4 g_3}{4 \sqrt{2}
   F_0^2}-\frac{3 g_2 g_4^2}{2 \sqrt{2} F_0^2}$&$-\frac{3 g_3 g_4 g_5}{4 \sqrt{2} F_0^2}$\\
\hline
\multirow{3}{*}{$g_{\Xi_c^+\Omega_c^0}$}
    &$K\textrm{-loop}$&$\frac{6 g_2^3}{F_0^2}-\frac{3 g_1^2 g_2}{F_0^2}$&$\frac{3 g_2 g_3^2}{4 F_0^2}+\frac{3 g_1 g_4 g_3}{4 F_0^2}-\frac{3
   g_2 g_4^2}{2 F_0^2}$&$-\frac{3 g_3 g_4 g_5}{4 F_0^2}$\\
    &$\eta\textrm{-loop}$&$-\frac{2 g_1^2 g_2}{F_0^2}$&$\frac{g_2 g_3^2}{2 F_0^2}+\frac{g_1 g_4 g_3}{2 F_0^2}$&$-\frac{g_3 g_4 g_5}{2 F_0^2}$\\
    &$\pi\textrm{-loop}$&$-$&$-$&$-$\\
\hline
\multirow{3}{*}{$g_{\Sigma_c^+\Xi_c^+}$}
    &$K\textrm{-loop}$&$\frac{\sqrt{2} g_1^2 g_2}{F_0^2}-\frac{2 \sqrt{2} g_2^3}{F_0^2}$&$-\frac{g_2 g_3^2}{2 \sqrt{2} F_0^2}-\frac{g_1 g_4 g_3}{2 \sqrt{2}
   F_0^2}+\frac{g_2 g_4^2}{\sqrt{2} F_0^2}$&$\frac{g_3 g_4 g_5}{2 \sqrt{2} F_0^2}$\\
    &$\eta\textrm{-loop}$&$\frac{g_1^2 g_2}{\sqrt{2} F_0^2}$&$-\frac{g_2 g_3^2}{4 \sqrt{2} F_0^2}-\frac{g_1 g_4 g_3}{4 \sqrt{2}
   F_0^2}$&$\frac{g_3 g_4 g_5}{4 \sqrt{2} F_0^2}$\\
    &$\pi\textrm{-loop}$&$\frac{\sqrt{2} g_1^2 g_2}{F_0^2}-\frac{\sqrt{2} g_2^3}{F_0^2}$&$-\frac{g_2 g_3^2}{2 \sqrt{2} F_0^2}-\frac{g_1 g_4 g_3}{2 \sqrt{2}
   F_0^2}+\frac{g_2 g_4^2}{2 \sqrt{2} F_0^2}$&$\frac{g_3 g_4 g_5}{2 \sqrt{2} F_0^2}$\\
\hline
\multirow{3}{*}{$g_{\Sigma_c^{++}\Xi_c^+}$}
    &$K\textrm{-loop}$&$\frac{2 g_1^2 g_2}{F_0^2}-\frac{4 g_2^3}{F_0^2}$&$-\frac{g_2 g_3^2}{2 F_0^2}-\frac{g_1 g_4 g_3}{2 F_0^2}+\frac{g_2
   g_4^2}{F_0^2}$&$\frac{g_3 g_4 g_5}{2 F_0^2}$\\
    &$\eta\textrm{-loop}$&$\frac{g_1^2 g_2}{F_0^2}$&$-\frac{g_2 g_3^2}{4 F_0^2}-\frac{g_1 g_4 g_3}{4 F_0^2}$&$\frac{g_3 g_4 g_5}{4 F_0^2}$\\
    &$\pi\textrm{-loop}$&$\frac{2 g_1^2 g_2}{F_0^2}-\frac{2 g_2^3}{F_0^2}$&$-\frac{g_2 g_3^2}{2 F_0^2}-\frac{g_1 g_4 g_3}{2 F_0^2}+\frac{g_2
   g_4^2}{2 F_0^2}$&$\frac{g_3 g_4 g_5}{2 F_0^2}$\\
\Xhline{1.2pt}
\end{tabular}
\end{table}

\begin{table}[!h]
\bigskip
  \centering
  \renewcommand{\arraystretch}{1.5}
  \caption{The coefficients $g_{5(ij)}^a$ of the axial current from diagram (a).}\label{loopa-g5}
\begin{tabular}{c|ccp{35mm}<{\centering}p{30mm}<{\centering}}
\Xhline{1.2pt}
&&Type I&Type II&Type III\\
\hline\hline
\multirow{3}{*}{$g_{{\Xi_c^*}'^+{\Xi_c^*}'^0}$}
    &$K\textrm{-loop}$&$\frac{g_1 g_3^2}{2 F_0^2}-\frac{g_2 g_3 g_4}{F_0^2}$&$\frac{g_4^2 g_5}{F_0^2}-\frac{g_3^2 g_5}{F_0^2}$&$\frac{2 g_5^3}{F_0^2}$\\
    &$\eta\textrm{-loop}$&$\frac{g_1 g_3^2}{48 F_0^2}-\frac{g_2 g_3 g_4}{4 F_0^2}$&$\frac{g_4^2 g_5}{4 F_0^2}-\frac{g_3^2 g_5}{24 F_0^2}$&$\frac{g_5^3}{12 F_0^2}$\\
    &$\pi\textrm{-loop}$&$-\frac{g_1 g_3^2}{16 F_0^2}-\frac{g_2 g_4 g_3}{4 F_0^2}$&$\frac{g_5 g_3^2}{8 F_0^2}+\frac{g_4^2 g_5}{4 F_0^2}$&$-\frac{g_5^3}{4 F_0^2}$\\
\hline
\multirow{3}{*}{$g_{\Sigma_c^{*+}\Sigma_c^{*0}}$}
    &$K\textrm{-loop}$&$\frac{g_1 g_3^2}{4 \sqrt{2} F_0^2}-\frac{g_2 g_3 g_4}{\sqrt{2}
   F_0^2}$&$\frac{g_4^2 g_5}{\sqrt{2} F_0^2}-\frac{g_3^2 g_5}{2 \sqrt{2} F_0^2}$&$\frac{g_5^3}{\sqrt{2} F_0^2}$\\
    &$\eta\textrm{-loop}$&$\frac{g_1 g_3^2}{6 \sqrt{2} F_0^2}$&$-\frac{g_3^2 g_5}{3 \sqrt{2} F_0^2}$&$\frac{\sqrt{2} g_5^3}{3 F_0^2}$\\
    &$\pi\textrm{-loop}$&$\frac{g_1 g_3^2}{2 \sqrt{2} F_0^2}-\frac{\sqrt{2} g_2 g_3
   g_4}{F_0^2}$&$\frac{\sqrt{2} g_4^2 g_5}{F_0^2}-\frac{g_3^2 g_5}{\sqrt{2} F_0^2}$&$\frac{\sqrt{2} g_5^3}{F_0^2}$\\
\hline
\multirow{3}{*}{$g_{{\Xi_c^*}'^+\Omega_c^{*0}}$}
    &$K\textrm{-loop}$&$\frac{3 g_1 g_3^2}{4 \sqrt{2} F_0^2}-\frac{\sqrt{2} g_2 g_3
   g_4}{F_0^2}$&$\frac{\sqrt{2} g_4^2 g_5}{F_0^2}-\frac{3 g_3^2 g_5}{2 \sqrt{2}
   F_0^2}$&$\frac{3 g_5^3}{\sqrt{2} F_0^2}$\\
    &$\eta\textrm{-loop}$&$\frac{g_1 g_3^2}{6 \sqrt{2} F_0^2}-\frac{g_2 g_3 g_4}{\sqrt{2}
   F_0^2}$&$\frac{g_4^2 g_5}{\sqrt{2} F_0^2}-\frac{g_3^2 g_5}{3 \sqrt{2} F_0^2}$&$\frac{\sqrt{2} g_5^3}{3 F_0^2}$\\
    &$\pi\textrm{-loop}$&$-$&$-$&$-$\\
\hline
\multirow{3}{*}{$g_{\Sigma_c^{^*+}{\Xi_c^*}'^0}$}
    &$K\textrm{-loop}$&$\frac{g_1 g_3^2}{4 F_0^2}-\frac{g_2 g_3 g_4}{2 F_0^2}$&$\frac{g_4^2 g_5}{2 F_0^2}-\frac{g_3^2 g_5}{2 F_0^2}$&$\frac{g_5^3}{F_0^2}$\\
    &$\eta\textrm{-loop}$&$-\frac{g_1 g_3^2}{24 F_0^2}-\frac{g_2 g_4 g_3}{4 F_0^2}$&$\frac{g_5 g_3^2}{12 F_0^2}+\frac{g_4^2 g_5}{4 F_0^2}$&$-\frac{g_5^3}{6 F_0^2}$\\
    &$\pi\textrm{-loop}$&$\frac{g_1 g_3^2}{4 F_0^2}-\frac{3 g_2 g_3 g_4}{4 F_0^2}$&$\frac{3 g_4^2 g_5}{4 F_0^2}-\frac{g_3^2 g_5}{2 F_0^2}$&$\frac{g_5^3}{F_0^2}$\\
\hline
\multirow{3}{*}{$g_{\Sigma_c^{*++}{\Xi_c^*}'^+}$}
    &$K\textrm{-loop}$&$\frac{g_1 g_3^2}{2 \sqrt{2} F_0^2}-\frac{g_2 g_3 g_4}{\sqrt{2}
   F_0^2}$&$\frac{g_4^2 g_5}{\sqrt{2} F_0^2}-\frac{g_3^2 g_5}{\sqrt{2} F_0^2}$&$\frac{\sqrt{2} g_5^3}{F_0^2}$\\
    &$\eta\textrm{-loop}$&$-\frac{g_1 g_3^2}{12 \sqrt{2} F_0^2}-\frac{g_2 g_4 g_3}{2 \sqrt{2}
   F_0^2}$&$\frac{g_5 g_3^2}{6 \sqrt{2} F_0^2}+\frac{g_4^2 g_5}{2 \sqrt{2}
   F_0^2}$&$-\frac{g_5^3}{3 \sqrt{2} F_0^2}$\\
    &$\pi\textrm{-loop}$&$\frac{g_1 g_3^2}{2 \sqrt{2} F_0^2}-\frac{3 g_2 g_3 g_4}{2 \sqrt{2}
   F_0^2}$&$\frac{3 g_4^2 g_5}{2 \sqrt{2} F_0^2}-\frac{g_3^2 g_5}{\sqrt{2}
   F_0^2}$&$\frac{\sqrt{2} g_5^3}{F_0^2}$\\
\Xhline{1.2pt}
\end{tabular}
\end{table}

\begin{table}[!h]
  \centering
  \renewcommand{\arraystretch}{1.5}
  \caption{The coefficients $g_{3(ij)}^a$ of the axial current from diagram (a).}\label{loopa-g3}
\begin{tabular}{c|ccccc}
\Xhline{1.2pt}
&&Type I&Type II&Type III&Type IV\\
\hline\hline
\multirow{3}{*}{$g_{\Xi_c'^+{\Xi_c^*}'^0}$}
    &$K\textrm{-loop}$&$\frac{g_3 g_1^2}{F_0^2}-\frac{g_2 g_4 g_1}{F_0^2}-\frac{g_2^2
   g_3}{F_0^2}$&$\frac{g_3 g_4^2}{4 F_0^2}-\frac{g_3^3}{4 F_0^2}$&$\frac{g_2 g_4 g_5}{F_0^2}-\frac{g_1 g_3 g_5}{F_0^2}$&$\frac{g_3 g_5^2}{F_0^2}$\\
    &$\eta\textrm{-loop}$&$\frac{g_3 g_1^2}{24 F_0^2}-\frac{g_2 g_4 g_1}{4 F_0^2}-\frac{g_2^2
   g_3}{4 F_0^2}$&$\frac{g_3 g_4^2}{16 F_0^2}-\frac{g_3^3}{96 F_0^2}$&$\frac{g_2 g_4 g_5}{4 F_0^2}-\frac{g_1 g_3 g_5}{24 F_0^2}$&$\frac{g_3 g_5^2}{24 F_0^2}$\\
    &$\pi\textrm{-loop}$&$-\frac{g_3 g_1^2}{8 F_0^2}-\frac{g_2 g_4 g_1}{4 F_0^2}-\frac{g_2^2
   g_3}{4 F_0^2}$&$\frac{g_3^3}{32 F_0^2}+\frac{g_4^2 g_3}{16 F_0^2}$&$\frac{g_1 g_3 g_5}{8 F_0^2}+\frac{g_2 g_4 g_5}{4 F_0^2}$&$-\frac{g_3 g_5^2}{8 F_0^2}$\\
\hline
\multirow{3}{*}{$g_{\Sigma_c^+\Sigma_c^{*0}}$}
    &$K\textrm{-loop}$&$\frac{g_3 g_1^2}{2 \sqrt{2} F_0^2}-\frac{g_2 g_4 g_1}{\sqrt{2}
   F_0^2}-\frac{g_2^2 g_3}{\sqrt{2} F_0^2}$&$\frac{g_3 g_4^2}{4 \sqrt{2} F_0^2}-\frac{g_3^3}{8 \sqrt{2} F_0^2}$&$\frac{g_2 g_4 g_5}{\sqrt{2} F_0^2}-\frac{g_1 g_3 g_5}{2 \sqrt{2}
   F_0^2}$&$\frac{g_3 g_5^2}{2 \sqrt{2} F_0^2}$\\
    &$\eta\textrm{-loop}$&$\frac{g_1^2 g_3}{3 \sqrt{2} F_0^2}$&$-\frac{g_3^3}{12 \sqrt{2} F_0^2}$&$-\frac{g_1 g_3 g_5}{3 \sqrt{2} F_0^2}$&$\frac{g_3 g_5^2}{3 \sqrt{2} F_0^2}$\\
    &$\pi\textrm{-loop}$&$\frac{g_3 g_1^2}{\sqrt{2} F_0^2}-\frac{\sqrt{2} g_2 g_4
   g_1}{F_0^2}-\frac{\sqrt{2} g_2^2 g_3}{F_0^2}$&$\frac{g_3 g_4^2}{2 \sqrt{2} F_0^2}-\frac{g_3^3}{4 \sqrt{2} F_0^2}$&$\frac{\sqrt{2} g_2 g_4 g_5}{F_0^2}-\frac{g_1 g_3 g_5}{\sqrt{2}
   F_0^2}$&$\frac{g_3 g_5^2}{\sqrt{2} F_0^2}$\\
\hline
\multirow{3}{*}{$g_{\Sigma_c^{++}\Sigma_c^{*+}}$}
    &$K\textrm{-loop}$&$\frac{g_3 g_1^2}{2 \sqrt{2} F_0^2}-\frac{g_2 g_4 g_1}{\sqrt{2}
   F_0^2}-\frac{g_2^2 g_3}{\sqrt{2} F_0^2}$&$\frac{g_3 g_4^2}{4 \sqrt{2} F_0^2}-\frac{g_3^3}{8 \sqrt{2} F_0^2}$&$\frac{g_2 g_4 g_5}{\sqrt{2} F_0^2}-\frac{g_1 g_3 g_5}{2 \sqrt{2}
   F_0^2}$&$\frac{g_3 g_5^2}{2 \sqrt{2} F_0^2}$\\
    &$\eta\textrm{-loop}$&$\frac{g_1^2 g_3}{3 \sqrt{2} F_0^2}$&$-\frac{g_3^3}{12 \sqrt{2} F_0^2}$&$-\frac{g_1 g_3 g_5}{3 \sqrt{2} F_0^2}$&$\frac{g_3 g_5^2}{3 \sqrt{2} F_0^2}$\\
    &$\pi\textrm{-loop}$&$\frac{g_3 g_1^2}{\sqrt{2} F_0^2}-\frac{\sqrt{2} g_2 g_4
   g_1}{F_0^2}-\frac{\sqrt{2} g_2^2 g_3}{F_0^2}$&$\frac{g_3 g_4^2}{2 \sqrt{2} F_0^2}-\frac{g_3^3}{4 \sqrt{2} F_0^2}$&$\frac{\sqrt{2} g_2 g_4 g_5}{F_0^2}-\frac{g_1 g_3 g_5}{\sqrt{2}
   F_0^2}$&$\frac{g_3 g_5^2}{\sqrt{2} F_0^2}$\\
\hline
\multirow{3}{*}{$g_{{\Xi_c}'^+\Omega_c^{*0}}$}
    &$K\textrm{-loop}$&$\frac{3 g_3 g_1^2}{2 \sqrt{2} F_0^2}-\frac{3 g_2 g_4 g_1}{\sqrt{2}
   F_0^2}-\frac{g_2^2 g_3}{\sqrt{2} F_0^2}$&$\frac{3 g_3 g_4^2}{4 \sqrt{2} F_0^2}-\frac{3 g_3^3}{8 \sqrt{2}
   F_0^2}$&$\frac{g_2 g_4 g_5}{\sqrt{2} F_0^2}-\frac{3 g_1 g_3 g_5}{2 \sqrt{2}
   F_0^2}$&$\frac{3 g_3 g_5^2}{2 \sqrt{2} F_0^2}$\\
    &$\eta\textrm{-loop}$&$\frac{g_1^2 g_3}{3 \sqrt{2} F_0^2}-\frac{\sqrt{2} g_2^2 g_3}{F_0^2}$&$-\frac{g_3^3}{12 \sqrt{2} F_0^2}$&$\frac{\sqrt{2} g_2 g_4 g_5}{F_0^2}-\frac{g_1 g_3 g_5}{3 \sqrt{2}
   F_0^2}$&$\frac{g_3 g_5^2}{3 \sqrt{2} F_0^2}$\\
    &$\pi\textrm{-loop}$&$-$&$-$&$-$&$-$\\
\hline
\multirow{3}{*}{$g_{\Sigma_c^+{\Xi_c^*}'^+}$}
    &$K\textrm{-loop}$&$\frac{g_1^2 g_3}{2 F_0^2}-\frac{g_2^2 g_3}{F_0^2}$&$-\frac{g_3^3}{8 F_0^2}$&$\frac{g_2 g_4 g_5}{F_0^2}-\frac{g_1 g_3 g_5}{2 F_0^2}$&$\frac{g_3 g_5^2}{2 F_0^2}$\\
    &$\eta\textrm{-loop}$&$-\frac{g_3 g_1^2}{12 F_0^2}-\frac{g_2 g_4 g_1}{2 F_0^2}$&$\frac{g_3^3}{48 F_0^2}+\frac{g_4^2 g_3}{8 F_0^2}$&$\frac{g_1 g_3 g_5}{12 F_0^2}$&$-\frac{g_3 g_5^2}{12 F_0^2}$\\
    &$\pi\textrm{-loop}$&$\frac{g_3 g_1^2}{2 F_0^2}-\frac{g_2 g_4 g_1}{F_0^2}-\frac{g_2^2
   g_3}{2 F_0^2}$&$\frac{g_3 g_4^2}{4 F_0^2}-\frac{g_3^3}{8 F_0^2}$&$\frac{g_2 g_4 g_5}{2 F_0^2}-\frac{g_1 g_3 g_5}{2 F_0^2}$&$\frac{g_3 g_5^2}{2 F_0^2}$\\
\hline
\multirow{3}{*}{$g_{\Sigma_c^{++}{\Xi_c^*}'^+}$}
    &$K\textrm{-loop}$&$\frac{g_1^2 g_3}{\sqrt{2} F_0^2}-\frac{\sqrt{2} g_2^2 g_3}{F_0^2}$&$-\frac{g_3^3}{4 \sqrt{2} F_0^2}$&$\frac{\sqrt{2} g_2 g_4 g_5}{F_0^2}-\frac{g_1 g_3 g_5}{\sqrt{2}
   F_0^2}$&$\frac{g_3 g_5^2}{\sqrt{2} F_0^2}$\\
    &$\eta\textrm{-loop}$&$-\frac{g_3 g_1^2}{6 \sqrt{2} F_0^2}-\frac{g_2 g_4 g_1}{\sqrt{2}
   F_0^2}$&$\frac{g_3^3}{24 \sqrt{2} F_0^2}+\frac{g_4^2 g_3}{4 \sqrt{2} F_0^2}$&$\frac{g_1 g_3 g_5}{6 \sqrt{2} F_0^2}$&$-\frac{g_3 g_5^2}{6 \sqrt{2} F_0^2}$\\
    &$\pi\textrm{-loop}$&$\frac{g_3 g_1^2}{\sqrt{2} F_0^2}-\frac{\sqrt{2} g_2 g_4
   g_1}{F_0^2}-\frac{g_2^2 g_3}{\sqrt{2} F_0^2}$&$\frac{g_3 g_4^2}{2 \sqrt{2} F_0^2}-\frac{g_3^3}{4 \sqrt{2} F_0^2}$&$\frac{g_2 g_4 g_5}{\sqrt{2} F_0^2}-\frac{g_1 g_3 g_5}{\sqrt{2}
   F_0^2}$&$\frac{g_3 g_5^2}{\sqrt{2} F_0^2}$\\
\hline
\multirow{3}{*}{$g_{{\Xi_c^*}'^+\Omega_c^0}$}
    &$K\textrm{-loop}$&$\frac{3 g_3 g_1^2}{2 \sqrt{2} F_0^2}-\frac{g_2 g_4 g_1}{\sqrt{2}
   F_0^2}-\frac{3 g_2^2 g_3}{\sqrt{2} F_0^2}$&$\frac{g_3 g_4^2}{4 \sqrt{2} F_0^2}-\frac{3 g_3^3}{8 \sqrt{2} F_0^2}$&$\frac{3 g_2 g_4 g_5}{\sqrt{2} F_0^2}-\frac{3 g_1 g_3 g_5}{2 \sqrt{2}
   F_0^2}$&$\frac{3 g_3 g_5^2}{2 \sqrt{2} F_0^2}$\\
    &$\eta\textrm{-loop}$&$\frac{g_1^2 g_3}{3 \sqrt{2} F_0^2}-\frac{\sqrt{2} g_1 g_2
   g_4}{F_0^2}$&$\frac{g_3 g_4^2}{2 \sqrt{2} F_0^2}-\frac{g_3^3}{12 \sqrt{2} F_0^2}$&$-\frac{g_1 g_3 g_5}{3 \sqrt{2} F_0^2}$&$\frac{g_3 g_5^2}{3 \sqrt{2} F_0^2}$\\
    &$\pi\textrm{-loop}$&$-$&$-$&$-$&$-$\\
\hline
\multirow{3}{*}{$g_{\Sigma_c^{*+}\Xi_c'^+}$}
    &$K\textrm{-loop}$&$\frac{g_1^2 g_3}{2 F_0^2}-\frac{g_1 g_2 g_4}{F_0^2}$&$\frac{g_3 g_4^2}{4 F_0^2}-\frac{g_3^3}{8 F_0^2}$&$-\frac{g_1 g_3 g_5}{2 F_0^2}$&$\frac{g_3 g_5^2}{2 F_0^2}$\\
    &$\eta\textrm{-loop}$&$-\frac{g_3 g_1^2}{12 F_0^2}-\frac{g_2^2 g_3}{2 F_0^2}$&$\frac{g_3^3}{48 F_0^2}$&$\frac{g_1 g_3 g_5}{12 F_0^2}+\frac{g_2 g_4 g_5}{2 F_0^2}$&$-\frac{g_3 g_5^2}{12 F_0^2}$\\
    &$\pi\textrm{-loop}$&$\frac{g_3 g_1^2}{2 F_0^2}-\frac{g_2 g_4 g_1}{2 F_0^2}-\frac{g_2^2
   g_3}{F_0^2}$&$\frac{g_3 g_4^2}{8 F_0^2}-\frac{g_3^3}{8 F_0^2}$&$\frac{g_2 g_4 g_5}{F_0^2}-\frac{g_1 g_3 g_5}{2 F_0^2}$&$\frac{g_3 g_5^2}{2 F_0^2}$\\
\hline
\multirow{3}{*}{$g_{\Sigma_c^{*++}\Xi_c'^+}$}
    &$K\textrm{-loop}$&$\frac{g_1^2 g_3}{\sqrt{2} F_0^2}-\frac{\sqrt{2} g_1 g_2 g_4}{F_0^2}$&$\frac{g_3 g_4^2}{2 \sqrt{2} F_0^2}-\frac{g_3^3}{4 \sqrt{2} F_0^2}$&$-\frac{g_1 g_3 g_5}{\sqrt{2} F_0^2}$&$\frac{g_3 g_5^2}{\sqrt{2} F_0^2}$\\
    &$\eta\textrm{-loop}$&$-\frac{g_3 g_1^2}{6 \sqrt{2} F_0^2}-\frac{g_2^2 g_3}{\sqrt{2} F_0^2}$&$\frac{g_3^3}{24 \sqrt{2} F_0^2}$&$\frac{g_1 g_3 g_5}{6 \sqrt{2} F_0^2}+\frac{g_2 g_4 g_5}{\sqrt{2}
   F_0^2}$&$-\frac{g_3 g_5^2}{6 \sqrt{2} F_0^2}$\\
    &$\pi\textrm{-loop}$&$\frac{g_3 g_1^2}{\sqrt{2} F_0^2}-\frac{g_2 g_4 g_1}{\sqrt{2}
   F_0^2}-\frac{\sqrt{2} g_2^2 g_3}{F_0^2}$&$\frac{g_3 g_4^2}{4 \sqrt{2} F_0^2}-\frac{g_3^3}{4 \sqrt{2} F_0^2}$&$\frac{\sqrt{2} g_2 g_4 g_5}{F_0^2}-\frac{g_1 g_3 g_5}{\sqrt{2}
   F_0^2}$&$\frac{g_3 g_5^2}{\sqrt{2} F_0^2}$\\
\Xhline{1.2pt}
\end{tabular}
\end{table}

\begin{table}[!h]
  \centering
  \renewcommand{\arraystretch}{1.5}
  \caption{The coefficients $g_{4(ij)}^a$ of the axial current from diagram (a).}\label{loopa-g4}
\begin{tabular}{c|cccp{25mm}<{\centering}p{25mm}<{\centering}}
\Xhline{1.2pt}
&&Type I&Type II&Type III&Type IV\\
\hline\hline
\multirow{3}{*}{$g_{\Lambda_c^+\Sigma_c^{*0}}$}
    &$K\textrm{-loop}$&$\frac{g_2^2 g_4}{F_0^2}-\frac{g_1 g_2 g_3}{2 F_0^2}$&$\frac{g_3^2 g_4}{8 F_0^2}-\frac{g_4^3}{4 F_0^2}$&$\frac{g_2 g_3 g_5}{2 F_0^2}$&$-\frac{g_4 g_5^2}{2 F_0^2}$\\
    &$\eta\textrm{-loop}$&$-$&$-$&$-$&$-$\\
    &$\pi\textrm{-loop}$&$\frac{2 g_2^2 g_4}{F_0^2}-\frac{2 g_1 g_2 g_3}{F_0^2}$&$\frac{g_3^2 g_4}{2 F_0^2}-\frac{g_4^3}{2 F_0^2}$&$\frac{2 g_2 g_3 g_5}{F_0^2}$&$-\frac{2 g_4 g_5^2}{F_0^2}$\\
\hline
\multirow{3}{*}{$g_{\Xi_c^+{\Xi_c^*}'^0}$}
    &$K\textrm{-loop}$&$\frac{\sqrt{2} g_2^2 g_4}{F_0^2}-\frac{\sqrt{2} g_1 g_2 g_3}{F_0^2}$&$\frac{g_3^2 g_4}{2 \sqrt{2} F_0^2}-\frac{g_4^3}{2 \sqrt{2} F_0^2}$&$\frac{\sqrt{2} g_2 g_3 g_5}{F_0^2}$&$-\frac{\sqrt{2} g_4 g_5^2}{F_0^2}$\\
    &$\eta\textrm{-loop}$&$\frac{3 g_2^2 g_4}{2 \sqrt{2} F_0^2}-\frac{g_1 g_2 g_3}{4 \sqrt{2} F_0^2}$&$\frac{g_3^2 g_4}{16 \sqrt{2} F_0^2}-\frac{3 g_4^3}{8 \sqrt{2} F_0^2}$&$\frac{g_2 g_3 g_5}{4 \sqrt{2} F_0^2}$&$-\frac{g_4 g_5^2}{4 \sqrt{2} F_0^2}$\\
    &$\pi\textrm{-loop}$&$-\frac{g_4 g_2^2}{2 \sqrt{2} F_0^2}-\frac{g_1 g_3 g_2}{4 \sqrt{2} F_0^2}$&$\frac{g_4^3}{8 \sqrt{2} F_0^2}+\frac{g_3^2 g_4}{16 \sqrt{2} F_0^2}$&$\frac{g_2 g_3 g_5}{4 \sqrt{2} F_0^2}$&$-\frac{g_4 g_5^2}{4 \sqrt{2} F_0^2}$\\
\hline
\multirow{3}{*}{$g_{\Lambda_c^+{\Xi_c^*}'^0}$}
    &$K\textrm{-loop}$&$-\frac{g_1 g_2 g_3}{\sqrt{2} F_0^2}$&$\frac{g_3^2 g_4}{4 \sqrt{2} F_0^2}$&$\frac{g_2 g_3 g_5}{\sqrt{2} F_0^2}$&$-\frac{g_4 g_5^2}{\sqrt{2} F_0^2}$\\
    &$\eta\textrm{-loop}$&$-$&$-$&$-$&$-$\\
    &$\pi\textrm{-loop}$&$\frac{3 g_2^2 g_4}{\sqrt{2} F_0^2}-\frac{3 g_1 g_2 g_3}{2 \sqrt{2} F_0^2}$&$\frac{3 g_3^2 g_4}{8 \sqrt{2} F_0^2}-\frac{3 g_4^3}{4 \sqrt{2} F_0^2}$&$\frac{3 g_2 g_3 g_5}{2 \sqrt{2} F_0^2}$&$-\frac{3 g_4 g_5^2}{2 \sqrt{2} F_0^2}$\\
\hline
\multirow{3}{*}{$g_{\Xi_c^+\Omega_c^{*0}}$}
    &$K\textrm{-loop}$&$\frac{3 g_2^2 g_4}{F_0^2}-\frac{3 g_1 g_2 g_3}{2 F_0^2}$&$\frac{3 g_3^2 g_4}{8 F_0^2}-\frac{3 g_4^3}{4 F_0^2}$&$\frac{3 g_2 g_3 g_5}{2 F_0^2}$&$-\frac{3 g_4 g_5^2}{2 F_0^2}$\\
    &$\eta\textrm{-loop}$&$-\frac{g_1 g_2 g_3}{F_0^2}$&$\frac{g_3^2 g_4}{4 F_0^2}$&$\frac{g_2 g_3 g_5}{F_0^2}$&$-\frac{g_4 g_5^2}{F_0^2}$\\
    &$\pi\textrm{-loop}$&$-$&$-$&$-$&$-$\\
\Xhline{1.2pt}
\end{tabular}
\bigskip
  \centering
  \renewcommand{\arraystretch}{1.5}
  \caption{The coefficients $g_{(ij)}^{b}$ of the axial current from diagram (b).}\label{loopb}
\begin{minipage}{0.5\textwidth}
\begin{tabular}{cccc}
\Xhline{1.2pt}
&$K$-loop&$\eta$-loop&$\pi$-loop\\
\hline\hline
$g_{\Xi_c'^+\Xi_c'^0}$&$-\frac{g_1}{2 F_0^2}$&$-\frac{g_1}{12 F_0^2}$&$-\frac{3 g_1}{4 F_0^2}$\\
$g_{\Sigma_c^+\Sigma_c^0}$&$-\frac{g_1}{\sqrt{2} F_0^2}$&$-\frac{g_1}{6 \sqrt{2} F_0^2}$&$-\frac{3 g_1}{2 \sqrt{2} F_0^2}$\\
$g_{\Xi_c'^+\Omega_c^0}$&$-\frac{3 g_1}{2 \sqrt{2} F_0^2}$&$-\frac{5 g_1}{12 \sqrt{2} F_0^2}$&$-\frac{3 g_1}{4 \sqrt{2} F_0^2}$\\
$g_{\Sigma_c^+\Xi_c'^0}$&$-\frac{3 g_1}{4 F_0^2}$&$-\frac{5 g_1}{24 F_0^2}$&$-\frac{3 g_1}{8 F_0^2}$\\
$g_{\Sigma_c^{++}\Xi_c'^+}$&$-\frac{3 g_1}{2 \sqrt{2} F_0^2}$&$-\frac{5 g_1}{12 \sqrt{2} F_0^2}$&$-\frac{3 g_1}{4 \sqrt{2} F_0^2}$\\
\hline\hline
$g_{\Lambda_c^+\Sigma_c^0}$&$-\frac{g_2}{F_0^2}$&$-\frac{g_2}{6 F_0^2}$&$-\frac{3 g_2}{2 F_0^2}$\\
$g_{\Xi_c^+\Xi_c'^0}$&$-\frac{g_2}{\sqrt{2} F_0^2}$&$-\frac{g_2}{6 \sqrt{2} F_0^2}$&$-\frac{3 g_2}{2 \sqrt{2} F_0^2}$\\
$g_{\Lambda_c^+\Xi_c'^0}$&$-\frac{3 g_2}{2 \sqrt{2} F_0^2}$&$-\frac{5 g_2}{12 \sqrt{2} F_0^2}$&$-\frac{3 g_2}{4 \sqrt{2} F_0^2}$\\
$g_{\Xi_c^+\Omega_c^0}$&$-\frac{3 g_2}{2 F_0^2}$&$-\frac{5 g_2}{12 F_0^2}$&$-\frac{3 g_2}{4 F_0^2}$\\
$g_{\Sigma_c^+\Xi_c^0}$&$\frac{3 g_2}{2 \sqrt{2} F_0^2}$&$\frac{5 g_2}{12 \sqrt{2} F_0^2}$&$\frac{3 g_2}{4 \sqrt{2} F_0^2}$\\
$g_{\Sigma_c^{++}\Xi_c^+}$&$\frac{3 g_2}{2 F_0^2}$&$\frac{5 g_2}{12 F_0^2}$&$\frac{3 g_2}{4 F_0^2}$\\
\hline\hline
$g_{{\Xi_c^*}'^+{\Xi_c^*}'^0}$&$-\frac{g_5}{2 F_0^2}$&$-\frac{g_5}{12 F_0^2}$&$-\frac{3 g_5}{4 F_0^2}$\\
$g_{\Sigma_c^{*+}\Sigma_c^{*0}}$&$-\frac{g_5}{\sqrt{2} F_0^2}$&$-\frac{g_5}{6 \sqrt{2} F_0^2}$&$-\frac{3 g_5}{2 \sqrt{2} F_0^2}$\\
$g_{{\Xi_c^*}'^+\Omega_c^{*0}}$&$-\frac{3 g_5}{2 \sqrt{2} F_0^2}$&$-\frac{5 g_5}{12 \sqrt{2} F_0^2}$&$-\frac{3 g_5}{4 \sqrt{2} F_0^2}$\\
\end{tabular}
\end{minipage}
\begin{minipage}{0.45\textwidth}
\begin{tabular}{cccc}
$g_{\Sigma_c^{*+}{\Xi_c^*}'^0}$&$-\frac{3 g_5}{4 F_0^2}$&$-\frac{5 g_5}{24 F_0^2}$&$-\frac{3 g_5}{8 F_0^2}$\\
$g_{\Sigma_c^{*++}{\Xi_c^*}'^+}$&$-\frac{3 g_5}{2 \sqrt{2} F_0^2}$&$-\frac{5 g_5}{12 \sqrt{2} F_0^2}$&$-\frac{3 g_5}{4 \sqrt{2} F_0^2}$\\
\hline\hline
$g_{\Xi_c'^+{\Xi_c^*}'^0}$&$-\frac{g_3}{4 F_0^2}$&$-\frac{g_3}{24 F_0^2}$&$-\frac{3 g_3}{8 F_0^2}$\\
$g_{\Sigma_c^+\Sigma_c^{*0}}$&$-\frac{g_3}{2 \sqrt{2} F_0^2}$&$-\frac{g_3}{12 \sqrt{2} F_0^2}$&$-\frac{3 g_3}{4 \sqrt{2} F_0^2}$\\
$g_{\Sigma_c^{++}\Sigma_c^{*+}}$&$-\frac{g_3}{2 \sqrt{2} F_0^2}$&$-\frac{g_3}{12 \sqrt{2} F_0^2}$&$-\frac{3 g_3}{4 \sqrt{2} F_0^2}$\\
$g_{\Xi_c'^+\Omega_c^{*0}}$&$-\frac{3 g_3}{4 \sqrt{2} F_0^2}$&$-\frac{5 g_3}{24 \sqrt{2} F_0^2}$&$-\frac{3 g_3}{8 \sqrt{2} F_0^2}$\\
$g_{\Sigma_c^+{\Xi_c^*}'^0}$&$-\frac{3 g_3}{8 F_0^2}$&$-\frac{5 g_3}{48 F_0^2}$&$-\frac{3 g_3}{16 F_0^2}$\\
$g_{\Sigma_c^{++}{\Xi_c^*}'^+}$&$-\frac{3 g_3}{4 \sqrt{2} F_0^2}$&$-\frac{5 g_3}{24 \sqrt{2} F_0^2}$&$-\frac{3 g_3}{8 \sqrt{2} F_0^2}$\\
$g_{{\Xi_c^*}'^+\Omega_c^0}$&$-\frac{3 g_3}{4 \sqrt{2} F_0^2}$&$-\frac{5 g_3}{24 \sqrt{2} F_0^2}$&$-\frac{3 g_3}{8 \sqrt{2} F_0^2}$\\
$g_{\Sigma_c^{*+}\Xi_c'^0}$&$-\frac{3 g_3}{8 F_0^2}$&$-\frac{5 g_3}{48 F_0^2}$&$-\frac{3 g_3}{16 F_0^2}$\\
$g_{\Sigma_c^{*++}\Xi_c'^+}$&$-\frac{3 g_3}{4 \sqrt{2} F_0^2}$&$-\frac{5 g_3}{24 \sqrt{2} F_0^2}$&$-\frac{3 g_3}{8 \sqrt{2} F_0^2}$\\
\hline\hline
$g_{\Lambda_c^+\Sigma_c^{*0}}$&$-\frac{g_4}{2 F_0^2}$&$-\frac{g_4}{12 F_0^2}$&$-\frac{3 g_4}{4 F_0^2}$\\
$g_{\Xi_c^+{\Xi_c^*}'^0}$&$-\frac{g_4}{2 \sqrt{2} F_0^2}$&$-\frac{g_4}{12 \sqrt{2} F_0^2}$&$-\frac{3 g_4}{4 \sqrt{2} F_0^2}$\\
$g_{\Lambda_c^+{\Xi_c^*}'^0}$&$-\frac{3 g_4}{4 \sqrt{2} F_0^2}$&$-\frac{5 g_4}{24 \sqrt{2} F_0^2}$&$-\frac{3 g_4}{8 \sqrt{2} F_0^2}$\\
$g_{\Xi_c^+\Omega_c^{*0}}$&$-\frac{3 g_4}{4 F_0^2}$&$-\frac{5 g_4}{24 F_0^2}$&$-\frac{3 g_4}{8 F_0^2}$\\
\Xhline{1.2pt}
\end{tabular}
\end{minipage}
\end{table}

\begin{table}[!h]
  \centering
  \renewcommand{\arraystretch}{1.5}
  \caption{The coefficients $\lambda_{1(ij)}$ of the axial
  current from the wave function renormalization.}\label{wav-1}
\begin{tabular}{c|ccc}
\Xhline{1.2pt}
&&Case I&Case II\\
\hline\hline
\multirow{3}{*}{$\lambda_{\Xi_c'^+\Xi_c'^0}$}
    &$K\textrm{-loop}$&$\frac{5 g_1^2}{2 F_0^2}+\frac{g_2^2}{F_0^2}$&$-\frac{5 g_3^2}{8 F_0^2}$\\
    &$\eta\textrm{-loop}$&$\frac{g_1^2}{12 F_0^2}+\frac{3 g_2^2}{2 F_0^2}$&$-\frac{g_3^2}{48 F_0^2}$\\
    &$\pi\textrm{-loop}$&$\frac{3 g_1^2}{4 F_0^2}+\frac{3 g_2^2}{2 F_0^2}$&$-\frac{3 g_3^2}{16 F_0^2}$\\
\hline
\multirow{3}{*}{$\lambda_{\Sigma_c^+\Sigma_c^0}$}
    &$K\textrm{-loop}$&$\frac{g_1^2}{F_0^2}+\frac{2 g_2^2}{F_0^2}$&$-\frac{g_3^2}{4 F_0^2}$\\
    &$\eta\textrm{-loop}$&$\frac{g_1^2}{3 F_0^2}$&$-\frac{g_3^2}{12 F_0^2}$\\
    &$\pi\textrm{-loop}$&$\frac{2 g_1^2}{F_0^2}+\frac{2 g_2^2}{F_0^2}$&$-\frac{g_3^2}{2 F_0^2}$\\
\hline
\multirow{3}{*}{$\lambda_{\Xi_c'^+\Omega_c^0}$}
    &$K\textrm{-loop}$&$\frac{9 g_1^2}{4 F_0^2}+\frac{5 g_2^2}{2 F_0^2}$&$-\frac{9 g_3^2}{16 F_0^2}$\\
    &$\eta\textrm{-loop}$&$\frac{17 g_1^2}{24 F_0^2}+\frac{3 g_2^2}{4 F_0^2}$&$-\frac{17 g_3^2}{96 F_0^2}$\\
    &$\pi\textrm{-loop}$&$\frac{3 g_1^2}{8 F_0^2}+\frac{3 g_2^2}{4 F_0^2}$&$-\frac{3 g_3^2}{32 F_0^2}$\\
\hline
\multirow{3}{*}{$\lambda_{\Sigma_c^+\Xi_c'^0}$}
    &$K\textrm{-loop}$&$\frac{7 g_1^2}{4 F_0^2}+\frac{3 g_2^2}{2 F_0^2}$&$-\frac{7 g_3^2}{16 F_0^2}$\\
    &$\eta\textrm{-loop}$&$\frac{5 g_1^2}{24 F_0^2}+\frac{3 g_2^2}{4 F_0^2}$&$-\frac{5 g_3^2}{96 F_0^2}$\\
    &$\pi\textrm{-loop}$&$\frac{11 g_1^2}{8 F_0^2}+\frac{7 g_2^2}{4 F_0^2}$&$-\frac{11 g_3^2}{32 F_0^2}$\\
\hline
\multirow{3}{*}{$\lambda_{\Sigma_c^{++}\Xi_c'^+}$}
    &$K\textrm{-loop}$&$\frac{7 g_1^2}{4 F_0^2}+\frac{3 g_2^2}{2 F_0^2}$&$-\frac{7 g_3^2}{16 F_0^2}$\\
    &$\eta\textrm{-loop}$&$\frac{5 g_1^2}{24 F_0^2}+\frac{3 g_2^2}{4 F_0^2}$&$-\frac{5 g_3^2}{96 F_0^2}$\\
    &$\pi\textrm{-loop}$&$\frac{11 g_1^2}{8 F_0^2}+\frac{7 g_2^2}{4 F_0^2}$&$-\frac{11 g_3^2}{32 F_0^2}$\\
\Xhline{1.2pt}
\end{tabular}
\end{table}

\begin{table}[!h]
  \centering
  \renewcommand{\arraystretch}{1.5}
  \caption{The coefficients $\lambda_{2(ij)}$ of the axial current from the wave function renormalization.}\label{wav-2}
\begin{tabular}{c|ccc}
\Xhline{1.2pt}
&&Case I&Case II\\
\hline\hline
\multirow{3}{*}{$\lambda_{\Lambda_c^+\Sigma_c^0}$}
    &$K\textrm{-loop}$&$\frac{g_1^2}{2 F_0^2}+\frac{2 g_2^2}{F_0^2}+\frac{2 g_6^2}{F_0^2}$&$-\frac{g_3^2}{8 F_0^2}-\frac{g_4^2}{4 F_0^2}$\\
    &$\eta\textrm{-loop}$&$\frac{g_1^2}{6 F_0^2}+\frac{2 g_6^2}{3 F_0^2}$&$-\frac{g_3^2}{24 F_0^2}$\\
    &$\pi\textrm{-loop}$&$\frac{g_1^2}{F_0^2}+\frac{4 g_2^2}{F_0^2}$&$-\frac{g_3^2}{4 F_0^2}-\frac{3 g_4^2}{4 F_0^2}$\\
\hline
\multirow{3}{*}{$\lambda_{\Xi_c^+\Xi_c'^0}$}
    &$K\textrm{-loop}$&$\frac{5 g_1^2}{4 F_0^2}+\frac{3 g_2^2}{F_0^2}+\frac{g_6^2}{F_0^2}$&$-\frac{5 g_3^2}{16 F_0^2}-\frac{5 g_4^2}{8 F_0^2}$\\
    &$\eta\textrm{-loop}$&$\frac{g_1^2}{24 F_0^2}+\frac{3 g_2^2}{2 F_0^2}+\frac{g_6^2}{6 F_0^2}$&$-\frac{g_3^2}{96 F_0^2}-\frac{3 g_4^2}{16 F_0^2}$\\
    &$\pi\textrm{-loop}$&$\frac{3 g_1^2}{8 F_0^2}+\frac{3 g_2^2}{2 F_0^2}+\frac{3 g_6^2}{2 F_0^2}$&$-\frac{3 g_3^2}{32 F_0^2}-\frac{3 g_4^2}{16 F_0^2}$\\
\hline
\multirow{3}{*}{$\lambda_{\Lambda_c^+\Xi_c'^0}$}
    &$K\textrm{-loop}$&$\frac{5 g_1^2}{4 F_0^2}+\frac{3 g_2^2}{2 F_0^2}+\frac{2 g_6^2}{F_0^2}$&$-\frac{5 g_3^2}{16 F_0^2}-\frac{g_4^2}{4 F_0^2}$\\
    &$\eta\textrm{-loop}$&$\frac{g_1^2}{24 F_0^2}+\frac{3 g_2^2}{4 F_0^2}+\frac{2 g_6^2}{3 F_0^2}$&$-\frac{g_3^2}{96 F_0^2}$\\
    &$\pi\textrm{-loop}$&$\frac{3 g_1^2}{8 F_0^2}+\frac{15 g_2^2}{4 F_0^2}$&$-\frac{3 g_3^2}{32 F_0^2}-\frac{3 g_4^2}{4 F_0^2}$\\
\hline
\multirow{3}{*}{$\lambda_{\Xi_c^+\Omega_c^0}$}
    &$K\textrm{-loop}$&$\frac{g_1^2}{F_0^2}+\frac{9 g_2^2}{2 F_0^2}+\frac{g_6^2}{F_0^2}$&$-\frac{g_3^2}{4 F_0^2}-\frac{5 g_4^2}{8 F_0^2}$\\
    &$\eta\textrm{-loop}$&$\frac{2 g_1^2}{3 F_0^2}+\frac{3 g_2^2}{4 F_0^2}+\frac{g_6^2}{6 F_0^2}$&$-\frac{g_3^2}{6 F_0^2}-\frac{3 g_4^2}{16 F_0^2}$\\
    &$\pi\textrm{-loop}$&$\frac{3 g_2^2}{4 F_0^2}+\frac{3 g_6^2}{2 F_0^2}$&$-\frac{3 g_4^2}{16 F_0^2}$\\
\hline
\multirow{3}{*}{$\lambda_{\Sigma_c^+\Xi_c^+}$}
    &$K\textrm{-loop}$&$\frac{g_1^2}{2 F_0^2}+\frac{7 g_2^2}{2 F_0^2}+\frac{g_6^2}{F_0^2}$&$-\frac{g_3^2}{8 F_0^2}-\frac{5 g_4^2}{8 F_0^2}$\\
    &$\eta\textrm{-loop}$&$\frac{g_1^2}{6 F_0^2}+\frac{3 g_2^2}{4 F_0^2}+\frac{g_6^2}{6 F_0^2}$&$-\frac{g_3^2}{24 F_0^2}-\frac{3 g_4^2}{16 F_0^2}$\\
    &$\pi\textrm{-loop}$&$\frac{g_1^2}{F_0^2}+\frac{7 g_2^2}{4 F_0^2}+\frac{3 g_6^2}{2 F_0^2}$&$-\frac{g_3^2}{4 F_0^2}-\frac{3 g_4^2}{16 F_0^2}$\\
\hline
\multirow{3}{*}{$\lambda_{\Sigma_c^{++}\Xi_c^+}$}
    &$K\textrm{-loop}$&$\frac{g_1^2}{2 F_0^2}+\frac{7 g_2^2}{2 F_0^2}+\frac{g_6^2}{F_0^2}$&$-\frac{g_3^2}{8 F_0^2}-\frac{5 g_4^2}{8 F_0^2}$\\
    &$\eta\textrm{-loop}$&$\frac{g_1^2}{6 F_0^2}+\frac{3 g_2^2}{4 F_0^2}+\frac{g_6^2}{6 F_0^2}$&$-\frac{g_3^2}{24 F_0^2}-\frac{3 g_4^2}{16 F_0^2}$\\
    &$\pi\textrm{-loop}$&$\frac{g_1^2}{F_0^2}+\frac{7 g_2^2}{4 F_0^2}+\frac{3 g_6^2}{2 F_0^2}$&$-\frac{g_3^2}{4 F_0^2}-\frac{3 g_4^2}{16 F_0^2}$\\
\Xhline{1.2pt}
\end{tabular}
\end{table}

\begin{table}[!h]
  \centering
  \renewcommand{\arraystretch}{1.5}
  \caption{The coefficients $\lambda_{5(ij)}$ of the axial current from the wave function renormalization.}\label{wav-5}
\begin{tabular}{c|cccc}
\Xhline{1.2pt}
&&Case I&Case II\\
\hline\hline
\multirow{3}{*}{$\lambda_{{\Xi_c^*}'^+{\Xi_c^*}'^0}$}
    &$K\textrm{-loop}$&$\frac{5 g_5^2}{2 F_0^2}$&$-\frac{5 g_3^2}{8 F_0^2}-\frac{g_4^2}{4 F_0^2}$\\
    &$\eta\textrm{-loop}$&$\frac{g_5^2}{12 F_0^2}$&$-\frac{g_3^2}{48 F_0^2}-\frac{3 g_4^2}{8 F_0^2}$\\
    &$\pi\textrm{-loop}$&$\frac{3 g_5^2}{4 F_0^2}$&$-\frac{3 g_3^2}{16 F_0^2}-\frac{3 g_4^2}{8 F_0^2}$\\
\hline
\multirow{3}{*}{$\lambda_{\Sigma_c^{*+}\Sigma_c^{*0}}$}
    &$K\textrm{-loop}$&$\frac{g_5^2}{F_0^2}$&$-\frac{g_3^2}{4 F_0^2}-\frac{g_4^2}{2 F_0^2}$\\
    &$\eta\textrm{-loop}$&$\frac{g_5^2}{3 F_0^2}$&$-\frac{g_3^2}{12 F_0^2}$\\
    &$\pi\textrm{-loop}$&$\frac{2 g_5^2}{F_0^2}$&$-\frac{g_3^2}{2 F_0^2}-\frac{g_4^2}{2 F_0^2}$\\
\hline
\multirow{3}{*}{$\lambda_{{\Xi_c^*}'^+\Omega_c^{*0}}$}
    &$K\textrm{-loop}$&$\frac{9 g_5^2}{4 F_0^2}$&$-\frac{9 g_3^2}{16 F_0^2}-\frac{5 g_4^2}{8 F_0^2}$\\
    &$\eta\textrm{-loop}$&$\frac{17 g_5^2}{24 F_0^2}$&$-\frac{17 g_3^2}{96 F_0^2}-\frac{3 g_4^2}{16 F_0^2}$\\
    &$\pi\textrm{-loop}$&$\frac{3 g_5^2}{8 F_0^2}$&$-\frac{3 g_3^2}{32 F_0^2}-\frac{3 g_4^2}{16 F_0^2}$\\
\hline
\multirow{3}{*}{$\lambda_{\Sigma_c^{^*+}{\Xi_c^*}'^0}$}
    &$K\textrm{-loop}$&$\frac{7 g_5^2}{4 F_0^2}$&$-\frac{7 g_3^2}{16 F_0^2}-\frac{3 g_4^2}{8 F_0^2}$\\
    &$\eta\textrm{-loop}$&$\frac{5 g_5^2}{24 F_0^2}$&$-\frac{5 g_3^2}{96 F_0^2}-\frac{3 g_4^2}{16 F_0^2}$\\
    &$\pi\textrm{-loop}$&$\frac{11 g_5^2}{8 F_0^2}$&$-\frac{11 g_3^2}{32 F_0^2}-\frac{7 g_4^2}{16 F_0^2}$\\
\hline
\multirow{3}{*}{$\lambda_{\Sigma_c^{*++}{\Xi_c^*}'^+}$}
    &$K\textrm{-loop}$&$\frac{7 g_5^2}{4 F_0^2}$&$-\frac{7 g_3^2}{16 F_0^2}-\frac{3 g_4^2}{8 F_0^2}$\\
    &$\eta\textrm{-loop}$&$\frac{5 g_5^2}{24 F_0^2}$&$-\frac{5 g_3^2}{96 F_0^2}-\frac{3 g_4^2}{16 F_0^2}$\\
    &$\pi\textrm{-loop}$&$\frac{11 g_5^2}{8 F_0^2}$&$-\frac{11 g_3^2}{32 F_0^2}-\frac{7 g_4^2}{16 F_0^2}$\\
\Xhline{1.2pt}
\end{tabular}
\end{table}

\begin{table}[!h]
  \centering
  \renewcommand{\arraystretch}{1.5}
  \caption{The coefficients $\lambda_{3(ij)}$ of the axial current from the wave function renormalization.}\label{wav-3}
\begin{tabular}{c|ccc}
\Xhline{1.2pt}
&&Case I&Case II\\
\hline\hline
\multirow{3}{*}{$\lambda_{\Xi_c'^+{\Xi_c^*}'^0}$}
    &$K\textrm{-loop}$&$\frac{5 g_1^2}{4 F_0^2}+\frac{g_2^2}{2 F_0^2}+\frac{5 g_5^2}{4 F_0^2}$&$-\frac{5 g_3^2}{8 F_0^2}-\frac{g_4^2}{8 F_0^2}$\\
    &$\eta\textrm{-loop}$&$\frac{g_1^2}{24 F_0^2}+\frac{3 g_2^2}{4 F_0^2}+\frac{g_5^2}{24 F_0^2}$&$-\frac{g_3^2}{48 F_0^2}-\frac{3 g_4^2}{16 F_0^2}$\\
    &$\pi\textrm{-loop}$&$\frac{3 g_1^2}{8 F_0^2}+\frac{3 g_2^2}{4 F_0^2}+\frac{3 g_5^2}{8 F_0^2}$&$-\frac{3 g_3^2}{16 F_0^2}-\frac{3 g_4^2}{16 F_0^2}$\\
\hline
\multirow{3}{*}{$\lambda_{\Sigma_c^+\Sigma_c^{*0}}$}
    &$K\textrm{-loop}$&$\frac{g_1^2}{2 F_0^2}+\frac{g_2^2}{F_0^2}+\frac{g_5^2}{2 F_0^2}$&$-\frac{g_3^2}{4 F_0^2}-\frac{g_4^2}{4 F_0^2}$\\
    &$\eta\textrm{-loop}$&$\frac{g_1^2}{6 F_0^2}+\frac{g_5^2}{6 F_0^2}$&$-\frac{g_3^2}{12 F_0^2}$\\
    &$\pi\textrm{-loop}$&$\frac{g_1^2}{F_0^2}+\frac{g_2^2}{F_0^2}+\frac{g_5^2}{F_0^2}$&$-\frac{g_3^2}{2 F_0^2}-\frac{g_4^2}{4 F_0^2}$\\
\hline
\multirow{3}{*}{$\lambda_{\Sigma_c^{++}\Sigma_c^{*+}}$}
    &$K\textrm{-loop}$&$\frac{g_1^2}{2 F_0^2}+\frac{g_2^2}{F_0^2}+\frac{g_5^2}{2 F_0^2}$&$-\frac{g_3^2}{4 F_0^2}-\frac{g_4^2}{4 F_0^2}$\\
    &$\eta\textrm{-loop}$&$\frac{g_1^2}{6 F_0^2}+\frac{g_5^2}{6 F_0^2}$&$-\frac{g_3^2}{12 F_0^2}$\\
    &$\pi\textrm{-loop}$&$\frac{g_1^2}{F_0^2}+\frac{g_2^2}{F_0^2}+\frac{g_5^2}{F_0^2}$&$-\frac{g_3^2}{2 F_0^2}-\frac{g_4^2}{4 F_0^2}$\\
\hline
\multirow{3}{*}{$\lambda_{{\Xi_c}'^+\Omega_c^{*0}}$}
    &$K\textrm{-loop}$&$\frac{5 g_1^2}{4 F_0^2}+\frac{g_2^2}{2 F_0^2}+\frac{g_5^2}{F_0^2}$&$-\frac{9 g_3^2}{16 F_0^2}-\frac{g_4^2}{2 F_0^2}$\\
    &$\eta\textrm{-loop}$&$\frac{g_1^2}{24 F_0^2}+\frac{3 g_2^2}{4 F_0^2}+\frac{2 g_5^2}{3 F_0^2}$&$-\frac{17 g_3^2}{96 F_0^2}$\\
    &$\pi\textrm{-loop}$&$\frac{3 g_1^2}{8 F_0^2}+\frac{3 g_2^2}{4 F_0^2}$&$-\frac{3 g_3^2}{32 F_0^2}$\\
\hline
\multirow{3}{*}{$\lambda_{\Sigma_c^+{\Xi_c^*}'^+}$}
    &$K\textrm{-loop}$&$\frac{g_1^2}{2 F_0^2}+\frac{g_2^2}{F_0^2}+\frac{5 g_5^2}{4 F_0^2}$&$-\frac{7 g_3^2}{16 F_0^2}-\frac{g_4^2}{8 F_0^2}$\\
    &$\eta\textrm{-loop}$&$\frac{g_1^2}{6 F_0^2}+\frac{g_5^2}{24 F_0^2}$&$-\frac{5 g_3^2}{96 F_0^2}-\frac{3 g_4^2}{16 F_0^2}$\\
    &$\pi\textrm{-loop}$&$\frac{g_1^2}{F_0^2}+\frac{g_2^2}{F_0^2}+\frac{3 g_5^2}{8 F_0^2}$&$-\frac{11 g_3^2}{32 F_0^2}-\frac{3 g_4^2}{16 F_0^2}$\\
\hline
\multirow{3}{*}{$\lambda_{\Sigma_c^{++}{\Xi_c^*}'^+}$}
    &$K\textrm{-loop}$&$\frac{g_1^2}{2 F_0^2}+\frac{g_2^2}{F_0^2}+\frac{5 g_5^2}{4 F_0^2}$&$-\frac{7 g_3^2}{16 F_0^2}-\frac{g_4^2}{8 F_0^2}$\\
    &$\eta\textrm{-loop}$&$\frac{g_1^2}{6 F_0^2}+\frac{g_5^2}{24 F_0^2}$&$-\frac{5 g_3^2}{96 F_0^2}-\frac{3 g_4^2}{16 F_0^2}$\\
    &$\pi\textrm{-loop}$&$\frac{g_1^2}{F_0^2}+\frac{g_2^2}{F_0^2}+\frac{3 g_5^2}{8 F_0^2}$&$-\frac{11 g_3^2}{32 F_0^2}-\frac{3 g_4^2}{16 F_0^2}$\\
\hline
\multirow{3}{*}{$\lambda_{{\Xi_c^*}'^+\Omega_c^0}$}
    &$K\textrm{-loop}$&$\frac{g_1^2}{F_0^2}+\frac{2 g_2^2}{F_0^2}+\frac{5 g_5^2}{4 F_0^2}$&$-\frac{9 g_3^2}{16 F_0^2}-\frac{g_4^2}{8 F_0^2}$\\
    &$\eta\textrm{-loop}$&$\frac{2 g_1^2}{3 F_0^2}+\frac{g_5^2}{24 F_0^2}$&$-\frac{17 g_3^2}{96 F_0^2}-\frac{3 g_4^2}{16 F_0^2}$\\
    &$\pi\textrm{-loop}$&$\frac{3 g_5^2}{8 F_0^2}$&$-\frac{3 g_3^2}{32 F_0^2}-\frac{3 g_4^2}{16 F_0^2}$\\
\hline
\multirow{3}{*}{$\lambda_{\Sigma_c^{*+}\Xi_c'^+}$}
    &$K\textrm{-loop}$&$\frac{5 g_1^2}{4 F_0^2}+\frac{g_2^2}{2 F_0^2}+\frac{g_5^2}{2 F_0^2}$&$-\frac{7 g_3^2}{16 F_0^2}-\frac{g_4^2}{4 F_0^2}$\\
    &$\eta\textrm{-loop}$&$\frac{g_1^2}{24 F_0^2}+\frac{3 g_2^2}{4 F_0^2}+\frac{g_5^2}{6 F_0^2}$&$-\frac{5 g_3^2}{96 F_0^2}$\\
    &$\pi\textrm{-loop}$&$\frac{3 g_1^2}{8 F_0^2}+\frac{3 g_2^2}{4 F_0^2}+\frac{g_5^2}{F_0^2}$&$-\frac{11 g_3^2}{32 F_0^2}-\frac{g_4^2}{4 F_0^2}$\\
\hline
\multirow{3}{*}{$\lambda_{\Sigma_c^{*++}\Xi_c'^+}$}
    &$K\textrm{-loop}$&$\frac{5 g_1^2}{4 F_0^2}+\frac{g_2^2}{2 F_0^2}+\frac{g_5^2}{2 F_0^2}$&$-\frac{7 g_3^2}{16 F_0^2}-\frac{g_4^2}{4 F_0^2}$\\
    &$\eta\textrm{-loop}$&$\frac{g_1^2}{24 F_0^2}+\frac{3 g_2^2}{4 F_0^2}+\frac{g_5^2}{6 F_0^2}$&$-\frac{5 g_3^2}{96 F_0^2}$\\
    &$\pi\textrm{-loop}$&$\frac{3 g_1^2}{8 F_0^2}+\frac{3 g_2^2}{4 F_0^2}+\frac{g_5^2}{F_0^2}$&$-\frac{11 g_3^2}{32 F_0^2}-\frac{g_4^2}{4 F_0^2}$\\
\Xhline{1.2pt}
\end{tabular}
\end{table}

\begin{table}[!h]
  \centering
  \renewcommand{\arraystretch}{1.5}
  \caption{The coefficients $\lambda_{4(ij)}$ of the axial current from the wave function renormalization.}\label{wav-4}
\begin{tabular}{c|cccc}
\Xhline{1.2pt}
&&Case I&Case II\\
\hline\hline
\multirow{3}{*}{$\lambda_{\Lambda_c^+\Sigma_c^{*0}}$}
    &$K\textrm{-loop}$&$\frac{g_2^2}{F_0^2}+\frac{g_5^2}{2 F_0^2}+\frac{2 g_6^2}{F_0^2}$&$-\frac{g_3^2}{8 F_0^2}-\frac{g_4^2}{2 F_0^2}$\\
    &$\eta\textrm{-loop}$&$\frac{g_5^2}{6 F_0^2}+\frac{2 g_6^2}{3 F_0^2}$&$-\frac{g_3^2}{24 F_0^2}$\\
    &$\pi\textrm{-loop}$&$\frac{3 g_2^2}{F_0^2}+\frac{g_5^2}{F_0^2}$&$-\frac{g_3^2}{4 F_0^2}-\frac{g_4^2}{F_0^2}$\\
\hline
\multirow{3}{*}{$\lambda_{\Xi_c^+{\Xi_c^*}'^0}$}
    &$K\textrm{-loop}$&$\frac{5 g_2^2}{2 F_0^2}+\frac{5 g_5^2}{4 F_0^2}+\frac{g_6^2}{F_0^2}$&$-\frac{5 g_3^2}{16 F_0^2}-\frac{3 g_4^2}{4 F_0^2}$\\
    &$\eta\textrm{-loop}$&$\frac{3 g_2^2}{4 F_0^2}+\frac{g_5^2}{24 F_0^2}+\frac{g_6^2}{6 F_0^2}$&$-\frac{g_3^2}{96 F_0^2}-\frac{3 g_4^2}{8 F_0^2}$\\
    &$\pi\textrm{-loop}$&$\frac{3 g_2^2}{4 F_0^2}+\frac{3 g_5^2}{8 F_0^2}+\frac{3 g_6^2}{2 F_0^2}$&$-\frac{3 g_3^2}{32 F_0^2}-\frac{3 g_4^2}{8 F_0^2}$\\
\hline
\multirow{3}{*}{$\lambda_{\Lambda_c^+{\Xi_c^*}'^0}$}
    &$K\textrm{-loop}$&$\frac{g_2^2}{F_0^2}+\frac{5 g_5^2}{4 F_0^2}+\frac{2 g_6^2}{F_0^2}$&$-\frac{5 g_3^2}{16 F_0^2}-\frac{3 g_4^2}{8 F_0^2}$\\
    &$\eta\textrm{-loop}$&$\frac{g_5^2}{24 F_0^2}+\frac{2 g_6^2}{3 F_0^2}$&$-\frac{g_3^2}{96 F_0^2}-\frac{3 g_4^2}{16 F_0^2}$\\
    &$\pi\textrm{-loop}$&$\frac{3 g_2^2}{F_0^2}+\frac{3 g_5^2}{8 F_0^2}$&$-\frac{3 g_3^2}{32 F_0^2}-\frac{15 g_4^2}{16 F_0^2}$\\
\hline
\multirow{3}{*}{$\lambda_{\Xi_c^+\Omega_c^{*0}}$}
    &$K\textrm{-loop}$&$\frac{5 g_2^2}{2 F_0^2}+\frac{g_5^2}{F_0^2}+\frac{g_6^2}{F_0^2}$&$-\frac{g_3^2}{4 F_0^2}-\frac{9 g_4^2}{8 F_0^2}$\\
    &$\eta\textrm{-loop}$&$\frac{3 g_2^2}{4 F_0^2}+\frac{2 g_5^2}{3 F_0^2}+\frac{g_6^2}{6 F_0^2}$&$-\frac{g_3^2}{6 F_0^2}-\frac{3 g_4^2}{16 F_0^2}$\\
    &$\pi\textrm{-loop}$&$\frac{3 g_2^2}{4 F_0^2}+\frac{3 g_6^2}{2 F_0^2}$&$-\frac{3 g_4^2}{16 F_0^2}$\\
\Xhline{1.2pt}
\end{tabular}
\end{table}

\clearpage

\subsection{INTEGRALS AND FUNCTIONS}\label{sec6.3}

\noindent1. The integral with one meson line and one baryon line in Fig. \ref{Figure-self}:

\begin{eqnarray}\label{eq33}
\left\{
\begin{aligned}
&J(m,\omega)=\frac{1}{8\pi^2}\left[\omega(R-1)+\omega\ln\frac{m^2}{\mu^2}+K\right],\omega=v\cdot k+\delta\\
&I(m)=\frac{m^2}{16\pi^2}\left(R+\ln\frac{m^2}{\mu^2}\right)\\
\end{aligned}
\right.
\end{eqnarray}

and
\[K=
\left\{
\begin{aligned}
&2\sqrt{\omega^2-m^2}\textrm{arccosh}\frac{\omega}{m}-2i\pi\sqrt{\omega^2-m^2},\quad\omega>m\\
&-2\sqrt{\omega^2-m^2}\textrm{arccosh}\frac{-\omega}{m},\quad\omega<-m\\
&2\sqrt{m^2-\omega^2}\arccos\frac{-\omega}{m},\quad\omega^2<m^2\\
\end{aligned}
\right.
\]

\[J_{\alpha\beta}=C_{21}g_{\alpha\beta}+C_{20}v_\alpha v_\beta\]
where
\[C_{21}=\frac{1}{d-1}[(m^2-\omega^2)J(m,\omega)+\omega I(m)]=\frac{1}{d-1}f(m,\omega)\]
The definition of $f$ can be read from above easily.

\noindent2. The integrals with one meson line and two baryon lines
in Figs. \ref{fig-g1}--\ref{fig-g4}:

When the masses of the two baryons in diagram (a) are the same, we
introduce the integrals
\begin{eqnarray}\label{eq34}
\{L,L_\mu,L_{\mu\nu}\}&=&\frac{1}{i}\int
d^dq\frac{\{1,q_\mu,q_\mu q_\nu\}}{(m^2-q^2-i\varepsilon)[v\cdot
q+\omega+i\varepsilon]^2}
\end{eqnarray}
using
\[\frac{1}{[v\cdot q+\omega]^2}=-\frac{\partial}{\partial\omega}\frac{1}{[v\cdot q+\omega]}\]
There is a relation between $L$ and $J$
\begin{eqnarray}\label{eq35}
\{L,L_\mu,L_{\mu\nu}\}&=&-\frac{\partial}{\partial\alpha}\{J,J_\mu,J_{\mu\nu}\}
\end{eqnarray}
When the masses of the two baryons in diagram (a) are different, we
define the integrals
\begin{eqnarray}\label{eq36}
\{F,F_\mu,F_{\mu\nu}\}&=&\frac{1}{i}\int
d^dq\frac{\{1,q_\mu,q_\mu q_\nu\}}{(m^2-q^2-i\varepsilon)[v\cdot
q+\omega_1+i\varepsilon][v\cdot q+\omega_2+i\varepsilon]}
\end{eqnarray}
using
\[\frac{1}{[v\cdot q+\omega_1][v\cdot q+\omega_2]}=-\frac{1}{\omega_1-\omega_2}(\frac{1}{[v\cdot q+\omega_1]}-\frac{1}{[v\cdot q+\omega_2]})\]
The relation between $F$ and $J$ is
\begin{eqnarray}\label{eq37}
\{F,F_\mu,F_{\mu\nu}\}&=&-\frac{1}{\omega_1-\omega_2}\{J(\omega_1)-J(\omega_2),J_\mu(\omega_1)-J_\mu(\omega_2),J_{\mu\nu}(\omega_1)-J_{\mu\nu}(\omega_2)\}
\end{eqnarray}

Especially, for the second-order tensor formula, $F_{\alpha\beta}$
and $L_{\alpha\beta}$ can be expressed as a sum of the two Lorentz
structures. $F_{\alpha\beta}^{20}$ and $L_{\alpha\beta}^{20}$ are
proportional to $v^\alpha v^\beta$ and vanish when contracted with
$S^\mu$ and $\mathcal{T}^\mu$. So, we are concerned about the remaining part
only:
\[F_{\alpha\beta}=-\frac{1}{\omega_1-\omega_2}(J_{\alpha\beta}(\omega_1)-J_{\alpha\beta}(\omega_2))=F_{\alpha\beta}^{21}+F_{\alpha\beta}^{20}\]
\begin{eqnarray}\label{eq38}
F_{\alpha\beta}^{21}&=&g_{\alpha\beta}\frac{-1}{d-1}\frac{f(m,\omega_1)-f(m,\omega_2)}{\omega_1-\omega_2}
\end{eqnarray}
\[L_{\alpha\beta}=-\frac{\partial}{\partial\omega}J_{\alpha\beta}(\omega)=L_{\alpha\beta}^{21}+L_{\alpha\beta}^{20}\]
\begin{eqnarray}\label{eq39}
L_{\alpha\beta}^{21}&=&g_{\alpha\beta}\frac{-1}{d-1}\frac{\partial
f(m,\omega)}{\partial\omega}
\end{eqnarray}

$F_{\alpha\beta}^{21}$ and $L_{\alpha\beta}^{21}$ can be uniformed as
\[g_{\alpha\beta}\frac{-1}{d-1}\frac{\Delta f}{\Delta\omega}\]
where
\[
\frac{\Delta f}{\Delta\omega}= \left\{
\begin{aligned}
&m^2\frac{\Delta J(\omega)}{\Delta\omega}-\frac{\Delta(\omega^2J(\omega))}{\Delta\omega}+I(m),&\omega_1\neq\omega_2\\
&(m^2-\omega^2)\frac{\partial J(\omega)}{\partial\omega}-2\omega J(\omega)+I(m),&\omega_1=\omega_2=\omega\\
\end{aligned}
\right.
\]
and $\frac{\Delta J(\omega)}{\Delta\omega}$ denotes
$\frac{J(\omega_1)-J(\omega_2)}{\omega_1-\omega_2}$, and similar
conventions hold for
$\frac{\Delta(\omega^2J(\omega))}{\Delta\omega}$ and
$\frac{\Delta(\omega^3)}{\Delta\omega}$. Combining with the
parameters $a$ and $b$, the integral from diagram (a) can be written
as
\begin{eqnarray*}
(a+b\epsilon)\frac{\Delta f}{\Delta\omega}&=&a\left\{
\begin{aligned}
&m^2\frac{\Delta J}{\Delta\omega}-\frac{\Delta(\omega^2J)}{\Delta\omega}+I,&(\omega_1\neq\omega_2)\\
&(m^2-\omega^2)\frac{\partial J}{\partial\omega}-2\omega J(\omega)+I,&(\omega_1=\omega_2=\omega)\\
\end{aligned}
\right\}-\frac{b}{4\pi^2}\left[\frac{3}{2}m^2-\frac{\Delta(\omega^3)}{\Delta\omega}\right]
\end{eqnarray*}

\subsection{THE CONTRACTION FORMULAS FOR $S^\mu$ and $P^{\frac{3}{2}}_{(33)\mu\nu}$}\label{sec6.4}
The Pauli-Lubanski vector $S^\mu$ and projection operator
$P^{\frac{3}{2}}_{(33)\mu\nu}$ are defined as follows:
\begin{eqnarray}\label{eq40}
S^\mu&=&-\frac{1}{2}\gamma_5(\gamma^\mu\slashed v-v^\mu)\nonumber\\
P^{\frac{3}{2}}_{(33)\mu\nu}&=&g_{\mu\nu}-v_\mu v_\nu+\frac{4}{d-1}S_\mu S_\nu
\end{eqnarray}
In the calculation of the loop correction of the self-energy
function and axial charges, the following formulas are very useful.
{\allowdisplaybreaks
\begin{eqnarray}\label{eq41-50}
P^{\rho\sigma}g_{\rho\sigma}&=&d-2\\
S^\sigma P^{\rho\mu}&=&\frac{4}{d-1}S^\rho S^\mu S^\sigma+g^{\rho\mu}_\perp S^\sigma+\frac{2}{d-1}S^\rho g^{\mu\sigma}_\perp-\frac{2}{d-1}g^{\rho\sigma}_\perp S^\mu\\
P^{\mu\sigma}S^\rho&=&\frac{4}{d-1}S^\rho S^\mu S^\sigma+S^\rho g^{\mu\sigma}_\perp+\frac{2}{d-1}g^{\rho\mu}_\perp S^\sigma-\frac{2}{d-1}g^{\rho\sigma}_\perp S^\mu\\
S_\alpha P^{\mu\alpha}&=&\frac{2(d-2)}{d-1}S^\mu\\
S^\alpha P^{\rho\sigma}S_\alpha&=&\left(\frac{1-d}{4}+\frac{1}{d-1}\right)g^{\rho\sigma}_\perp+\frac{5-d}{d-1}S^\rho S^\sigma\\
P^{\rho\lambda}S^\mu{P_\lambda}^\sigma&=&-\frac{4(d+1)}{(d-1)^2}S^\rho S^\mu S^\sigma+S^\mu g^{\rho\sigma}_\perp-\frac{2}{d-1}(g^{\rho\mu}_\perp S^\sigma+S^\rho g^{\mu\sigma}_\perp)\\
P^{\rho\lambda}S^\mu P_{\lambda\rho}&=&\frac{(d-3)(d-2)(d+1)}{(d-1)^2}S^\mu\\
P^{\alpha\lambda}S^\mu{P_\lambda}^\sigma S_\alpha&=&\frac{2(d-3)(d+1)}{(d-1)^2}S^\mu S^\sigma+\frac{(d-3)(d+1)}{2(d-1)^2}g^{\mu\sigma}_\perp\\
S^\alpha P^{\rho\lambda}S^\mu{P_\lambda}^\sigma S_\alpha&=&-\frac{(d+1)(d-7)}{(d-1)^2}S^\rho S^\mu S^\sigma-\frac{d^2-6d+1}{2(d-1)^2}(S^\rho g^{\mu\sigma}_\perp+g^{\rho\mu}_\perp S^\sigma)\\
&&+\frac{d^3-5d^2+3d-7}{4(d-1)^2}S^\mu g^{\rho\sigma}_\perp
\end{eqnarray}
}

\end{document}